\RequirePackage{ifpdf}
\ifpdf 
\documentclass[pdftex]{sigma}
\else
\documentclass{sigma}
\fi

\numberwithin{equation}{section}

\begin{document}

\allowdisplaybreaks

\renewcommand{\PaperNumber}{021}

\FirstPageHeading

\renewcommand{\thefootnote}{$\star$}

\ShortArticleName{On Parametrization of $GL(4,C)$ and $SU(4)$}

\ArticleName{On Parametrization of the Linear $\boldsymbol{GL(4,C)}$ and\\
Unitary $\boldsymbol{SU(4)}$ Groups in Terms of Dirac
Matrices\footnote{This paper is a contribution to the Proceedings
of the Seventh International Conference ``Symmetry in Nonlinear
Mathematical Physics'' (June 24--30, 2007, Kyiv, Ukraine). The
full collection is available at
\href{http://www.emis.de/journals/SIGMA/symmetry2007.html}{http://www.emis.de/journals/SIGMA/symmetry2007.html}}}

\Author{Victor M. RED'KOV, Andrei A. BOGUSH and Natalia G.
TOKAREVSKAYA}

\AuthorNameForHeading{V.M. Red'kov, A.A. Bogush and N.G.
Tokarevskaya}

\Address{B.I.~Stepanov Institute of Physics, National Academy of Sciences of Belarus, Minsk, Belarus}

\Email{\href{mailto:redkov@dragon.bas-net.by}{redkov@dragon.bas-net.by},
\href{mailto:bogush@dragon.bas-net.by}{bogush@dragon.bas-net.by},
\href{mailto:tokarev@dragon.bas-net.by}{tokarev@dragon.bas-net.by}}

\ArticleDates{Received September 19, 2007, in f\/inal form January
24, 2008; Published online February 19, 2008}

\Abstract{Parametrization of $4 \times 4$-matrices $G$ of the
complex linear group $GL(4,C)$ in terms of four complex 4-vector
parameters $(k,m,n,l)$ is investigated. Additional restrictions
separating some subgroups of $GL(4,C)$ are given explicitly. In
the given parametrization, the problem of inverting any $4\times
4$ matrix $G$ is solved. Expression for determinant of any matrix
$G$ is found: $\det G = F(k,m,n,l)$.
 Unitarity conditions $G^{+} = G^{-1}$ have been formulated in the form of
non-linear cubic algebraic equations including complex
conjugation. Several simplest solutions of these unitarity
equations have been found:
 three 2-parametric subgroups $G_{1}$, $G_{2}$, $G_{3}$ --
each of subgroups consists of two commuting Abelian unitary
groups; 4-parametric unitary subgroup consisting of a product of a
3-parametric group isomorphic $SU(2)$ and 1-pa\-rametric Abelian
group. The Dirac basis of generators $\Lambda_{k}$, being of
Gell-Mann type, substantially dif\/fers from the basis
 $\lambda_{i}$ used in the literature on $SU(4)$ group, formulas relating them are found --
 they permit to separate $SU(3)$ subgroup in $SU(4)$.
Special way to list 15 Dirac generators of $GL(4,C)$ can be used
$\{ \Lambda_{k} \} = \{ \alpha_{i} \oplus \beta_{j} \oplus (
\alpha_{i} V \beta_{j} =
 {\boldsymbol K} \oplus {\boldsymbol L} \oplus {\boldsymbol M} ) \}$, which permit to factorize $SU(4)$ transformations
 according
 to
 $S =
e^ {i \vec{a} \vec{\alpha} } e^ {i \vec{b} \vec{\beta} } e^{i
{\boldsymbol k} {\boldsymbol K} } e^{i {\boldsymbol l}
{\boldsymbol L} } e^{i {\boldsymbol m}
 {\boldsymbol M} } $, where two f\/irst factors commute with each
other and are isomorphic to $SU(2)$ group, the three last ones are
3-parametric groups, each of them consisting of three Abelian
commuting unitary subgroups. Besides, the structure of f\/ifteen
Dirac matrices $\Lambda_{k}$ permits to separate twenty
3-parametric subgroups in $SU(4)$ isomorphic to $SU(2)$; those
subgroups might be used as bigger elementary blocks in
constructing of a general
 transformation $SU(4)$.
It is shown how one can specify the present approach for the
pseudounitary group $SU(2,2)$ and~$SU(3,1)$.}


\Classification{20C35; 20G45; 22E70; 81R05}

\renewcommand{\thefootnote}{\arabic{footnote}}
\setcounter{footnote}{0}

\section{Introduction}

 The unitary groups play an important
 role in numerous research areas: quantum theory
 of massless particles, cosmology models, quantum systems with dynamical symmetry,
 nano-scale physics, numerical calculations concerning entanglement and other quantum
 information parameters, high-energy particle theory~-- let us just specify these
 several points:

\begin{itemize}\itemsep=0pt
\item $SU(2,2)$ and conformal symmetry, massless particles
\cite{barut-1971,Bracken-2005,Bracken-75,keane-1999,vlasov-2003}; \item
classical Yang--Mills equations and gauge f\/ields
\cite{Sanchez-Monroy-2006};

\item quantum computation and control, density matrices for
entangled states \cite{Barenco-1995,Deutsch-1989,Schirmer-2002};

\item geometric phases and invariants for multi-level quantum
systems \cite{Mukunda-2001};

\item high-temperature superconductivity and antiferromagnets
\cite{Guidry-2000,Mishra-2002};

\item composite structure of quarks and leptons
\cite{Terazawa-1977,Terazawa-1980,Marchuk-1998};

\item $SU(4)$ gauge models \cite{Volovik-2003,Dahiya-2003};

\item classif\/ication of hadrons and their interactions
\cite{Dahm-1996,Gonzalez-2006,Chaturvedi-2000}.

\end{itemize}

Because of so many applications in physics, various
parametrizations for the group elements of unitary group $SU(4)$
and related to it deserve special attention.
 Our ef\/forts will be given to extending some classical technical approaches proving their ef\/fectiveness
 in simple cases of the linear and unitary groups $SL(2,C)$ and $SU(2)$, so that we will
work with objects known by every physicist, such as Pauli and
Dirac matrices. This paper, written for physicists, is
self-contained in that it does not require any previous
 knowledge of the subject nor any advanced mathematics.

Let us start with the known example of spinor covering for complex
Lorentz group: consider the 8-parametric $4\times 4$ matrices in
the quasi diagonal form \cite{Bogush-2006,Fedorov, Macfarlane-1966} 
\begin{gather*}
G = \left | \begin{array}{cc}
k_{0} + {\boldsymbol k} \vec{\sigma} & 0 \\[3mm]
0 & m_{0} - {\boldsymbol m} \vec{\sigma}
\end{array} \right | .
\end{gather*}
 The composition rules for parameters $k=(k_{0},{\boldsymbol
k})$ and $m=(m_{0},{\boldsymbol m})$ are
 \begin{gather*}
k_{0}'' = k_{0}' k_{0} + {\boldsymbol k}' {\boldsymbol k} , \qquad
{\boldsymbol k} '' = k_{0}' {\boldsymbol k} + {\boldsymbol k}'
k_{0} + i {\boldsymbol k}' \times {\boldsymbol k} , \nonumber
\\
m_{0}'' = m_{0}' m_{0} + {\boldsymbol m}' {\boldsymbol m} , \qquad
{\boldsymbol m} '' = m_{0}' {\boldsymbol m} + {\boldsymbol m}'
m_{0} - i {\boldsymbol m}' \times
{\boldsymbol m} . 
\end{gather*}
With two additional constraints on 8 quantities $ k_{0}^{2} -
{\boldsymbol k}^{2} = +1$, $m_{0}^{2} - {\boldsymbol m}^{2} = +1$,
 we will arrive at a~def\/inite way to parameterize
a double (spinor) covering for complex Lorentz group~$SO(4,C)$. At
this, the problem of inverting of the $G$ matrices with unit
determinant $\det G $ is
 solved straightforwardly:
$G=G(k_{0}, {\boldsymbol k}, m_{0},{\boldsymbol m})$,
$G^{-1}=G(k_{0}, -{\boldsymbol k}, m_{0},-{\boldsymbol m})$.
 Transition from covering 4-spinor transformations to
4-vector ones is performed through the known relationship $
G\gamma^{a}G^{-1} = \gamma^{c}L_{c}^{\;\;a} $
 which determine $2 \Longrightarrow 1$ map from~$\pm G$
to~$L$.

There exists a direct connection between the above 4-dimensional
vector parametrization of the spinor group $G (k_{a},m_{a})$ and
 the Fedorov's parametrization~\cite{Fedorov}
 of the group of complex orthogonal Lorentz transformations
 in terms of 3-dimensional vectors
${\boldsymbol Q} = {\boldsymbol k} / k_{0} $, ${\boldsymbol M} =
 {\boldsymbol m} / m_{0}$, with the simple composition rules for vector
parameters
\begin{gather*}
 {\boldsymbol Q} '' = { {\boldsymbol Q} + {\boldsymbol Q}' + i {\boldsymbol Q}'
\times {\boldsymbol Q} \over 1 + {\boldsymbol Q}' {\boldsymbol Q}}
,\qquad
 {\boldsymbol M} '' = { {\boldsymbol M} + {\boldsymbol M}' - i {\boldsymbol M}' \times
{\boldsymbol M} \over 1 + {\boldsymbol M}' {\boldsymbol M} } . 
\end{gather*}

Evidently, the pair $({\boldsymbol Q}, {\boldsymbol M})$ provides
us with possibility to parameterize correctly orthogonal matrices
only. Instead, the $(k_{a},m_{a})$ represent correct parameters
for the spinor covering group. When we are interested only in
local properties of the spinor representations, no substantial
dif\/ferences between orthogonal groups and their spinor coverings
exist. However, in opposite cases global dif\/ference between
orthogonal and spinor groups may be very substantial as well as
correct parametrization of them.

Restrictions specifying the spinor coverings for
 orthogonal subgroups are well known~\cite{Fedorov}. In particular,
restriction to real Lorentz group $O(3,1)$ is achieved by imposing
one condition
 (including complex conjugation)
$ (k,m) \Longrightarrow (k, k^{*}) $. The case of real orthogonal
group $O(4)$ is achieved by a formal change (transition to real
parameters) $(k_{0}, {\boldsymbol k}) \Longrightarrow (k_{0}, i
{\boldsymbol k})$, $(m_{0}, {\boldsymbol m}) \Longrightarrow
(m_{0}, i {\boldsymbol m}) $,
 and the real orthogonal group $O(2,2)$ corresponds to
transition to real parameters according to $(k_{0}, k_{1}, k_{2},
k_{3} ) \Longrightarrow (k_{0}, k_{1}, k_{2}, ik_{3} )$, $(m_{0},
m_{1}, m_{2}, m_{3} ) \Longrightarrow
 (m_{0}, m_{1}, m_{2}, im_{3} ) $.

To parameterize 4-spinor and 4-vector transformations of the
complex Lorentz
 group one may use curvilinear coordinates.
The simplest and widely used ones are Euler's complex angles
(see~\cite{Fedorov} and references in~\cite{Bogush-2006}). In
general, on the basis of the analysis given by Olevskiy~\cite{3}
about coordinates in the real
 Lobachevski space, one can
 propose 34 dif\/ferent complex coordinate systems appropriate to
 parameterize the complex Lorentz group and its double
 covering.

 A particular, Euler angle parametrization is closely connected with cylindrical coordinates on the complex
 3-sphere, one of 34 possible coordinates.
Such complex cylindrical coordinates can be introduced by the
following relations \cite{Bogush-2006}:
\begin{gather*}
k_{0} = \cos \rho \cos z , \qquad k_{3} = i \cos \rho \sin z
 , \qquad
 k_{1} = i \sin \rho \cos \phi , \qquad k _{2} = i \sin \rho
\sin \phi , \nonumber
\\
m_{0} = \cos R \cos Z , \qquad m_{3} = i \cos R \sin Z , \qquad
m_{1} = i \sin R \Phi , \qquad m _{2} = i \sin R \sin \Phi
 . 
\end{gather*}
Here 6 complex variables are independent, $ (\rho, z,\phi)$, $(R,
Z ,\Phi) $, additional restrictions are satisf\/ied identically by
def\/inition. Instead of cylindrical coordinates in $(\rho, z,
\phi)$ and $(R, Z ,\Phi)$ one can introduce Euler's complex
variables $(\alpha,\beta,\gamma)$ and $(A,B,\Gamma)$ through the
simple linear formulas:
\begin{gather*}
\alpha = \phi + z , \qquad \beta = 2 \rho , \qquad \gamma = \phi -
z , \qquad A = \Phi + Z , \qquad B = 2 R , \qquad \Gamma = \Phi -
Z .
\end{gather*}
Euler's angles $(\alpha, \beta, \gamma)$ and $(A,B,\Gamma)$ are
referred to $k_{a},m_{a}$-parameters by the formulas (see
in~\cite{Fedorov})
\begin{gather*}
\cos \beta = k_{0}^{2} - k_{3}^{2} + k_{1}^{2} + k_{2}^{2} ,
\qquad \sin \beta = 2 \sqrt{k_{0}^{2} - k_{3}^{2}}
\sqrt{-k_{1}^{2} -k_{2}^{2}} , \nonumber
\\
\cos \alpha = { -i k_{0} k_{1} + k_{2} k_{3} \over \sqrt{k_{0}^{2}
- k_{3}^{2}} \sqrt{-k_{1}^{2} - k_{2}^{2}} } , \qquad \sin \alpha
= { -ik_{0} k_{2} - k_{1} k_{3} \over \sqrt{k_{0}^{2} - k_{3}^{2}}
\sqrt{-k_{1}^{2} -k_{2}^{2}} } , \nonumber
\\
\cos \gamma = { -i k_{0} k_{1} - k_{2} k_{3} \over \sqrt{k_{0}^{2}
- k_{3}^{2}} \sqrt{-k_{1}^{2} - k_{2}^{2}} } , \qquad \sin \gamma
= { -i k_{0} k_{2} + k_{1} k_{3} \over \sqrt{k_{0}^{2} -
k_{3}^{2}} \sqrt{-k_{1}^{2} - k_{2}^{2}} }, \nonumber
\\
\cos B = m_{0}^{2} - m_{3}^{2} + m_{1}^{2} + m_{2}^{2} , \qquad
\sin B = 2 \sqrt{m_{0}^{2} - m_{3}^{2}} \sqrt{-m_{1}^{2}
-m_{2}^{2}} , \nonumber
\\
\cos A = { +i m_{0} m_{1} + m_{2} m_{3} \over \sqrt{m_{0}^{2} -
m_{3}^{2}} \sqrt{-m_{1}^{2} - m_{2}^{2}} } , \qquad \sin A = {
+im_{0} m_{2} - m_{1} m_{3} \over \sqrt{m_{0}^{2} - m_{3}^{2}}
 \sqrt{-m_{1}^{2} -m_{2}^{2}} } ,
 \nonumber
\\
\cos \Gamma = { +i m_{0} m_{1} - m_{2} m_{3} \over \sqrt{m_{0}^{2}
- m_{3}^{2}} \sqrt{-m_{1}^{2} - m_{2}^{2}} } , \qquad \sin \Gamma
= { +i m_{0} m_{2} + m_{1} m_{3} \over \sqrt{m_{0}^{2} -
m_{3}^{2}} \sqrt{-m_{1}^{2} - m_{2}^{2}} } . 
\end{gather*}

Complex Euler's angles as parameters for complex Lorentz group
$SO(4,C)$ have a distinguished feature: 2-spinor constituents are
factorized into three elementary Euler's transforms ($\sigma^{i}$
stands for the known Pauli matrices):
\begin{gather*}
B(k) = e^{-i\sigma^{3} \alpha/2} e^{i \sigma^{1}\beta/2}
e^{+i\sigma^{3} \gamma/2} \in SL(2,C) , \nonumber
\\
B(\bar{m}) = e^{-i\sigma^{3} \Gamma/2} e^{i \sigma^{1}B/2}
e^{+i\sigma^{3} A/2} \in SL(2,C)' . 
\end{gather*}

The main question is how to extend possible parameterizations of
small orthogonal group $SO(4,C)$ and its double covering to bigger
orthogonal and unitary groups\footnote{In this subject, especially
concerned with generalized Euler angles, we have found out much
from Murnaghan's book \cite{Murnaghan}.}. To be concrete we are
going to focus attention mainly
 on the group $SU(4)$ and its counterparts $SU(2,2)$, $SU(3,1)$.

There exist many publications on the subject, a great deal of
facts are known~-- in the following we will be turning to them. A good
classif\/ication of dif\/ferent approaches in parameterizing
f\/inite transformations of $SU(4)$ was done in the recent paper
by A.~Gsponer~\cite{Gsponer-1}. Recalling it, we will try to cite
publications in appropriate places though many of them should be
placed in several dif\/ferent subclasses~-- it is natural because
all approaches are closely connected to each other.

\begin{itemize}\itemsep=0pt
\item {\bf Canonical form} \cite{ Moler, Moshinsky-62, Murnaghan,
Nelson-67, Costa, Exponential, Wilcox-67}.
They use explicitly the full set of the Lie generators\footnote{In
the paper we will designate
 generators in Dirac basis by $\Lambda_{i}$ whereas another set of
 generators mainly used in the literature will be referred as $\lambda_{i}$.}
 so that the group element is expressed as the exponential
of the linear combination
\begin{gather*}
G = \exp_{n} [ i(a_{1}\lambda_{1}+ \cdots +
a_{n}\lambda_{N}) ] 
\end{gather*}
the inf\/inite series of terms implied be $\exp$~-- symbol is
usually very dif\/f\/icult to be summed in closed form~-- though
there exists many interesting examples of those:

\item {\bf Non-canonical forms} \cite{Barnes1, Barnes2, Barut-1,
Barut-2, Bincer-90, Gsponer-1, Holland, Kihlberg, Kusnezov,
Nelson-67, Raghunathan-1989, Exponential, Rosen, 2}.
As a consequence of the Baker--Campbell--Hausdorf\/f theorem
\cite{Baker, Campbell,Hausdorff} it is possible to break-down the
canonical form into a product
\begin{gather*}
G = \exp_{n^{(1)} } \times \cdots \times
\exp_{n^{(k)} } , \qquad n^{(1)} + \cdots +n^{(k)} = N . 
\end{gather*}
with the hope that $\exp_{n^{(i)} }$ could be summed in closed
form and also that these factors
 have simple properties. This possibility for the groups
$SU(4)$ and $SU(2,2)$ will be discussed in more detail in
 sections below.

\item {\bf Product form} \cite{Bogush-73, Bogush-74', Bogush-72',
Bogush-74, Moshinski, Holland, Murnaghan, Nelson-67}.
An extreme non-canonical form is to factorize the general
exponential into a product of $n$ simplest 1-parametric
exponentials
\begin{gather*}
G = \exp [ia_{1} \lambda_{1}] \times \cdots \times
\exp [ia_{N} \lambda_{N}] .
\end{gather*}

\item {\bf Basic elements } (the main approach in the present
treatment) \cite{Barut-1,Barut-2,Gsponer-1,
Kihlberg,Macfarlane,Macfarlane-1966,Macfarlane-80,Macfarlane et
al, 2}.
This way is to expand the elements of the group (matrices or
quaternions) into a sum over basis elements and to work with a
linear decomposition of the matrices over basic ones:
\begin{gather}
 G' = x_{n}' \lambda_{m} , \qquad G = x_{n} \lambda_{m} ,
 \qquad \lambda_{0} = I , \qquad k \in \{0, 1, \dots,N \} ,
\nonumber
\\
G'' = G' G , \qquad x_{k}'' \lambda_{k} = x'_{m} \lambda_{m}
 x_{n} \lambda_{n}= x'_{m} x_{n} \lambda_{m}
\lambda_{n}, \label{I.13}
\end{gather}
as by def\/inition the relationships $\lambda_{m} \lambda_{n} =
e_{mnk} \lambda_{k}$ must hold, the group multiplication rule for
parameters $x_{k}$ looks
\begin{gather}
x_{k}'' = e_{mnk} x'_{m} x_{n} . \label{I.14}
\end{gather}
The main claim is that the all properties of any matrix group are
straightforwardly determined by the bilinear function,
 the latter is described by structure constants $e_{mnk}$ entering the
 multiplication rule
 $\lambda_{m} \lambda_{n} = e_{mnk} \lambda_{k}$.

\item {\bf Hamilton--Cayley form}
\cite{Barut-1,Barut-2,Bincer-90,Bogush-73, Bogush-74', Bogush-72',
Bogush-72, Bogush-74,Bogush-76,Fedorovykh-75, Rosen}.
It is possible to expand the elements of the group into a power
series of linear combination of generators:
\begin{gather*}
 \lambda (a) = i(a_{1}\lambda_{1}+ \cdots + a_{N}\lambda_{N}),
 \end{gather*}
because of Hamilton--Cayley theorem this series has three terms
for $SU(3)$ and four terms for $SU(4)$:
\begin{gather*}
SU(3), \quad G(a) = e_{0} (a) I + e_{1} (a) \lambda (a) + e_{2}
(a) \lambda ^{2}(a) , \nonumber
\\
SU(4), \quad G(a) = e_{0} (a) I + e_{1} (a) \lambda (a) + e_{2}
(a) \lambda^{2}(a) + e_{3} (a) \lambda^{3}(a).
\end{gather*}

\item {\bf Euler-angles representations} \cite{Beg,Bracken-1984,Byrd-2, -3, Byrd-1,
Byrd-4, Gsponer-1, Kusnezov, Murnaghan, Nelson-67, Tilma-1}.
In Euler-angles representations only a sub-set $\{ \lambda_{n} \}
\subset \{\lambda_{N}\}$ of the Lie generators are suf\/f\/icient
to produce the whole set (for $SU(N)$ we need only $2(N-1)$
generators). In that sense all other way to obtain the whole set
of elements are not minimal.
\end{itemize}

In our opinion, we should search the most simplicity in
mathematical sense while working with basic elements $\lambda_{k}$
and the structure constants determining the group multiplication
rule~(\ref{I.13}),~(\ref{I.14}).

The material of this paper is arranged as follows.

In Section~\ref{sec2}
 an arbitrary $4\times 4 $ matrix $G \in GL(4,C)$
is decomposed into sixteen Dirac matrices\footnote{That Dirac
matrices-based approach was widely used in physical context (see
\cite{Barut-70,Barut-1,Barut-2,Bracken-81,Macfarlane-1966,Mack',Salam,
Mack, Kate} and especially \cite{Kihlberg}).}
\begin{gather}
G = A I + iB \gamma^{5} + iA_{l} \gamma^{l} + B_{l} \gamma^{l}
\gamma^{5} + F_{mn} \sigma_{mn} =
 \left | \begin{array}{cc} k_{0} + {\boldsymbol k} \vec{\sigma}
& n_{0} - {\boldsymbol n}
\vec{\sigma} \vspace{1mm} \\
 - l_{0} - {\boldsymbol l} \vec{\sigma} & m_{0} - {\boldsymbol m} \vec{\sigma}
\end{array} \right | ,
\label{I.17}
\end{gather}

\noindent for def\/initeness we will use the Weyl spinor basis;
four 4-dimensional vectors $(k,m,l,n)$ are def\/inite linear
combinations of $A$, $B$, $A_{l}$, $B_{l}$, $F_{mn}$~-- see
(\ref{6}). In such parameters (\ref{5}), the group multiplication
law $G''=G'G$ is found in explicit form.

Then we turn to the following problem: at given $G=G (k,m,n,l ) $
one should f\/ind parameters of the inverse matrix: $ G^{-1} = G(
k',m ',n',l' )$ -- expressions for $( k',m ',n',l' )$ have been
found explicitly (for details of calculation see \cite{Redkov}).
Also, several equivalent expressions for determinant $\det G$ have
been obtained, which is essential when going to special groups
$SL(4,C)$ and its subgroups.

In Section~\ref{sec3}, with the help of the expression for the
inverse matrix $G^{-1}(k',m',l',n')$ we begin to consider the
unitary group $SU(4)$. To this end, one should specify the
requirement of unitarity $ G^{+} = G^{-1} $ to the above vector
parametrization --
 so that unitarity conditions are given as
non-linear cubic algebraic equations for parameters $(k,m,l,n)$
including complex conjugation.

In Section \ref{sec4} we have constructed three 2-parametric
solutions of the produced equations of unitarity\footnote{At this,
the unitarity equations may be considered as special eigenvalue
problems in 2-dimensional space.}, these subgroups $G_{1}$,
$G_{2}$, $G_{3}$ consist of two commuting Abelian unitary
subgroups.

In Section~\ref{sec5} we have constructed a 4-parametric
solution\footnote{The problem again is reduced to solving of a
special eigenvalue problem in 2-dimensional space.}~-- it may be
factorized into two commuting unitary factors: $G = G_{0} \otimes
SU(2)$ -- see~(\ref{B.15}).

The task of complete solving of the unitarity conditions seems to
be rather complicated. In remaining part of the present paper we
describe some relations of the above treatment to other
considerations of the problem in the literature. We hope that the
full general solution of the unitary equations obtained can be
constructed on the way of combining dif\/ferent techniques used in
the theory of the unitary group $SU(4)$ and it will be considered
elsewhere.

We turn again to the explicit form of the Dirac basis and note
that all 15 matrices are of Gell-Mann type: they have a
zero-trace, they are Hermitian, besides their squares are unite:
\begin{gather*}
 \mbox{Sp}\Lambda =0 , \qquad
 (\Lambda )^{2}= I , \qquad (\Lambda )^{+} = \Lambda , \qquad \Lambda \in \{ \Lambda_{k}, \
 k =1,\dots,15 \} .
\end{gather*}
Exponential function of any of them equals to
\begin{gather*}
U_{j} = e^{ia_{j} \Lambda_{j}} = \cos a _{j} + i \sin a_{j}
\Lambda_{j} , \qquad \det e^{ia_{j}\Lambda_{j}} = +1 , \qquad
U^{+}_{j} = U^{-1}_{j} , \qquad a_{i} \in {R} . \nonumber
\end{gather*}
Evidently, multiplying such 15 elementary unitary matrices (at
real parameters $x_{i}$) gives again an unitary matrix
\begin{gather*}
U = e^{ia_{1}\Lambda_{1}} e^{ia_{2}\Lambda_{2}} \cdots
 e^{ia_{14}\Lambda_{14}} e^{ia_{l5}\Lambda_{l5}} , \qquad
U^{+} = e^{-ia_{l}\Lambda_{l}} e^{-ia_{k}\Lambda_{k}} \cdots
 e^{-ia_{j}\Lambda_{j}} e^{-ia_{i}\Lambda_{i}} . \nonumber
\end{gather*}
 At this there arises one special possibility to
determine extended Euler angles $a_{1} , \dots, a_{15}$. For the
group $SU(4)$ the Euler parametrization of that type was found in
\cite{Tilma-1}. A method to solve the problem in~\cite{Tilma-1}
 was based on the use yet known Euler parametrization for
$SU(3)$~-- the latter problem was solved in \cite{Byrd-4}.
Extension to $SU(N)$ group was done in \cite{Tilma-2,Tilma-3}.
Evident advantage of the Euler angles approach is its simplicity,
and evident defect consists in the following: we do not know any
simple group multiplication rule for these angles -- even the
known solution for~$SU(2)$ is too complicated and cannot be used
ef\/fectively in calculation.

In Section~\ref{sec6} the main question is
 how in Dirac parametrization one can
distinguish~$SU(3)$, the subgroup in~$SU(4)$. In this connection,
it should be noted that the basis $\lambda_{i}$ used
in~\cite{Byrd-4} substantially dif\/fers from the above Dirac
basis~$\Lambda_{i}$~-- this peculiarity is closely connected with
distinguishing the $SU(3)$ in $SU(4)$. In order to have
possibility to compare two approaches we need exact connection
between $\lambda_{i}$ and $\Lambda_{i}$~-- we have found required
formulas\footnote{This problem evidently is related to the task of
distinguishing $GL(3,C)$ in $GL(4,C)$ as well.}. The separation of
$SL(3,C)$ in $SL(4,C)$ is given explicitly, at this
 $3\times3$ matrix group is described with the
help of $4\times4$ matrices\footnote{Interesting arguments related
to this point but in the quaternion approach are given in
\cite{Gsponer-1}.}. The group law for parameters of $SL(3,C)$ is
specif\/ied.

In Section~\ref{sec7} one dif\/ferent way to list 15 generators of
$GL(4,C)$ is examined\footnote{Such a possibility is well-known --
see~\cite{Kihlberg}; our approach looks simpler and more
symmetrical because we use the Weyl basis for Dirac matrices
instead of the standard one as in~\cite{Kihlberg}.}
\begin{gather*}
 \alpha_{1} = \gamma^{0}\gamma^{2} , \qquad
 \alpha_{2} = i \gamma^{0}\gamma^{5} , \qquad \alpha_{3} = \gamma^{5}\gamma^{2},
\qquad
 \beta_{1} = i \gamma^{3}\gamma^{1} , \qquad
 \beta_{2} = i \gamma^{3} , \qquad
 \beta_{3} = i\gamma^{1} ,
\end{gather*}
these two set commute with each others $\alpha_{j} \beta_{k} =
\beta_{k} \alpha_{j} $, and their multiplications provides us with
9 remaining basis elements of f\/ifteen:
\begin{alignat*}{4}
& A_{1}= \alpha_{1} \beta_{1} , \qquad &&
 B_{1}= \alpha_{1} \beta_{2} , \qquad &&
 C_{1}=\alpha_{1} \beta_{3} , & \\
& A_{2}= \alpha_{2} \beta_{1} , &&
 B_{2}= \alpha_{2} \beta_{2} , &&
 C_{2}= \alpha_{2} \beta_{3} , & \\
& A_{3}= \alpha_{3} \beta_{1} , &&
 B_{3}= \alpha_{3} \beta_{2} , &&
C_{3}= \alpha_{3} \beta_{3} .&
\end{alignat*}
We turn to the rule of multiplying 15 generators $\alpha_{i}$,
$\beta_{i}$, $A_{i}$, $B_{i}$, $C_{i}$ and derive its explicit
form (see (\ref{B.7})).

Section \ref{sec8} adds some facts to a factorized structure of
$SU(4)$. To this end, between 9~generators we distinguish three
sets of commuting ones
\begin{gather*}
{\boldsymbol K} = \{ A_{1}, B_{2} , C_{3} \} , \qquad {\boldsymbol
L} = \{ C_{1} , A_{2} , B_{3} \} , \qquad {\boldsymbol M} = \{
B_{1} , C_{2} , A_{3} \} , \nonumber
\end{gather*}
an arbitrary element from $GL(4,C)$ can be factorized as
follows\footnote{These facts were described in main parts
in~\cite{Kihlberg}.}
\begin{gather}
S = e^ {i \vec{a} \vec{\alpha} } e^ {i \vec{b} \vec{\beta} } e^{i
{\boldsymbol k} {\boldsymbol K} } e^{i {\boldsymbol l}
 {\boldsymbol L} } e^{i {\boldsymbol m} {\boldsymbol M} } , \label{I.31}
\end{gather}
where ${\boldsymbol K}$, ${\boldsymbol L}$, ${\boldsymbol M}$ are
3-parametric groups, each of them consists of three Abelian
commuting unitary subgroups\footnote{Note that existence of three
Abelian commuting unitary subgroups was shown in~\cite{Kihlberg}
as well.}. On the basis of 15 matrices one can easily see 20 ways
to separate $SU(2)$ subgroups, which
 might be used as bigger elementary blocks in constructing a general
 transformation\footnote{This possibility was studied partly in \cite{Bogush-74',Bogush-72'} on the basis of the
 Hamilton--Cayley approach.}.

In Sections \ref{sec9} and \ref{sec10} we specify our approach for
pseudounitary groups $SU(2,2)$ and $SU(3,1)$ respectively. All
generators $\Lambda'_{k} $ of these groups can readily be
constructed on the basis of the known Dirac generators of $SU(4)$
(see~(\ref{D.5})).

\section[On parameters of inverse transformations $G^{-1}$]{On parameters of inverse transformations $\boldsymbol{G^{-1}}$}\label{sec2}

 Arbitrary $4\times 4 $ matrix $G \in GL(4,C)$
can be decomposed in terms of 16 Dirac matrices (such an approach
to the group $L(4,C)$ was discussed and partly developed in
\cite{Barut-70,Barut-1,Barut-2,Bracken-81,Macfarlane-1966,Mack',Salam,
Mack, Kate} and especially in \cite{Kihlberg}):
\begin{gather}
G = A I + iB \gamma^{5} + iA_{l} \gamma^{l} + B_{l} \gamma^{l}
\gamma^{5} + F_{mn} \sigma_{mn} , \label{1}
\end{gather}
where
\begin{gather*}
 \gamma^{a} \gamma^{b} +
\gamma^{b} \gamma^{a} = 2 g^{ab}, \qquad \gamma^{5} = -i
\gamma^{0} \gamma^{1} \gamma^{2} \gamma^{3} , \nonumber
\\
\sigma^{ab} = \tfrac 14 ( \gamma^{a} \gamma^{b} - \gamma^{b}
\gamma^{a}) , \qquad g^{ab} = \mbox{diag} (+1,-1,-1,-1) .
\nonumber
\end{gather*}
Taking 16 coef\/f\/icients $A$, $B$, $A_{l}$, $B_{l}$, $F_{mn}$ as
parameters in the group $G = G(A,B, A_{l}, B_{l}, F_{mn})$ one can
establish the corresponding multiplication law for these
parameters:
\begin{gather*}
G' = A' I + iB ' \gamma^{5} + iA_{l}' \gamma^{l} + B_{l}'
\gamma^{l} \gamma^{5} + F'_{mn} \sigma_{mn} , \nonumber
\\
G = A I + iB \gamma^{5} + iA_{l} \gamma^{l} + B_{l} \gamma^{l}
\gamma^{5} + F_{mn} \sigma_{mn} ,
\\
G'' = G'G = A'' I + iB '' \gamma^{5} + iA_{l}'' \gamma^{l} +
B_{l}'' \gamma^{l} \gamma^{5} + F''_{mn} \sigma_{mn} ,
\end{gather*}
 where
\begin{gather}
A''= A' A -B' B -A'_{l} A^{l} -B'_{l} B^{l} - \tfrac 12 F'_{kl}
F^{kl} , \nonumber
\\
B''= A' B + B' A + A'_{l} B^{l} - B'_{l} A^{l} + \tfrac 14 F'_{mn}
F_{cd} \epsilon^{mncd} , \nonumber
\\
A''_{l} = A' A_{l} - B' B_{l} + A'_{l }A + B'_{l} B + A'^{k}F_{kl}
\nonumber
\\
\phantom{A''_{l} =}{} + F'_{lk}A^{k} + \tfrac 12 B'_{k} F_{mn}
\epsilon_{l}^{\; kmn} + \tfrac 12 F'_{mn} B_{k} \epsilon_{l}^{\;
mnk} , \nonumber
\\
 B''_{l} = A' B_{l} + B' A_{l} - A'_{l} B + B'_{l} A + B'^{k} F_{kl}
 \nonumber
 \\
 \phantom{B''_{l} =}{} + F'_{lk} B^{k} + \tfrac 12 A'_{k} F_{mn} \epsilon^{kmn}_{\;\;\;\;\;\;\;l} +
 \tfrac 12 F'_{mn} A_{k} \epsilon^{mnk}_{\;\;\;\;\;\;\;l} ,
\nonumber
\\
F''_{mn} = A' F_{mn} + F'_{mn} A - (A'_{m} A_{n} -A'_{n} A_{m}) -
(B'_{m} B_{n} -B'_{n} B_{m}) \label{3}
\\
\phantom{F''_{mn} =}{} + A'_{l} B_{k} \epsilon^{lkmn} - B'_{l}
A_{k} \epsilon^{lkmn} +\tfrac 12 B' F_{kl} \epsilon^{kl}_{
\;\;\;mn} + \tfrac 12 F'_{kl} B \epsilon^{kl}_{ \;\;\;mn} +
(F'_{mk} F^{k}_{\;\;n} - F'_{nk} F^{k}_{\;\;m} ) .\nonumber
\end{gather}
The latter formulas are correct in any basis for Dirac matrices.
Below we will use mainly Weyl spinor basis:
\begin{gather*}
\gamma^{a} = \left | \begin{array}{cc} 0 & \bar{\sigma}^{a} \\
\sigma^{a} & 0
\end{array} \right | , \qquad \sigma^{a}= (I , \sigma^{j}) , \qquad \bar{\sigma}^{a}= (I , -\sigma^{j}) ,\qquad
\gamma^{5}= \left | \begin{array}{cc} -I & 0 \\ 0 & +I
\end{array} \right | .
\end{gather*}
With this choice, let us make $3+1$-splitting in all the formulas:
\begin{gather}
 G \in GL(4,C) , \qquad
G = \left | \begin{array}{cc} k_{0} + {\boldsymbol k} \vec{\sigma}
& n_{0} - {\boldsymbol n}
\vec{\sigma} \vspace{1mm}\\
 - l_{0} - {\boldsymbol l} \vec{\sigma} & m_{0} - {\boldsymbol m} \vec{\sigma}
\end{array} \right | ,
\label{5}
\end{gather}
where complex 4-vector parameters $(k,l,m,n)$ are def\/ined by
\cite{Bogush-2006}:
\begin{gather}
k_{0} = A - i B , \qquad k_{j} = a_{j} - i b_{j} , \qquad m_{0} =
A + i B , \qquad m_{j} = a_{j} + i b_{j} , \nonumber
\\
 l_{0} = B_{0} - i A_{0} , \qquad l_{j} = B_{j} - i A_{j}
, \qquad
 n_{0} = B_{0} + i A_{0} , \qquad n_{j} = B_{j} + i A_{j}
. \label{6}
\end{gather}
For such parameters (\ref{5}), the composition rule (\ref{3}) will
look as follows:
\begin{gather}
(k'',m'';n'',l'') =(k',m';n',l') (k,m;n,l) , \nonumber
\\
k_{0}'' = k_{0}' k_{0} + {\boldsymbol k}' {\boldsymbol k}
 - n'_{0} l_{0} + {\boldsymbol n}' {\boldsymbol l} ,
\nonumber
\\
{\boldsymbol k}'' = k'_{0} {\boldsymbol k} + {\boldsymbol k}'
k_{0} + i {\boldsymbol k}' \times {\boldsymbol k} - n_{0}'
{\boldsymbol l} + {\boldsymbol n}' l_{0} + i {\boldsymbol n}'
\times {\boldsymbol l} , \nonumber
\\
m_{0}'' = m_{0}' m_{0} + {\boldsymbol m}' {\boldsymbol m}
 - l'_{0} n_{0} + {\boldsymbol l}' {\boldsymbol n} ,
\nonumber
\\
{\boldsymbol m}'' = m'_{0} {\boldsymbol m} + {\boldsymbol m}'
m_{0} - i {\boldsymbol m}' \times {\boldsymbol m} - l_{0}'
{\boldsymbol n} + {\boldsymbol l}' n_{0} - i {\boldsymbol l}'
\times {\boldsymbol n} , \nonumber
\\
n_{0}'' = k_{0}' n_{0} - {\boldsymbol k}' {\boldsymbol n}
 + n'_{0} m_{0} + {\boldsymbol n}' {\boldsymbol m} ,
\nonumber
\\
{\boldsymbol n}'' = k'_{0} {\boldsymbol n} - {\boldsymbol k}'
n_{0} + i {\boldsymbol k}' \times {\boldsymbol n} + n_{0}'
{\boldsymbol m} + {\boldsymbol n}' m_{0} - i {\boldsymbol n}'
\times {\boldsymbol m} , \nonumber
\\
l_{0}'' = l_{0}' k_{0} + {\boldsymbol l}' {\boldsymbol k}
 + m'_{0} l_{0} - {\boldsymbol m}' {\boldsymbol l} ,
\nonumber
\\
{\boldsymbol l}'' = l'_{0} {\boldsymbol k} + {\boldsymbol l}'
k_{0} + i {\boldsymbol l}' \times {\boldsymbol k} + m_{0}'
{\boldsymbol l} - {\boldsymbol m}' l_{0} - i {\boldsymbol m}'
\times {\boldsymbol l } . \label{7}
\end{gather}

Now let us turn to the following problem: with given $G=G (k,m,n,l
) $ one should f\/ind para\-me\-ters of the inverse matrix: $
G^{-1} = G( k',m ',n',l' )$. In other words, starting from
\begin{gather}
G(k,m,n,l ) = \left | \begin{array}{lccr}
+(k_{0} + k_{3}) & +(k_{1} - ik_{2}) & \hspace{3mm} +(n_{0} - n_{3}) & -(n_{1} - in_{2}) \\
+(k_{1} + ik_{2} ) & + (k_{0} - k_{3} ) & \hspace{3mm}-(n_{1} + in_{2}) & + (n_{0} + n_{3} ) \\
-(l_{0} + l_{3}) & -(l_{1} - il_{2}) & \hspace{3mm} +(m_{0} - m_{3}) & -(m_{1} - im_{2}) \\
-(l_{1} + il_{2}) & -(l_{0} - l_{3}) & \hspace{3mm} -(m_{1} +
im_{2}) & +(m_{0} + m_{3})
\end{array} \right | ,
\label{9a}
\end{gather}
one should calculate parameters of the inverse matrix $G^{-1}$.
The problem turns to be rather complicated\footnote{For more
details see \cite{Redkov}; also see a preceding  paper~\cite{Kleefeld}.}, the f\/inal result is ($D = \det G$,
$ (mn) \equiv m_{0}n_{0} - {\boldsymbol m } {\boldsymbol n} $, and
so on)
\begin{gather}
k_{0}' = D ^{-1} [ k_{0} (mm) + m_{0} (ln) + l_{0} (nm) - n_{0}
(lm) +
 i {\boldsymbol l} ({\boldsymbol m} \times {\boldsymbol n} ) ] ,
\nonumber
\\
{\boldsymbol k} ' = D^{-1} [ - {\boldsymbol k} (mm) - {\boldsymbol
m} (ln) - {\boldsymbol l} (nm) +
 {\boldsymbol n } (lm) +
 2 {\boldsymbol l} \times ({\boldsymbol n} \times {\boldsymbol m})
 \nonumber
 \\
 \phantom{{\boldsymbol k} ' =}{} +
i m_{0} ( {\boldsymbol n} \times {\boldsymbol l} ) + i l_{0} (
{\boldsymbol n} \times {\boldsymbol m} ) + i n_{0} ( {\boldsymbol
l} \times {\boldsymbol m} ) ] , \nonumber
\\
m_{0}' = D ^{-1} [ k_{0} (ln) + m_{0} (kk) - l_{0} (kn) + n_{0}
lk) +
 i {\boldsymbol n} ({\boldsymbol l} \times {\boldsymbol k} ) ] ,
\nonumber
\\
{\boldsymbol m}' = D^{-1} [ - {\boldsymbol k} (ln) - {\boldsymbol
m} (kk) + {\boldsymbol l } (kn) - {\boldsymbol n} (kl) + 2
{\boldsymbol n} \times ( {\boldsymbol l} \times {\boldsymbol k})
\nonumber
\\
\phantom{{\boldsymbol m}' =}{} +
 i n_{0} ({\boldsymbol k} \times {\boldsymbol l}) + i l_{0} ( {\boldsymbol k} \times {\boldsymbol n}) +
i k_{0} ({\boldsymbol n} \times {\boldsymbol l}) ] , \nonumber
\\
l_{0}' = D ^{-1} [ + k_{0} (ml) - m_{0} (kl) - l_{0} (km) - n_{0}
(ll) + i {\boldsymbol m} ( {\boldsymbol l} \times {\boldsymbol k}
) ] , \nonumber
\\
{\boldsymbol l} ' = D^{-1} [ + {\boldsymbol k} (ml) - {\boldsymbol
m} (kl) - {\boldsymbol l} (km) - {\boldsymbol n} (ll) +
 2 {\boldsymbol m} \times ( {\boldsymbol k} \times {\boldsymbol l} )
\nonumber
\\
\phantom{{\boldsymbol l} ' =}{} +
 i m_{0} ({\boldsymbol l} \times {\boldsymbol k} ) + i k_{0} ({\boldsymbol l} \times {\boldsymbol m} ) +
i l_{0} ({\boldsymbol m} \times {\boldsymbol k}) ] , \nonumber
\\
n_{0}' = D ^{-1} [ - k_{0} (nm) + m_{0} (kn) - l_{0} (nn) - n_{0}
(km) +
 i {\boldsymbol k} ({\boldsymbol m} \times {\boldsymbol n} ) ] ,
\nonumber
\\
{\boldsymbol n}' = D ^{-1} [ - {\boldsymbol k} (nm) + {\boldsymbol
m} (kn)
 - {\boldsymbol l} (nn) - {\boldsymbol n} (km) +
 2 {\boldsymbol k} \times ( {\boldsymbol m} \times {\boldsymbol n} ) \nonumber
\\
 \phantom{{\boldsymbol n}' =}{}+
 i k_{0} ({\boldsymbol m} \times {\boldsymbol n}) +
i m_{0} ({\boldsymbol k} \times {\boldsymbol n}) + i n_{0}
({\boldsymbol m} \times {\boldsymbol k}) ] . \label{10}
\end{gather}

 Substituting equations (\ref{10}) into equation
 $G^{-1} G= I$ one arrives at
\begin{gather*}
D = k_{0}'' = k_{0}' k_{0} + {\boldsymbol k}' {\boldsymbol k}
 - n'_{0} l_{0} + {\boldsymbol n}' {\boldsymbol l} ,
\nonumber
\\
0= {\boldsymbol k}'' = k'_{0} {\boldsymbol k} + {\boldsymbol k}'
k_{0} + i {\boldsymbol k}' \times {\boldsymbol k} - n_{0}'
{\boldsymbol l} + {\boldsymbol n}' l_{0} + i {\boldsymbol n}'
\times {\boldsymbol l} , \nonumber
\\
D = m_{0}'' = m_{0}' m_{0} + {\boldsymbol m}' {\boldsymbol m}
 - l'_{0} n_{0} + {\boldsymbol l}' {\boldsymbol n} ,
\nonumber
\\
0= {\boldsymbol m}'' = m'_{0} {\boldsymbol m} + {\boldsymbol m}'
m_{0} - i {\boldsymbol m}' \times {\boldsymbol m} - l_{0}'
{\boldsymbol n} + {\boldsymbol l}' n_{0} - i {\boldsymbol l}'
\times {\boldsymbol n} , \nonumber
\\
0= n_{0}'' = k_{0}' n_{0} - {\boldsymbol k}' {\boldsymbol n}
 + n'_{0} m_{0} + {\boldsymbol n}' {\boldsymbol m} ,
\nonumber
\\
0= {\boldsymbol n}'' = k'_{0} {\boldsymbol n} - {\boldsymbol k}'
n_{0} + i {\boldsymbol k}' \times {\boldsymbol n} + n_{0}'
{\boldsymbol m} + {\boldsymbol n}' m_{0} - i {\boldsymbol n}'
\times {\boldsymbol m} , \nonumber
\\
0= l_{0}'' = l_{0}' k_{0} + {\boldsymbol l}' {\boldsymbol k}
 + m'_{0} l_{0} - {\boldsymbol m}' {\boldsymbol l} ,
\nonumber
\\
0= {\boldsymbol l}'' = l'_{0} {\boldsymbol k} + {\boldsymbol l}'
k_{0} + i {\boldsymbol l}' \times {\boldsymbol k} + m_{0}'
{\boldsymbol l} - {\boldsymbol m}' l_{0} - i {\boldsymbol m}'
\times {\boldsymbol l } . \nonumber
\end{gather*}
 After calculation, one can prove these identities and
f\/ind the determinant:
\begin{gather}
D = \det G (k,m,n,l) = (kk) (mm) + (ll) (nn) + 2 (mk) (ln) +2 (lk)
(nm) -2 (nk) (lm) \nonumber
\\
\phantom{D =}{} + 2 i [ k_{0} {\boldsymbol l} ({\boldsymbol m}
\times {\boldsymbol n}) + m_{0} {\boldsymbol k} ({\boldsymbol n}
\times {\boldsymbol l}) + l_{0} {\boldsymbol k} ({\boldsymbol n}
\times {\boldsymbol m}) + n_{0} {\boldsymbol l} ({\boldsymbol m}
\times {\boldsymbol k}) ] \nonumber
\\
\phantom{D =}{} + 4 ({\boldsymbol k} {\boldsymbol n} )
({\boldsymbol m} {\boldsymbol l}) - 4 ({\boldsymbol k}
{\boldsymbol m} ) ({\boldsymbol n} {\boldsymbol l}) . \label{11}
\end{gather}

Let us specify several more simple subgroups.

\subsection*{Case A}

Let 0-components $k_{0}$, $m_{0}$, $l_{0}$, $n_{0}$ be
real-valued, and 3-vectors
 ${\boldsymbol k}$, ${\boldsymbol m}$, ${\boldsymbol l}$, ${\boldsymbol n}$ be imaginary.
Performing in (\ref{7}) the formal change (new vectors are
real-valued)
\begin{gather}
{\boldsymbol k} \Longrightarrow i {\boldsymbol k} ,\qquad
{\boldsymbol m} \Longrightarrow i {\boldsymbol m} ,\qquad
{\boldsymbol l} \Longrightarrow i {\boldsymbol l} ,\qquad
{\boldsymbol n} \Longrightarrow i {\boldsymbol n} , \nonumber
\\
G = \left | \begin{array}{cc}
k_{0} +i {\boldsymbol k} \vec{\sigma} & n_{0} - i {\boldsymbol n} \vec{\sigma} \\[3mm]
 - l_{0} - i {\boldsymbol l} \vec{\sigma} & m_{0} - i {\boldsymbol m} \vec{\sigma}
\end{array} \right | ,
\label{12b}
\end{gather}
then the multiplication rules (\ref{7}) for sixteen real variables
look as follows
\begin{gather*}
k_{0}'' = k_{0}' k_{0} - {\boldsymbol k}' {\boldsymbol k}
 - n'_{0} l_{0} - {\boldsymbol n}' {\boldsymbol l} ,
\nonumber
\\
{\boldsymbol k}'' = k'_{0} {\boldsymbol k} + {\boldsymbol k}'
k_{0} - {\boldsymbol k}' \times {\boldsymbol k} - n_{0}'
{\boldsymbol l} + {\boldsymbol n}' l_{0} - {\boldsymbol n}' \times
{\boldsymbol l} , \nonumber
\\
m_{0}'' = m_{0}' m_{0} - {\boldsymbol m}' {\boldsymbol m}
 - l'_{0} n_{0} - {\boldsymbol l}' {\boldsymbol n} ,
\nonumber
\\
{\boldsymbol m}'' = m'_{0} {\boldsymbol m} + {\boldsymbol m}'
m_{0} + {\boldsymbol m}' \times {\boldsymbol m} - l_{0}'
{\boldsymbol n} + {\boldsymbol l}' n_{0} + {\boldsymbol l}' \times
{\boldsymbol n} , \nonumber
\\
n_{0}'' = k_{0}' n_{0} + {\boldsymbol k}' {\boldsymbol n}
 + n'_{0} m_{0} - {\boldsymbol n}' {\boldsymbol m} ,
\nonumber
\\
{\boldsymbol n}'' = k'_{0} {\boldsymbol n} - {\boldsymbol k}'
n_{0} - {\boldsymbol k}' \times {\boldsymbol n} + n_{0}'
{\boldsymbol m} + {\boldsymbol n}' m_{0} + {\boldsymbol n}' \times
{\boldsymbol m} , \nonumber
\\
l_{0}'' = l_{0}' k_{0} - {\boldsymbol l}' {\boldsymbol k}
 + m'_{0} l_{0} + {\boldsymbol m}' {\boldsymbol l} ,
\nonumber
\\
{\boldsymbol l}'' = l'_{0} {\boldsymbol k} + {\boldsymbol l}'
k_{0} - {\boldsymbol l}' \times {\boldsymbol k} + m_{0}'
{\boldsymbol l} - {\boldsymbol m}' l_{0} + {\boldsymbol m}'
\times {\boldsymbol l } . 
\end{gather*}
Correspondingly, expression for determinant (\ref{11}) becomes
\begin{gather*}
D= [kk] [mm] + [ll] [nn] + 2 [mk] [ln] +2 [lk] [nm] -2 [nk] [lm]
\nonumber
\\
\phantom{D=}{} + 2 [ k_{0} {\boldsymbol l} ({\boldsymbol m} \times
{\boldsymbol n}) + m_{0} {\boldsymbol k} ({\boldsymbol n} \times
{\boldsymbol l}) + l_{0} {\boldsymbol k} ({\boldsymbol n} \times
{\boldsymbol m}) + n_{0} {\boldsymbol l} ({\boldsymbol m} \times
{\boldsymbol k}) ] \nonumber
\\
\phantom{D=}{}+ 4 ({\boldsymbol k} {\boldsymbol n} ) ({\boldsymbol
m} {\boldsymbol l}) - 4 ({\boldsymbol k} {\boldsymbol m}
) ({\boldsymbol n} {\boldsymbol l}) , 
\end{gather*}
where the notation is used:
 $[ab] = a_{0}b_{0}
+ {\boldsymbol a } {\boldsymbol b} $.

\subsection*{Case B}

Equations (\ref{7}) permit the following restrictions:
\begin{gather*}
m_{a} = k^{*}_{a} , \qquad l_{a} = n^{*}_{a} , 
\end{gather*}
and 
become
\begin{gather*}
k_{0}'' = k_{0}' k_{0} + {\boldsymbol k}' {\boldsymbol k}
 - n'_{0} n^{*}_{0} + {\boldsymbol n}' {\boldsymbol n}^{*} ,
\nonumber
\\
{\boldsymbol k}'' = k'_{0} {\boldsymbol k} + {\boldsymbol k}'
k_{0} + i {\boldsymbol k}' \times {\boldsymbol k} - n_{0}'
{\boldsymbol n}^{*} + {\boldsymbol n}' n_{0}^{*} + i {\boldsymbol
n}' \times {\boldsymbol n}^{*} , \nonumber
\\
n_{0}'' = k_{0}' n_{0} - {\boldsymbol k}' {\boldsymbol n}
 + n'_{0} k_{0}^{*} + {\boldsymbol n}' {\boldsymbol k}^{*} ,
\nonumber
\\
{\boldsymbol n}'' = k'_{0} {\boldsymbol n} - {\boldsymbol k}'
n_{0} + i {\boldsymbol k}' \times {\boldsymbol n} + n_{0}'
{\boldsymbol k}^{*} + {\boldsymbol n}' k^{*}_{0}
- i {\boldsymbol n}' \times {\boldsymbol k}^{*} . 
\end{gather*}
Determinant $D$ is given by
\begin{gather*}
D=
 (kk) (kk) ^{*} + (nn)^{*} (nn) + 2 (k^{*}k) (n^{*}n)
 +2 (n^{*}k) (nk^{*}) -2 (nk) (nk)^{*}
\nonumber
\\
\phantom{D=}{} + 2 i [ k_{0} {\boldsymbol k}^{*} ({\boldsymbol n}
\times {\boldsymbol n}^{*}) - k_{0}^{*} {\boldsymbol k}
({\boldsymbol n}^{*} \times {\boldsymbol n})
 + n_{0}^{*} {\boldsymbol n} ({\boldsymbol k} \times {\boldsymbol k}^{*}) -
n_{0} {\boldsymbol n}^{*} ({\boldsymbol k}^{*} \times {\boldsymbol
k}) ] \nonumber
\\
\phantom{D=}{}+ 4 ({\boldsymbol k} {\boldsymbol n} ) ({\boldsymbol
k}^{*} {\boldsymbol n}^{*}) -
 4 ({\boldsymbol k} {\boldsymbol k}^{*} ) ({\boldsymbol n} {\boldsymbol n}^{*}) .
\end{gather*}

\subsection*{Case C}

In (\ref{12b}) one can impose additional restrictions
\begin{gather}
m_{0} = k_{0} , \qquad l_{0} = n_{0} , \qquad {\boldsymbol m} = -
{\boldsymbol k} , \qquad {\boldsymbol l} = -{\boldsymbol n} ;
\label{14a}
\end{gather}
at this $G(k_{0},{\boldsymbol k},n_{0},{\boldsymbol n})$ looks
\begin{gather*}
G = \left | \begin{array}{cc}
(k_{0} +i {\boldsymbol k} \vec{\sigma}) & (n_{0} - i {\boldsymbol n} \vec{\sigma} ) \\[1mm]
 - (n_{0} - i {\boldsymbol n} \vec{\sigma} )& (k_{0} + i {\boldsymbol k} \vec{\sigma})
\end{array} \right | ;
\end{gather*}
and the composition rule is
\begin{gather*}
k_{0}'' = k_{0}' k_{0} - {\boldsymbol k}' {\boldsymbol k}
 - n'_{0} n_{0} + {\boldsymbol n}' {\boldsymbol n} ,
\nonumber
\\
{\boldsymbol k}'' = k'_{0} {\boldsymbol k} + {\boldsymbol k}'
k_{0} - {\boldsymbol k}' \times {\boldsymbol k} + n_{0}'
{\boldsymbol n} + {\boldsymbol n}' n_{0} + {\boldsymbol n}' \times
{\boldsymbol n} , \nonumber
\\
n_{0}'' = k_{0}' n_{0} + {\boldsymbol k}' {\boldsymbol n}
 + n'_{0} k_{0} + {\boldsymbol n}' {\boldsymbol k} ,
\nonumber
\\
{\boldsymbol n}'' = k'_{0} {\boldsymbol n} - {\boldsymbol k}'
n_{0} - {\boldsymbol k}' \times {\boldsymbol n} - n_{0}'
{\boldsymbol k} + {\boldsymbol n}' k_{0} - {\boldsymbol
n}' \times {\boldsymbol k} . 
\end{gather*}
 Determinant equals to
 \begin{gather*}
\det G =
 [kk] [kk] + [nn] [nn]
+ 2 (kk) (nn) +2 (nk) (nk) -2 [nk] [nk]) \nonumber
\\
\phantom{\det G =}{} + 4 ({\boldsymbol k} {\boldsymbol n} )
({\boldsymbol k} {\boldsymbol n}) - 4 ({\boldsymbol k}
{\boldsymbol k}
) ({\boldsymbol n} {\boldsymbol n}) . 
\end{gather*}

\subsection*{Case D}
There exists one other subgroup def\/ined by
\begin{gather*}
n_{a} =0 , \qquad l_{a} = 0 , \qquad G = \left |
\begin{array}{cc}
(k_{0} + {\boldsymbol k} \vec{\sigma}) & 0 \\[1mm]
 0 & (m_{0} - {\boldsymbol m} \vec{\sigma})
\end{array} \right | ,
\end{gather*}
the composition law (\ref{7}) becomes simpler
\begin{gather*}
k_{0}'' = k_{0}' k_{0} + {\boldsymbol k}' {\boldsymbol k}
 , \qquad
{\boldsymbol k}'' = k'_{0} {\boldsymbol k} + {\boldsymbol k}'
k_{0} + i {\boldsymbol k}' \times {\boldsymbol k} , \nonumber
\\
m_{0}'' = m_{0}' m_{0} + {\boldsymbol m}' {\boldsymbol m}
 ,
{\boldsymbol m}'' = m'_{0} {\boldsymbol m} + {\boldsymbol m}'
m_{0} - i {\boldsymbol
m}' \times {\boldsymbol m} , 
\end{gather*}
as well as the determinant $D$
\begin{gather*}
\det G =
 (kk) (mm) .
\end{gather*}
If one additionally imposes two requirements $(kk)=+1$,
 $(mm)= +1$, the Case~D describes spinor covering for
special complex rotation group $SO(4,C)$; this most simple case
was considered in detail in \cite{Bogush-2006}.

It should be noted that the above general expression (\ref{11})
for determinant can be transformed to a shorter form
\begin{gather*}
\det G =
 (kk) (mm) + (nn) (ll) + 2 [kn] [ml]
\nonumber
\\
\phantom{\det G =}{} - 2 ( k_{0} {\boldsymbol n} + n_{0}
{\boldsymbol k} - i {\boldsymbol k} \times {\boldsymbol n} ) (
m_{0} {\boldsymbol l} + l_{0} {\boldsymbol m} +i {\boldsymbol m}
\times {\boldsymbol l} ) , 
\end{gather*}
which for the three Cases A, B, C becomes yet simpler:
\begin{gather*}
{\rm (A)}: \quad  \det G =
 [kk] [mm] + [nn] [ll] + 2 (kn) (ml)
 \nonumber
 \\
\phantom{{\rm (A)}: \quad  \det G =}{} + 2 ( k_{0} {\boldsymbol n}
+ n_{0} {\boldsymbol k} + {\boldsymbol k} \times {\boldsymbol n} )
( m_{0} {\boldsymbol l} + l_{0} {\boldsymbol m} - {\boldsymbol m}
\times {\boldsymbol l} ) , \nonumber
\\
{\rm (B)}: \quad  \det G =
 (kk) (k^{*}k^{*}) + (nn) (n^{*}n^{*}) + 2 [kn] [k^{*} n^{*}]
\nonumber
\\
\phantom{{\rm (B)}: \quad  \det G =}{} - 2 ( k_{0} {\boldsymbol n}
+ n_{0} {\boldsymbol k} - i
 {\boldsymbol k} \times {\boldsymbol n} ) ( k^{*}_{0} {\boldsymbol n}^{*} +
n^{*}_{0} {\boldsymbol k}^{*} +i {\boldsymbol k}^{*} \times
{\boldsymbol n}^{*} ) , \nonumber
\\
{\rm (C)}: \quad  \det G =
 [kk]^{2} + [nn]^{2} + 2 (kn)^{2} -
 2 ( k_{0} {\boldsymbol n} + n_{0} {\boldsymbol k} + {\boldsymbol k} \times {\boldsymbol n} )^{2} .
\end{gather*}

\section{Unitarity condition}\label{sec3}

Now let us turn to consideration of the unitary group $SU(4)$. One
should specify the requirement of unitarity $ G^{+} = G^{-1} $ to
the above vector parametrization. Taking into account the formulas
 \begin{gather}
 G^{+} = \left | \begin{array}{rr}
k_{0}^{*} + {\boldsymbol k}^{*} \vec{\sigma} & - l_{0}^{*} - {\boldsymbol l}^{*} \vec{\sigma} \\[1mm]
n_{0}^{*} - {\boldsymbol n} ^{*} \vec{\sigma} & m_{0}^{*} -
 {\boldsymbol m}^{*} \vec{\sigma}
\end{array} \right | , \qquad
G ^{-1} = \left | \begin{array}{rr}
k'_{0} + {\boldsymbol k}' \vec{\sigma} & n'_{0} - {\boldsymbol n}' \vec{\sigma} \\[1mm]
- l'_{0} - {\boldsymbol l}' \vec{\sigma} & m'_{0} - {\boldsymbol
m} ' \vec{\sigma}
\end{array} \right | ,
\label{3a}
\end{gather}
which can be represented dif\/ferently
\begin{gather*}
 G^{+} =  G ( k_{0}^{*},  {\boldsymbol
k}^{*};  m_{0}^{*},  {\boldsymbol m}^{*};
  -l_{0}^{*},  {\boldsymbol l}^{*}, -n_{0}^{*},{\boldsymbol n}^{*}) ,\qquad
 G^{-1}=  G( k'_{0},  {\boldsymbol k}';  m'_{0},  {\boldsymbol m}';  n'_{0},
 {\boldsymbol n}',  l'_{0},{\boldsymbol l}' ) ,
\end{gather*}
 we arrive at
 \begin{gather}
 k_{0}^{*} = k'_{0} , \qquad {\boldsymbol k}^{*} = {\boldsymbol k}'
, \qquad m_{0}^{*} = m'_{0} , \qquad {\boldsymbol m}^{*} =
{\boldsymbol m}' , \nonumber
\\
-l_{0}^{*} = n'_{0} , \qquad {\boldsymbol l}^{*} = {\boldsymbol
n}' , \qquad -n_{0}^{*}= l'_{0} , \qquad {\boldsymbol n}^{*} =
{\boldsymbol l}' . \label{1.2}
\end{gather}
With the use of expressions for parameters of the inverse matrix
with additional restriction
 $ \det G = +1$ equations (\ref{1.2}) can be rewritten as
 \begin{gather}
k_{0}^{*} = + k_{0} (mm) + m_{0} (ln) + l_{0} (nm)
 - n_{0} (lm) + i {\boldsymbol l} ({\boldsymbol m} \times {\boldsymbol n} ) ,
\nonumber
\\
m_{0}^{*} = + m_{0} (kk) + k_{0} (nl) + n_{0} (lk)
 - l_{0} (nk) - i {\boldsymbol n} ( {\boldsymbol k} \times {\boldsymbol l} ) ,
\nonumber
\\
{\boldsymbol k}^{*} =
 - {\boldsymbol k} (mm) - {\boldsymbol m} (ln) - {\boldsymbol l} (nm) + {\boldsymbol n } (lm) +
 2 {\boldsymbol l} \times ({\boldsymbol n} \times {\boldsymbol m})
 \nonumber
 \\
\phantom{{\boldsymbol k}^{*} =}{} +
 i m_{0} ( {\boldsymbol n} \times {\boldsymbol l} ) + i l_{0} ( {\boldsymbol n} \times {\boldsymbol m} ) + i
n_{0} ( {\boldsymbol l} \times {\boldsymbol m} ) , \nonumber
\\
{\boldsymbol m}^{*} = - {\boldsymbol m} (kk) - {\boldsymbol k}
(nl) - {\boldsymbol n} (lk) + {\boldsymbol l } (nk) +
 2 {\boldsymbol n} \times ( {\boldsymbol l} \times {\boldsymbol k})
 \nonumber
 \\
\phantom{{\boldsymbol m}^{*} =}{}  - i k_{0} ({\boldsymbol l}
\times {\boldsymbol n}) - in_{0} ({\boldsymbol l} \times
{\boldsymbol k}) - i l_{0} ( {\boldsymbol n} \times {\boldsymbol
k}) , \nonumber
\\
 l_{0}^{*} = + k_{0} (nm) - m_{0}
(kn) + l_{0} (nn) + n_{0} (km) +
 i {\boldsymbol k} ({\boldsymbol n} \times {\boldsymbol m} ) ,
 \nonumber
 \\
n_{0}^{*} = + m_{0} (lk) - k_{0} (ml) + n_{0} (ll) + l_{0} (mk) -
i {\boldsymbol m} ( {\boldsymbol l} \times {\boldsymbol k} ) ,
\nonumber
\\
{\boldsymbol l}^{*} = - {\boldsymbol k} (nm) + {\boldsymbol m}
(kn) - {\boldsymbol l} (nn) - {\boldsymbol n} (km) + 2
{\boldsymbol k} \times ( {\boldsymbol m} \times {\boldsymbol n} )
\nonumber
\\
\phantom{n_{0}^{*} =}{} + i k_{0} ({\boldsymbol m} \times
{\boldsymbol n}) + i m_{0} ({\boldsymbol k} \times {\boldsymbol
n}) + i n_{0} ({\boldsymbol m} \times {\boldsymbol k}) , \nonumber
\\
{\boldsymbol n}^{*} = - {\boldsymbol m} (kl) + {\boldsymbol k}
(ml) - {\boldsymbol n} (ll) - {\boldsymbol l} (mk) + 2{\boldsymbol
m} \times ( {\boldsymbol k} \times {\boldsymbol l} ) - i m_{0}
({\boldsymbol k} \times {\boldsymbol l} ) \nonumber
\\
\phantom{{\boldsymbol n}^{*} =}{} - i k_{0} ({\boldsymbol m}
\times {\boldsymbol l} ) - i l_{0} ({\boldsymbol k} \times
{\boldsymbol m}) . \label{1.3}
\end{gather}
Thus, the known form for parameters of the inverse matrix
 $G^{-1}$ makes possible to write easily relations
(\ref{1.3}) representing the unitarity condition for group
$SU(4)$. Here there are 16 equations for 16 variables; evidently,
not all of them are independent.

Let us write down several simpler cases.

\subsection*{Case A}

With formal change\footnote{Let 0-components $k_{0}$, $m_{0}$,
$l_{0}$, $n_{0}$ be real-valued, and 3-vectors
 ${\boldsymbol k}$, ${\boldsymbol m}$, ${\boldsymbol l}$, ${\boldsymbol n}$ be imaginary.}
\begin{gather}
{\boldsymbol k} \Longrightarrow i {\boldsymbol k} ,\qquad
{\boldsymbol m} \Longrightarrow i {\boldsymbol m} ,\qquad
{\boldsymbol l} \Longrightarrow i {\boldsymbol l} ,\qquad
{\boldsymbol n} \Longrightarrow i {\boldsymbol n} , \label{1.4a}
\end{gather}
equations (\ref{1.3}) give
\begin{gather*}
k_{0} = + k_{0} [mm] + m_{0} [ln] + l_{0} [nm]
 - n_{0} [lm] + {\boldsymbol l} ({\boldsymbol m} \times {\boldsymbol n} ) ,
 \nonumber
 \\
m_{0} = + m_{0} [kk] + k_{0} [nl] + n_{0} [lk] - l_{0} [nk] -
{\boldsymbol n} ({\boldsymbol k} \times {\boldsymbol l} ) ,
\nonumber
\\
{\boldsymbol k} =
 {\boldsymbol k} [mm] + {\boldsymbol m} [ln] + {\boldsymbol l} [nm] - {\boldsymbol n } [lm] +
 2 {\boldsymbol l} \times ({\boldsymbol n} \times {\boldsymbol m})
 \nonumber
 \\
\phantom{{\boldsymbol k} =}{} +
 m_{0} ( {\boldsymbol n} \times {\boldsymbol l} ) + l_{0} ( {\boldsymbol n} \times {\boldsymbol m} ) +
n_{0} ( {\boldsymbol l} \times {\boldsymbol m} ) , \nonumber
\\
{\boldsymbol m} = + {\boldsymbol m} [kk] + {\boldsymbol k} [nl] +
{\boldsymbol n} [lk] - {\boldsymbol l } [nk] +
 2 {\boldsymbol n} \times ( {\boldsymbol l} \times {\boldsymbol k})
 \nonumber
 \\
\phantom{{\boldsymbol m} =}{} - k_{0} ({\boldsymbol l} \times
{\boldsymbol n}) - n_{0} ({\boldsymbol l} \times {\boldsymbol k})
- l_{0} ( {\boldsymbol n} \times {\boldsymbol k})
 ,
\nonumber
\\
l_{0} = + k_{0} [nm] - m_{0} [kn] + l_{0} [nn] + n_{0} [km] +
 {\boldsymbol k} ({\boldsymbol n} \times {\boldsymbol m} ) ,
\nonumber
\\
n_{0}= + m_{0} [lk] - k_{0} [ml] + n_{0} [ll] + l_{0} [mk]
 - {\boldsymbol m} ( {\boldsymbol l} \times {\boldsymbol k} ) ,
\nonumber
\\
{\boldsymbol l} = + {\boldsymbol k} [nm] - {\boldsymbol m} [kn] +
{\boldsymbol l} [nn] + {\boldsymbol n} [km] + 2 {\boldsymbol k}
\times ( {\boldsymbol m} \times {\boldsymbol n} ) \nonumber
\\
\phantom{{\boldsymbol l} =}{} + k_{0} ({\boldsymbol m} \times
{\boldsymbol n}) +
 m_{0} ({\boldsymbol k} \times {\boldsymbol n}) +
 n_{0} ({\boldsymbol m} \times {\boldsymbol k}) ,
\nonumber
\\
{\boldsymbol n} = + {\boldsymbol m} [kl] - {\boldsymbol k} [ml] +
{\boldsymbol n} [ll] + {\boldsymbol l} [mk] + 2{\boldsymbol m}
\times ( {\boldsymbol k} \times {\boldsymbol l} ) \nonumber
\\
\phantom{{\boldsymbol n} =}{}- m_{0}({\boldsymbol k} \times
{\boldsymbol l} ) - k_{0} ({\boldsymbol m} \times {\boldsymbol l}
) - l_{0} ({\boldsymbol k} \times {\boldsymbol m}). 
\end{gather*}
Here there are 16 equations for 16 real-valued variables.

\subsection*{Case B}

Let
\begin{gather*}
m_{0} = k^{*}_{0} , \qquad {\boldsymbol m} = {\boldsymbol k}^{*} ,
\qquad l_{0} = n^{*}_{0} , \qquad {\boldsymbol l} = {\boldsymbol
n}^{*} , \nonumber
\\
k_{0} = m^{*}_{0} , \qquad {\boldsymbol k} = {\boldsymbol m}^{*} ,
\qquad n_{0} = l^{*}_{0} , \qquad {\boldsymbol n} = {\boldsymbol
l}^{*} ,
\end{gather*}
or symbolically $ m = k^{*}$, $l= n^{*} $. The unitarity relations
become
\begin{gather*}
k_{0}^{*} =
 + k_{0} (k^{*}k^{*}) + k^{*}_{0} (n^{*}n) + n^{*}_{0} (nk^{*})
 - n_{0} (n^{*}k^{*}) + i {\boldsymbol n}^{*} ({\boldsymbol k}^{*} \times {\boldsymbol n} ) ,
\nonumber
\\
{\boldsymbol k}^{*} =
 - {\boldsymbol k} (k^{*}k^{*}) - {\boldsymbol k}^{*} (n^{*}n) - {\boldsymbol n}^{*} (nk^{*}) +
 {\boldsymbol n } (n^{*}k^{*})
\nonumber
\\
\phantom{{\boldsymbol k}^{*} =}{} +
 2 {\boldsymbol n}^{*} \times ({\boldsymbol n} \times {\boldsymbol k}^{*}) +
 i k^{*}_{0} ( {\boldsymbol n} \times {\boldsymbol n}^{*} ) + i n^{*}_{0} ( {\boldsymbol n} \times {\boldsymbol k}^{*} ) + i
n_{0} ( {\boldsymbol l} \times {\boldsymbol m} ) , \nonumber
\\
n_{0}^{*}=
 + k^{*}_{0} (n^{*}k)
 - k_{0} (k^{*}n^{*})
 + n_{0} (n^{*}n^{*})
 + n^{*}_{0} (k^{*}k)
 - i {\boldsymbol k}^{*} ( {\boldsymbol n}^{*} \times {\boldsymbol k} ) ,
\nonumber
\\
{\boldsymbol n}^{*} = - {\boldsymbol k}^{*} (kn^{*}) +
{\boldsymbol k} (k^{*}n^{*}) - {\boldsymbol n} (n^{*}n^{*}) -
{\boldsymbol n}^{*} (k^{*}k) \nonumber
\\
\phantom{{\boldsymbol n}^{*} =}{} + 2{\boldsymbol k}^{*} \times (
{\boldsymbol k} \times {\boldsymbol n}^{*} ) - i
k^{*}_{0}({\boldsymbol k} \times {\boldsymbol n}^{*} ) - i k_{0}
({\boldsymbol k}^{*} \times {\boldsymbol n}^{*} ) - i n^{*}_{0}
({\boldsymbol k} \times {\boldsymbol k}^{*}) ,
\end{gather*}
and 8 conjugated ones
\begin{gather*}
k_{0} = + k^{*}_{0} (kk) + k_{0} (nn^{*}) + n_{0} (n^{*}k) -
n^{*}_{0} (nk) - i {\boldsymbol n} ({\boldsymbol k} \times
{\boldsymbol n}^{*} ) , \nonumber
\\
{\boldsymbol k} = - {\boldsymbol k}^{*} (kk) - {\boldsymbol k}
(nn^{*}) - {\boldsymbol n} (n^{*}k) + {\boldsymbol n }^{*} (nk)
\nonumber
\\
\phantom{{\boldsymbol k} =}{} +
 2 {\boldsymbol n} \times ( {\boldsymbol n}^{*} \times {\boldsymbol k})
 - i k_{0} ({\boldsymbol n}^{*} \times {\boldsymbol n}) - in_{0} ({\boldsymbol n}^{*} \times {\boldsymbol k})
 - i n^{*}_{0} ( {\boldsymbol n} \times {\boldsymbol k})
 ,
\nonumber
\\
n_{0} = + k_{0} (nk^{*}) - k^{*}_{0} (kn) + n^{*}_{0} (nn) + n_{0}
(kk^{*}) +
 i {\boldsymbol k} ({\boldsymbol n} \times {\boldsymbol k}^{*} ) ,
\nonumber
\\
{\boldsymbol n} = - {\boldsymbol k} (nk^{*}) + {\boldsymbol k}^{*}
(kn) - {\boldsymbol n}^{*} (nn) - {\boldsymbol n} (kk^{*})
\nonumber
\\
\phantom{{\boldsymbol n} =}{} + 2 {\boldsymbol k} \times (
{\boldsymbol k}^{*} \times {\boldsymbol n} ) + i k_{0}
({\boldsymbol k}^{*} \times {\boldsymbol n}) + i k^{*}_{0}
({\boldsymbol k} \times {\boldsymbol
n}) + i n_{0} ({\boldsymbol k}^{*} \times {\boldsymbol k}) . 
\end{gather*}
It may be noted that latter relations are greatly simplif\/ied
when
 $n=0$, or when $k=0$. Firstly, let us consider the case $n=0$:
\begin{gather*}
k_{0}^{*} =
 + k_{0} (k^{*} k^{*}) , \qquad
 {\boldsymbol k}^{*} = - {\boldsymbol k} (k^{*}k^{*}) .
\end{gather*}
Taking in mind the identity
\begin{gather*}
\det G = (kk) (kk)^{*} = +1 \qquad \Longrightarrow \qquad
 (kk) = +1 ,\qquad (kk)^{*} = +1 ,
\nonumber
\end{gather*}
 we arrive at
 $ k_{0}^{*} = + k_{0}$, ${\boldsymbol
k}^{*} = - {\boldsymbol k} $. It has sense to introduce the
real-valued vector $c_{a}$:
\begin{gather*}
k_{0}^{*} = + k_{0} = c_{0} , \qquad {\boldsymbol k}^{*} = -
{\boldsymbol k} : \qquad {\boldsymbol k} = i {\boldsymbol c} ,
\nonumber
\end{gather*}
then matrix $G$ is
\begin{gather*}
G (k,m=k^{*},0,0) = \left | \begin{array}{cc}
c_{0} + i {\boldsymbol c} \vec{\sigma} & 0 \\
0 & c_{0} - i {\boldsymbol c} \vec{\sigma}
\end{array} \right | \sim SU(2) .
\end{gather*}

Another possibility is realized when $k=0$:
\begin{gather*}
n_{0}^{*}=
 + n_{0} (nn)^{*} , \qquad
{\boldsymbol n}^{*} = - {\boldsymbol n} (nn)^{*} . 
\end{gather*}
With the use of identity
\begin{gather*}
\det G = (nn) (nn)^{*} = +1  \qquad \Longrightarrow \qquad (nn)=
+1,   \qquad (nn)^{*} = +1 , \nonumber
\end{gather*}
we get
\begin{gather*}
n_{0}^{*}=
 + n_{0} = c_{0} , \qquad
{\boldsymbol n}^{*} = - {\boldsymbol n} , \qquad {\boldsymbol n}
\equiv i {\boldsymbol c}
 ,
\end{gather*}
Corresponding matrices $G(0,0,n,l=n^{*})$ make up a special set of
unitary matrices
\begin{gather}
G = \left | \begin{array}{cc}
0 & c_{0} - i {\boldsymbol c} \vec{\sigma} \\
 -( c_{0} + i {\boldsymbol c} \vec{\sigma} ) & 0
\end{array} \right | , \qquad
G^{+} = \left | \begin{array}{cc}
0 & -(c_{0} - i {\boldsymbol c} \vec{\sigma}) \\
 ( c_{0} + i {\boldsymbol c} \vec{\sigma} ) & 0
\end{array} \right | .
\label{1.7c}
\end{gather}
However, it must be noted that these matrices (\ref{1.7c}) do not
provide us with any subgroup because $ G^{2} = - I$.

\subsection*{Case C}

Now in equations (\ref{1.4a}) one should take
\begin{gather*}
m_{0} = k_{0} , \qquad l_{0} = n_{0} , \qquad
 {\boldsymbol m} = - {\boldsymbol k} , \qquad
 {\boldsymbol l} = -{\boldsymbol n} ,
\end{gather*}
then
\begin{gather}
k_{0} =
 + k_{0} [kk] + k_{0} (nn) + n_{0} (nk)
 - n_{0} [nk] ,
\nonumber
\\
{\boldsymbol k} =
 {\boldsymbol k} [kk] - {\boldsymbol k} (nn) - {\boldsymbol n} (nk) - {\boldsymbol n } [nk] +
 2 {\boldsymbol n} \times ({\boldsymbol n} \times {\boldsymbol k}) ,
\nonumber
\\
n_{0} = + k_{0} (nk) - k_{0} [kn] + n_{0} [nn] + n_{0} (kk) ,
\nonumber
\\
{\boldsymbol n} = - {\boldsymbol k} (kn) - {\boldsymbol k} [kn] +
{\boldsymbol n} [nn] - {\boldsymbol n} (kk) + 2{\boldsymbol k}
\times ( {\boldsymbol k} \times {\boldsymbol n} ) . \label{1.8b}
\end{gather}

\section[2-parametric subgroups in $SU(4)$]{2-parametric subgroups in $\boldsymbol{SU(4)}$}\label{sec4}

 To be certain in correctness of the produced equations of unitarity,
 one should try to solve them at least in several most simple particular cases.
For instance, let us turn to the Case~C and specify equations
(\ref{1.8b})
 for a subgroup arising when $k=(k_{0},k_{1},0,0)$ and $ n=(n_{0},n_{1},0,0)$:
\begin{gather}
k_{0} =
 + k_{0} [kk] + k_{0} (nn) + n_{0} (nk)
 - n_{0} [nk] ,
\nonumber
\\
 k_{1} = +
 k_{1} [kk] - k_{1} (nn) - n_{1} (nk) - n_{1 } [nk] ,
\nonumber
\\
n_{0} = + k_{0} (nk) - k_{0} [kn] + n_{0} [nn] + n_{0} (kk) ,
\nonumber
\\
n_{1} = - k_{1} (kn) - k_{1} [kn] + n_{1} [nn] - n_{1} (kk),
\label{2.1}
\end{gather}
they are four non-linear equations for four real variables. It may
be noted that equations~(\ref{2.1}) can be regarded as two
eigenvalue problems in two dimensional space (with
eigenvalue~$+1$):
\begin{gather*}
\left | \begin{array}{cc}
(k_{0}^{2} + n_{0}^{2})-1 +(k_{1}^{2} -n_{1}^{2} ) & -2n_{1} k_{1} \\
-2n_{1}k_{1} & (k_{0}^{2} + n_{0}^{2}) -1 -(k_{1}^{2} -n_{1}^{2})
\end{array} \right |
\left | \begin{array}{c} k_{0}\\n_{0}
\end{array} \right | =
\left | \begin{array}{c}
 0\\0
\end{array} \right | ,
\\
\left | \begin{array}{cc}
(k_{1}^{2} + n_{1}^{2})-1 +(k_{0}^{2} -n_{0}^{2} ) & -2n_{0} k_{0} \\
-2n_{0}k_{0} & (k_{1}^{2} + n_{1}^{2}) -1 -(k_{0}^{2} -n_{0}^{2})
\end{array} \right |
\left | \begin{array}{c} k_{1}\\n_{1}
\end{array} \right | =
\left | \begin{array}{c}
 0\\0
\end{array} \right | .
\end{gather*}
The determinants in both problems must be equated to zero:
\begin{gather*}
[(k_{0}^{2} + n_{0}^{2})-1]^{2} - (k_{1}^{2} -n_{1}^{2} )^{2} -
4n_{1}^{2} k_{1}^{2} = 0 , \nonumber
\\
[(k_{1}^{2} + n_{1}^{2})-1]^{2} - (k_{0}^{2} -n_{0}^{2} )^{2}
-4n^{2}_{0} k^{2}_{0} =0 , \nonumber
\end{gather*}
or
\begin{gather*}
[(k_{0}^{2} + n_{0}^{2})-1]^{2} - (k_{1}^{2} +n_{1}^{2} )^{2} = 0
 , \qquad
 [(k_{1}^{2} + n_{1}^{2})-1]^{2} - (k_{0}^{2}
+n_{0}^{2} )^{2} =0 .
\nonumber
\end{gather*}
The latter equations may be rewritten in factorized form:
\begin{gather*}
[(k_{0}^{2} + n_{0}^{2})-1 - (k_{1}^{2} +n_{1}^{2} ) ] [(k_{0}^{2}
+ n_{0}^{2})-1 + (k_{1}^{2} +n_{1}^{2} ) ] = 0 , \nonumber
\\
 [(k_{1}^{2} + n_{1}^{2})-1 - (k_{0}^{2} +n_{0}^{2} )]
[(k_{1}^{2} + n_{1}^{2})-1 + (k_{0}^{2} +n_{0}^{2} )] =0 .
\end{gather*}
 They have the structure: $ AC = 0$, $BC = 0 $. Four dif\/ferent cases arise.

(1) Let $C= 0 $, then
\begin{gather}
 k_{0}^{2} + n_{0}^{2} + k_{1}^{2} +n_{1}^{2}
= +1 . \label{2.6}
\end{gather}

(2) Now, let $A=0$, $B=0$, but a contradiction arises: $
 A+B =0$, $A+B = -2 $.

(3)--(4) There are two simples cases:
\begin{gather}
A=0, C=0 \qquad k_{0}^{2} + n_{0}^{2} =1 , \qquad k_{1}=0,n_{1} =
0 , \label{2.7a}
\\
B=0, C=0 \qquad k_{1}^{2} + n_{1}^{2} =1 , \qquad k_{0}=0,n_{0} =
0 . \label{2.7b}
\end{gather}
Evidently, (\ref{2.7a}) and (\ref{2.7b}) can be regarded as
particular cases of the above variant (\ref{2.6}). Now, one should
take into account additional relation
\begin{gather*}
\det G =
 [kk] [kk] + [nn] [nn] + 2 (kk) (nn)
\nonumber
\\
\phantom{\det G =}{} + 2 (nk) (nk) -2 [nk] [nk]) +
 4 ({\boldsymbol k} {\boldsymbol n} ) ({\boldsymbol k} {\boldsymbol n}) - 4 ({\boldsymbol k} {\boldsymbol k} ) ({\boldsymbol n} {\boldsymbol n}) = 1 ,
\nonumber
\end{gather*}
which can be transformed to
\begin{gather}
\det G = (k_{0}^{2} + k_{1}^{2} + n_{0}^{2} + n_{1}^{2})^{2} -4
(k_{1} n_{0} + k_{0} n_{1})^{2} =+1 . \label{2.8}
\end{gather}
Both equations (\ref{2.6}) and (\ref{2.8}) are to be satisf\/ied
\begin{gather*}
(k_{0}^{2} + n_{0}^{2} + k_{1}^{2} +n_{1}^{2}) = 1 ,\qquad
(k_{0}^{2} + k_{1}^{2} + n_{0}^{2} + n_{1}^{2})^{2} -1 -4 (k_{1}
n_{0} + k_{0} n_{1})^{2} = 0 ,
\end{gather*}
from where it follows
\begin{gather*}
k_{1} n_{0} + k_{0} n_{1} =0 , \qquad k_{0}^{2} +
n_{0}^{2} + k_{1}^{2} +n_{1}^{2} = +1 . 
\end{gather*}
They specify a 2-parametric unitary subgroup in $SU(4)$
\begin{gather}
G_{1}^{+}= G_{1}^{-1} , \qquad \det G_{1} = +1 , \nonumber
\\
k_{1} n_{0} + k_{0} n_{1} =0 , \qquad k_{0}^{2} + n_{0}^{2} +
k_{1}^{2} +n_{1}^{2} = +1 , \nonumber
\\
 G _{1} = \left | \begin{array}{rr}
k_{0} +i k_{1} \sigma^{1} & n_{0} - i n_{1} \sigma^{1} \\[3mm]
 - n_{0} + i n_{1} \sigma^{1} & k_{0} + i k_{1} \sigma^{1}
\end{array} \right | =
 \left | \begin{array}{cccc}
k_{0} & ik_{1} & n_{0} & -i n_{1} \\
ik_{1} & k_{0}& -i n_{1} & n_{0} \\
 -n_{0} & i n_{1} & k_{0} & ik_{1} \\
i n_{1} & -n_{0}& ik_{1} & k_{0}
\end{array} \right | .
\label{2.10b}
\end{gather}

Two analogous subgroups are possible:
\begin{gather}
G_{2}^{+}= G_{2}^{-1} , \qquad \det G_{2} = +1 , \nonumber
\\
k_{2} n_{0} + k_{0} n_{2} =0 , \qquad k_{0}^{2} + n_{0}^{2} +
k_{2}^{2} +n_{2}^{2} = +1 , \nonumber
\\
 G _{2} = \left | \begin{array}{rr}
k_{0} +i k_{2} \sigma^{2} & n_{0} - i n_{2} \sigma^{2} \\
 - n_{0} + i n_{2} \sigma^{2} & k_{0} + i k_{2} \sigma^{2}
\end{array} \right |
= \left | \begin{array}{rrrr}
k_{0} & k_{2} & n_{0} & - n_{2} \\
-k_{2} & k_{0}& n_{2} & n_{0} \\
 -n_{0} & n_{2} & k_{0} & k_{2} \\
 -n_{2} & -n_{0}& -k_{2} & k_{0}
\end{array} \right | ;
\label{2.11}
\\
G_{3}^{+}= G_{3}^{-1} , \qquad \det G_{3} = +1 , \nonumber
\\
k_{3} n_{0} + k_{0} n_{3} =0 , \qquad k_{0}^{2} + n_{0}^{2} +
k_{3}^{2} +n_{3}^{2} = +1 , \nonumber
\\
 G _{3} = \left | \begin{array}{cccc}
(k_{0} +i k_{3}) & 0 & (n_{0}- in_{3})& 0 \\
0 & (k_{0} -i k_{3}) & 0 & (n_{0} + in_{3})\\
-(n_{0}- in_{3} ) & 0 & (k_{0} +i k_{3}) & 0 \\
0 & -( n_{0} + in_{3}) & 0 & (k_{0} -i k_{3})
 \end{array} \right | .
\label{2.12}
\end{gather}

 Let us consider the latter subgroup (\ref{2.12}) in some detail.
The multiplication law for parame\-ters~is
\begin{gather*}
k_{0}'' = k_{0}' k_{0} - k_{3}' k_{3}
 - n'_{0} n_{0} + n_{3}' n_{3} ,
\qquad
 k_{3}'' = k'_{0} k_{3} + k_{3}' k_{0} +
n_{0}' n_{3} + n_{3}' n_{0} , \nonumber
\\
n_{0}'' = k_{0}' n_{0} + k_{3}' n_{3}
 + n'_{0} k_{0} + n_{3}' k_{3} ,
\qquad n_{3}'' = k'_{0} n_{3} - k_{3}' n_{0} - n_{0}'
k_{3} + n_{3}' k_{0} . 
\end{gather*}
For two particular cases (see (\ref{2.7a}) and (\ref{2.7b})),
these formulas take the form:
\begin{alignat}{4}
& \{ G_{3}^{'0} \} : \quad &&  k_{3}^{2} + n_{3}^{2} =1 ,\qquad \quad && k_{0}=0,\qquad n_{0} = 0 ,&\nonumber\\
&&& k_{0}'' = - k_{3}' k_{3}  + n_{3}' n_{3} ,\qquad
&& k_{3}'' = 0 ,& \nonumber\\
&&& n_{0}'' = + k_{3}' n_{3}
 + n_{3}' k_{3} , && n_{3}'' = 0 ,
&\nonumber\\
& \{ G^{0} \} : &&  k_{0}^{2} + n_{0}^{2} =1 , && k_{3}=0,\qquad n_{3} = 0 ,& \nonumber\\
& && k_{0}'' = k_{0}' k_{0} - n'_{0} n_{0} , & && \nonumber\\
& &&  n_{0}'' = k_{0}' n_{0}
 + n'_{0} k_{0} . &&&
\label{2.15}
\end{alignat}
Therefore, multiplying of any two elements from $G_{3}^{'0}$ does
not lead us to any element from $G_{3}^{'0}$, instead belonging to
$G^{0}$: $G_{3}^{0'} G_{3}^{0} \in
 G^{0} $. Similar result would be achieved for $G_{1}$ and
$G_{2}$: $ G_{1}^{0'} G_{1}^{0} \in G^{0}$, $G_{2}^{0'} G_{2}^{0}
\in G^{0}$. In the subgroup given by~(\ref{2.15}) one can easily
see the structure of the 1-parametric Abelian subgroup:
\begin{gather}
k_{0} = \cos \alpha , \qquad n_{0} = \sin \alpha , \nonumber
\\
G^{0} (\alpha) = \left | \begin{array}{cccc}
\cos \alpha & 0 & \sin \alpha & 0 \\
0 & \cos \alpha & 0 & \sin \alpha \\[2mm]
- \sin \alpha & 0 & \cos \alpha & 0 \\
0 & - \sin \alpha & 0 & \cos \alpha
 \end{array} \right | , \qquad \alpha \in [0, 2\pi ] .
\label{2.16}
\end{gather}

In the same manner, similar curvilinear parametrization can be
readily produced for 2-parametric groups
(\ref{2.10b})--(\ref{2.12}). For def\/initeness, for the subgroup
 $G_{3}$ such coordinates are given~by
\begin{gather*}
k_{0} = \cos \alpha \cos \rho , \qquad k_{3} = \cos \alpha \sin
\rho , \nonumber
\\
n_{0} = \sin \alpha \cos \rho , \qquad - n_{3} = \sin
\alpha \sin \rho , \alpha \in [0 , 2 \pi ] , 
\end{gather*}
and matrix $G_{3}$ is
\begin{gather}
G_{3}(\rho, \alpha) = \left | \begin{array}{cccc}
\cos \alpha e^{i\rho} & 0 & \sin \alpha e^{i\rho} & 0 \\
0 & \cos \alpha e^{-i\rho} & 0 & \sin \alpha e^{-i\rho}\\
-\sin \alpha e^{i\rho} & 0 & \cos \alpha e^{i\rho} & 0 \\
0 & -\sin \alpha e^{-i\rho} & 0 & \cos \alpha e^{-i\rho}
 \end{array} \right | .
\label{2.18}
\end{gather}
One may note that equation (\ref{2.18}) at $\rho = 0$ will
coincide with $G^{0} (\alpha)$ in (\ref{2.16}): $ G_{3}(\rho=0,
\alpha) = G^{0} (\alpha)  $. Similar curvilinear parametrization
may be introduced for two other subgroups, $G_{1}$~and~$G_{2}$.

One could try to obtain more general result just changing real
valued curvilinear coordinates on complex. However it is easily
verif\/ied that it is not the case: through that change though
there arise subgroups but they are not unitary. Indeed, let the
matrix \eqref{2.16} be complex: then unitarity condition gives
\begin{gather*}
\cos \alpha \cos \alpha^{*} + \sin \alpha \sin \alpha^{*} =1 ,
\qquad - \cos \alpha \sin \alpha^{*} + \sin \alpha \cos
\alpha^{*} =0 . 
\end{gather*}
These two equations can be satisf\/ied only by a real
valued~$\alpha$. In the same manner, the the formal change $
\{G_{1},G_{2},G_{3}\} \Longrightarrow
\{G_{1}^{C},G_{2}^{C},G_{3}^{C}\} $ again provides us with
non-unitary subgroups.

It should be noted that each of three 2-parametric subgroup
$G_{1}$, $G_{2}$, $G_{3} $,
 in addition to $G_{0}(\alpha)$, contains one additional Abelian unitary subgroup:
\begin{gather*}
K_{1} = \left | \begin{array}{cc}
k_{0} +i k_{1} \sigma^{1} & 0 \\
0 & k_{0} + i k_{1} \sigma^{1}
\end{array} \right | , \qquad
K_{1} \subset G _{1} , \qquad k_{0}^{2} + k_{1}^{2} = 1 ,
\nonumber
\\
K_{2} =\left | \begin{array}{cc}
k_{0} +i k_{2} \sigma^{2} & 0 \\
0 & k_{0} + i k_{2} \sigma^{2}
\end{array} \right | , \qquad
K_{2} \subset G _{2} , \qquad k_{0}^{2} + k_{2}^{2} = 1 ,
\nonumber
\\
K_{3} =\left | \begin{array}{cc}
k_{0} +i k_{3} \sigma^{3} & 0 \\
0 & k_{0} + i k_{3} \sigma^{3}
\end{array} \right | , \qquad
K_{3} \subset G _{3} , \qquad k_{0}^{2} + k_{3}^{2} = 1
. 
\end{gather*}
It may be easily verif\/ied that
\begin{gather*}
G_{1} = G_{0} K_{1} = K_{1} G_{0} , \qquad G_{2} = G_{0} K_{2} =
K_{2} G_{0} , \qquad G_{3} = G_{0}
K_{3} = K_{3} G_{0} . 
\end{gather*}
Indeed
\begin{gather*}
G_{0}(\alpha) K_{1} = K_{1} G_{0}(\alpha) = \left |
\begin{array}{cc}
\cos \alpha k_{0} +i \cos \alpha k_{1} \sigma^{1} & \sin \alpha k_{0} +i \sin \alpha k_{1} \\
-\sin \alpha k_{0} -i \sin \alpha k_{1} & \cos \alpha
 k_{0} + i \cos \alpha k_{1} \sigma^{1}
\end{array} \right | ,
\end{gather*}
and with notation
\begin{gather*}
\cos \alpha k_{0} = k_{0}' , \qquad \cos \alpha k_{1} = k_{1}' ,
\qquad \sin \alpha k_{0} = n_{0}' , \qquad \sin \alpha k_{1} = -
n'_{1} , \nonumber
\\
k_{0}'n_{1}' + k_{1}'n_{0}' = 0 , \qquad k_{0}^{'2} +k_{3}^{'2} +
n_{0}^{'2} + n_{3}^{'2} = 1 \nonumber
\end{gather*}
we arrive at
\begin{gather*}
G_{0} K_{1} = K_{1} G_{0} = \left | \begin{array}{rr}
 k_{0}' +i k_{1}' \sigma^{1} & n_{0}' - i n'_{1} \\[3mm]
-n' _{0} + i n'_{1} & k_{0} ' + i k_{1}' \sigma^{1}
\end{array} \right | \subset G_{1}.
\nonumber
\end{gather*}

\section{4-parametric unitary subgroup}\label{sec5}
 Let us turn again to the subgroup in $GL(4,C)$
given by Case C (see~(\ref{14a})):
\begin{gather*}
G = \left | \begin{array}{rr}
(k_{0} + {\boldsymbol k} \vec{\sigma} ) & (n_{0} - {\boldsymbol n} \vec{\sigma} )\\
- (n_{0} - {\boldsymbol n} \vec{\sigma}) & (k_{0} + {\boldsymbol
k}
 \vec{\sigma} )
\end{array} \right | ,
\end{gather*}
when the unitarity equations look as follows:
\begin{gather*}
k_{0} =
 + k_{0} [kk] + k_{0} (nn) + n_{0} (nk)
 - n_{0} [nk] ,
\nonumber
\\
n_{0} = + k_{0} (nk) - k_{0} [kn] + n_{0} [nn] + n_{0} (kk) ,
\nonumber
\\
{\boldsymbol k} =
 {\boldsymbol k} [kk] - {\boldsymbol k} (nn) - {\boldsymbol n} (nk) - {\boldsymbol n } [nk] +
 2 {\boldsymbol n} \times ({\boldsymbol n} \times {\boldsymbol k}) ,
\nonumber
\\
{\boldsymbol n} = - {\boldsymbol k} (kn) - {\boldsymbol k} [kn] +
{\boldsymbol n} [nn] - {\boldsymbol n}
(kk) + 2{\boldsymbol k} \times ( {\boldsymbol k} \times {\boldsymbol n} ) . 
\end{gather*}
They can be rewritten as four eigenvalue problems:
\begin{gather}
\left | \begin{array}{cc}
[kk]+(nn) & (nk) -[nk] \\
(nk) - [nk] & (kk) + [nn]
\end{array} \right |
\left | \begin{array}{c} k_{0}\\  n_{0}
\end{array} \right | =
(+1) \left | \begin{array}{c} k_{0}\\  n_{0}
\end{array} \right | ,
\label{B3}
\\
\left | \begin{array}{cc} +([kk]-[nn])  & - 2(nk) \\
 -2(nk) \qquad & - ([kk] - [nn])
\end{array} \right |
\left | \begin{array}{c} k_{1}\\  n_{1}
\end{array} \right | =
(+1) \left | \begin{array}{c} k_{1}\\ n_{1}
\end{array} \right | ,
\nonumber
\\
\left | \begin{array}{cc} +([kk]-[nn])  & - 2(nk) \\
 -2(nk) \qquad & - ([kk] - [nn])
\end{array} \right |
\left | \begin{array}{c} k_{2}\\  n_{2}
\end{array} \right | =
(+1) \left | \begin{array}{c} k_{2}\\ n_{2}
\end{array} \right | ,
\nonumber
\\
\left | \begin{array}{cc} +([kk]-[nn])  & - 2(nk) \\
 -2(nk) \qquad & - ([kk] - [nn])
\end{array} \right |
\left | \begin{array}{c} k_{3}\\  n_{3}
\end{array} \right | =
(+1) \left | \begin{array}{c} k_{3}\\ n_{3}
\end{array} \right | .
\label{B4}
\end{gather}

\noindent These equations have the same structure
\begin{gather*}
\left |\begin{array}{cc} A & C \\C & B \end{array} \right | \left
| \begin{array}{cc} Z_{1} \\ Z_{2} \end{array} \right | = \lambda
\left | \begin{array}{cc} Z_{1} \\ Z_{2} \end{array}
\right | , 
\end{gather*}
where $\lambda = +1$. Non-trivial solutions may exist only if
\begin{gather*}
\det \left |\begin{array}{cc} A - \lambda & C \\C & B -
\lambda \end{array} \right | = 0 , 
\end{gather*}
which gives two dif\/ferent eigenvalues
\begin{gather*}
\lambda _{1} = {A+B + \sqrt{(A-B)^{2} + 4C^{2} } \over 2} , \qquad
\lambda _{2} = {A+B + \sqrt{(A-B)^{2} + 4C^{2} } \over 2}
 . 
\end{gather*}

In explicit form, equations (\ref{B3}) looks as follows:
\begin{gather}
\left |\begin{array}{cc} A & C \\C & B \end{array} \right | \left
| \begin{array}{cc} k_{0} \\ n_{0} \end{array} \right | = \lambda
\left | \begin{array}{cc} k_{0} \\ n_{0} \end{array} \right | ,
\nonumber
\\
A = (k_{0}^{2} + n_{0}^{2} ) + ( {\boldsymbol k}^{2} -
{\boldsymbol n}^{2}) , \qquad
 B = (k_{0}^{2} + n_{0}^{2} ) -( {\boldsymbol k}^{2} - {\boldsymbol
n}^{2}) , \qquad C = -2 {\boldsymbol k} {\boldsymbol n} ,
\label{B6a}
\\
\lambda_{1} = (k_{0}^{2} + n_{0}^{2} ) + \sqrt{( {\boldsymbol
k}^{2} - {\boldsymbol n}^{2})^{2} + 4({\boldsymbol k} {\boldsymbol
n})^{2}} , \nonumber
\\
\lambda_{2} = (k_{0}^{2} + n_{0}^{2} ) - \sqrt{( {\boldsymbol
k}^{2} - {\boldsymbol n}^{2})^{2} + 4({\boldsymbol k} {\boldsymbol
n})^{2}} . \nonumber
\end{gather}
The eigenvalue $\lambda = +1$ might be constructed by two ways:
\begin{gather}\lambda_{1} = +1 , \qquad k_{0}^{2} + n_{0}^{2}
 =1 - \sqrt{( {\boldsymbol k}^{2} - {\boldsymbol n}^{2})^{2} + 4({\boldsymbol k} {\boldsymbol n})^{2}}
 ,
\nonumber
\\
\lambda_{2} = +1 , \qquad k_{0}^{2} + n_{0}^{2}
 =1 + \sqrt{( {\boldsymbol k}^{2} - {\boldsymbol n}^{2})^{2} + 4({\boldsymbol k} {\boldsymbol n})^{2}}
 . \label{B6b}
\end{gather}

\noindent These two relations (\ref{B6b}) are equivalent to the
following one:
\begin{gather}
( 1 -k_{0}^{2} - n_{0}^{2} ) ^{2}= ( {\boldsymbol k}^{2} -
{\boldsymbol n}^{2})^{2} + 4({\boldsymbol k} {\boldsymbol n})^{2}
. \nonumber
\end{gather}

Thus, equations (\ref{B6a}) have two dif\/ferent types:

Type I
\begin{gather}
(A-1) k_{0} + C n_{0} =0 ,\qquad C k_{0} + (B-1) n_{0} = 0 ,
\nonumber
\\
k_{0}^{2} + n_{0}^{2}
 =1 - \sqrt{( {\boldsymbol k}^{2} - {\boldsymbol n}^{2})^{2} + 4({\boldsymbol k} {\boldsymbol n})^{2}}
 , \label{B6c}
\\
 k_{0}^{2} + n_{0}^{2} < +1 ,
\qquad ( {\boldsymbol k}^{2} - {\boldsymbol n}^{2})^{2} +
4({\boldsymbol k} {\boldsymbol n})^{2} < +1 ; \nonumber
\end{gather}

Type II
\begin{gather}
 (A-1) k_{0} + C
n_{0} =0 ,\qquad C k_{0} + (B-1) n_{0} = 0 , \label{B6d}
\\
k_{0}^{2} + n_{0}^{2}
 =1 + \sqrt{( {\boldsymbol k}^{2} - {\boldsymbol n}^{2})^{2} + 4({\boldsymbol k} {\boldsymbol n})^{2}}
 , \qquad
 k_{0}^{2} + n_{0}^{2} > +1 .
\nonumber
\end{gather}

Now let us turn to equations (\ref{B4}). They have the form
\begin{gather*}
\left |\begin{array}{cc} A & C \\C & -A \end{array} \right | \left
| \begin{array}{cc} k_{i} \\ n_{i} \end{array} \right | = \lambda
\left | \begin{array}{cc} k_{i} \\ n_{i} \end{array} \right | ,
\qquad i = 1,2,3 , \nonumber
\\
A = k_{0}^{2} + {\boldsymbol k}^{2} - n_{0}^{2} - {\boldsymbol
n}^{2} , \qquad B = -A, \qquad C = - 2(k_{0}n_{0} - {\boldsymbol
k} {\boldsymbol n}) ,
\\
\lambda _{1} = +\sqrt { (k_{0}^{2} + {\boldsymbol k}^{2} -
n_{0}^{2} - {\boldsymbol n}^{2}) ^{2} + 4(k_{0}n_{0} -
{\boldsymbol k} {\boldsymbol n})^{2} } , \nonumber
\\
\lambda _{2} = -\sqrt { (k_{0}^{2} + {\boldsymbol k}^{2} -
n_{0}^{2} - {\boldsymbol n}^{2}) ^{2} + 4(k_{0}n_{0} -
{\boldsymbol k} {\boldsymbol n})^{2} } . \nonumber
\end{gather*}

As we are interested only in positive eigenvalue $\lambda = +1$,
we must use only one possibility $\lambda = +1 = \lambda_{1}$, so
that
\begin{gather}
(A-1) {\boldsymbol k} + C {\boldsymbol n} =0 ,\qquad C
{\boldsymbol k}- (A +1) {\boldsymbol n} = 0 , \label{B7b}
\\
1 = (k_{0}^{2} + {\boldsymbol k}^{2} - n_{0}^{2} - {\boldsymbol
n}^{2}) ^{2} + 4(k_{0}n_{0} - {\boldsymbol k} {\boldsymbol n})^{2}
. \nonumber
\end{gather}

Vector condition in (\ref{B7b}) says that
 ${\boldsymbol k}$ and ${\boldsymbol n}$ are (anti)collinear:
\begin{gather}
{\boldsymbol k} = K {\boldsymbol e} , \qquad {\boldsymbol n} = N
{\boldsymbol e}, \qquad {\boldsymbol e}^{2} = 1 ,\qquad
{\boldsymbol e} \in S_{2} , \label{B7c}
\end{gather}
 so that (\ref{B7b}) give
\begin{gather}
(A-1) K + C N =0 , \qquad C K - (A +1) N = 0 , \nonumber
\\
1 = (k_{0}^{2} +K^{2} - n_{0}^{2} - N^{2}) ^{2} + 4(k_{0}n_{0} - K
N )^{2} , \label{B7d}
\\
A = k_{0}^{2} + K^{2} - n_{0}^{2} - N^{2} ,
 C = - 2(k_{0}n_{0} - KN ) .
\nonumber
\end{gather}

With notation (\ref{B7c}), equations (\ref{B6c})--(\ref{B6d}) take
the form:

Type I
\begin{gather*}
 (A-1) k_{0} + C
n_{0} =0 , \qquad  C k_{0} + (B-1) n_{0} = 0 , \qquad k_{0}^{2} +
n_{0}^{2}
 =1 - ( K^{2} + N^{2}) ,
\end{gather*}

Type II
\begin{gather}
 (A-1) k_{0} + C
n_{0} =0 , \qquad
 C k_{0} + (B-1) n_{0} = 0 ,
\qquad
 k_{0}^{2} + n_{0}^{2}
 =1 + ( K^{2} + N^{2})
 , \label{B8a}
\end{gather}
where
\begin{gather}
A = (k_{0}^{2} + n_{0}^{2} ) + ( K^{2} - N^{2}) , \qquad B =
(k_{0}^{2} + n_{0}^{2} ) -( K^{2} - N^{2}) , \qquad C = -2 KN .
\label{B8c}
\end{gather}

Therefore, we have 8 variables $ {\boldsymbol e}$, $k_{0}$,
$n_{0}$, $K$, $N $ and the set of equations,
(\ref{B7d})--(\ref{B8c}) for them. Its solving turns to be rather
involving, so let us formulate only the f\/inal result:
\begin{gather}
k_{0} , \qquad {\boldsymbol k} = K {\boldsymbol e} , \qquad n_{0}
, \qquad {\boldsymbol n} = N {\boldsymbol e} , \nonumber
\\
 k_{0}^{2} + K^{2} + n_{0}^{2} + N^{2} = +1 , \qquad
k_{0}N + n_{0} K =0 ,
\nonumber
\\
G =
 \left | \begin{array}{rr}
(k_{0} +i K {\boldsymbol e} \vec{\sigma}) & (n_{0} - iN {\boldsymbol e} \vec{\sigma} ) \\
 -( n_{0} - iN {\boldsymbol e} \vec{\sigma}) & (k_{0} + iK {\boldsymbol e} \vec{\sigma})
\end{array} \right | .
\label{B9c}
\end{gather}
It should be noted that
\begin{gather*}
\det G =
 (k_{0}^{2} + K^{2} + n_{0}^{2} + N^{2} )^{2} = +1 .
\end{gather*}
The unitarity of the matrices (\ref{B9c}) may be verif\/ied by
direct calculation. Indeed,
\begin{gather}
G ^{+} = \left | \begin{array}{rr}
(k_{0} -i K {\boldsymbol e} \vec{\sigma} ) & -(n_{0} + iN {\boldsymbol e} \vec{\sigma} ) \\
 (n_{0} + iN {\boldsymbol e} \vec{\sigma} )& (k_{0} - iK {\boldsymbol e} \vec{\sigma})
\end{array} \right | ,
\nonumber
\end{gather}

\noindent and further for $G G^{+}=I$ we get (by $2 \times 2$
blocks)
\begin{gather}
(G G^{+} )_{11} = k_{0}^{2} + K^{2} + n_{0} ^{2} + N^{2} = +1 ,
\qquad (G G^{+} )_{12} = -2i (n_{0}K + k_{0}N ) ({\boldsymbol e}
 \vec{\sigma} ) = 0 , \nonumber
\\
(G G^{+} )_{22} = k_{0}^{2} + K^{2} + n_{0} ^{2} + N^{2} = +1 ,
\qquad (G G^{+} )_{21} = +2i (n_{0}K + k_{0}N ) ({\boldsymbol e}
\vec{\sigma} ) = 0 . \nonumber
\end{gather}

One dif\/ferent way to parameterize (\ref{B9c}) can be proposed.
Indeed, relations (\ref{B9c}) are
\begin{gather*}
k_{0} , \qquad {\boldsymbol k} = K {\boldsymbol e} , \qquad n_{0}
, \qquad {\boldsymbol n} = N {\boldsymbol e} , \nonumber
\\
 k_{0}^{2} ( 1+ {K^{2} \over k_{0}^{2} }) + n_{0}^{2}(1 + { N^{2} \over n_{0}^{2} } ) = +1 ,
\qquad {K \over k_{0}}= - {N \over n_{0} } \equiv W ,
\end{gather*}
or
\begin{gather*}
k_{0} , \qquad {\boldsymbol k} = k_{0} W {\boldsymbol e} , \qquad
n_{0} , \qquad {\boldsymbol n} = - n_{0} W {\boldsymbol e} ,
\nonumber
\\
( k_{0}^{2} + n_{0}^{2}) ( 1+ W^{2} ) = +1 , \qquad K = k_{0} W ,
\qquad N = - n_{0} W , \qquad 0\leq
k_{0}^{2} + n_{0}^{2} \leq 1 . 
\end{gather*}
Therefore, matrix $G$ can be presented as follows:
\begin{gather}
G = \left | \begin{array}{rr}
k_{0}(1 +i W {\boldsymbol e} \vec{\sigma} )& n_{0} (1 + iW {\boldsymbol e} \vec{\sigma} ) \\[3mm]
 - n_{0}(1 + i W {\boldsymbol e} \vec{\sigma} )& k_{0} (1 + i W {\boldsymbol e} \vec{\sigma})
\end{array} \right | ,
\label{B10c}
\\
( k_{0}^{2} + n_{0}^{2}) ( 1+ W^{2} ) = +1  \qquad \Longrightarrow
\qquad W = \pm \sqrt{ {1 \over k_{0}^{2} + n_{0}^{2}} -1 } .
\nonumber
\end{gather}
Evidently, it suf\/f\/ices to take positive values for $W$. The
constructed subgroup (\ref{B10c}) depends upon four parameters
$k_{0}$, $n_{0}$, ${\boldsymbol e}$:
\begin{gather*}
0\leq k_{0}^{2} + n_{0}^{2} \leq 1 , \qquad {\boldsymbol e}^{ 2} =
1 , \qquad ( k_{0}^{2} + n_{0}^{2}) ( 1+ W^{2} ) = +1 .
\end{gather*}

Let us establish the law of multiplication for four parameters
$k_{0}$, $n_{0}$, ${\boldsymbol W}= W {\boldsymbol e}$:
\begin{gather*}
G'' =
 \left | \begin{array}{rr}
k_{0}' (1 +i {\boldsymbol W} ' \vec{\sigma} )& n_{0}' (1 + i {\boldsymbol W}' \vec{\sigma} ) \\
 - n_{0}'(1 + i {\boldsymbol W} ' \vec{\sigma} )& k_{0}' (1 + i {\boldsymbol W}' \vec{\sigma})
\end{array} \right |
\left | \begin{array}{rr}
k_{0}(1 +i {\boldsymbol W} \vec{\sigma} )& n_{0} (1 + i {\boldsymbol W} \vec{\sigma} ) \\
 - n_{0}(1 + i {\boldsymbol W} \vec{\sigma} )& k_{0} (1 + i {\boldsymbol W} \vec{\sigma})
\end{array} \right |
\end{gather*}
or by $2 \times 2$ blocks
\begin{gather*}
(11) = (k_{0}' k_{0} - n_{0}' n_{0}) (1 +i {\boldsymbol W} '
\vec{\sigma} ) (1 +i {\boldsymbol W} \vec{\sigma} ) , \nonumber
\\
(12) = (k_{0}' n_{0} + n_{0}' k_{0}) (1 +i {\boldsymbol W} '
\vec{\sigma} ) (1 +i {\boldsymbol W} \vec{\sigma} ) , \nonumber
\\
(21) = - (k_{0}' n_{0} + n_{0}' k_{0}) (1 +i {\boldsymbol W} '
\vec{\sigma} ) (1 +i {\boldsymbol W} \vec{\sigma} ) , \nonumber
\\
(22) = (k_{0}' k_{0} - n_{0}' n_{0}) (1 +i {\boldsymbol W} '
\vec{\sigma} ) (1 +i {\boldsymbol W} \vec{\sigma} ) . \nonumber
\end{gather*}
As $ (11) = (22)$, $(12) = - (21) $; further one can consider only
two blocks:
\begin{gather*}
(11) = (k_{0}' k_{0} - n_{0}' n_{0}) (1 - {\boldsymbol
W}'{\boldsymbol W}) \left( 1 + i { {\boldsymbol W} ' +
{\boldsymbol W} - {\boldsymbol W}' \times {\boldsymbol W} \over 1
- {\boldsymbol W}'{\boldsymbol W} } \vec{\sigma} \right) ,
\nonumber
\\
(12) = (k_{0}' n_{0} + n_{0}' k_{0}) (1 - {\boldsymbol
W}'{\boldsymbol W}) \left( 1 + i { {\boldsymbol W} ' +
{\boldsymbol W} - {\boldsymbol W}' \times {\boldsymbol W} \over 1
- {\boldsymbol W}'{\boldsymbol W} } \vec{\sigma} \right) .
\nonumber
\end{gather*}
So the composition rules should be
\begin{gather*}
k_{0}'' = (k_{0}' k_{0} - n_{0}' n_{0}) (1 - {\boldsymbol
W}'{\boldsymbol W}) , \qquad n_{0}'' = (k_{0}' n_{0} + n_{0}'
k_{0}) (1 - {\boldsymbol W}'{\boldsymbol W}) , \nonumber
\\
{\boldsymbol W}'' = { {\boldsymbol W} ' + {\boldsymbol W} -
{\boldsymbol W}' \times {\boldsymbol W} \over 1 - {\boldsymbol
W}'{\boldsymbol W} } .
\end{gather*}
The later formula coincides with the Gibbs multiplication rule
(see in \cite{Fedorov}) for 3-dimensional rotation group
$SO(3,R)$.
 It remains to prove the identity:
\begin{gather*}
( k_{0}^{''2} + n_{0}^{''2}) ( 1+ W^{''2} ) = +1
\nonumber
\end{gather*}
which reduces to
\begin{gather}
(k_{0}' k_{0} - n_{0}' n_{0})^{2} + (k_{0}' n_{0} + n_{0}'
k_{0})^{2} \left [ ( 1 - {\boldsymbol W}'{\boldsymbol W}) ^{2} +
({\boldsymbol W} ' + {\boldsymbol W} - {\boldsymbol W}' \times
{\boldsymbol W})^{2} \right ] . \label{B12b}
\end{gather}
First terms are
\begin{gather*}
(k_{0}' k_{0} - n_{0}' n_{0})^{2} + (k_{0}' n_{0} + n_{0}'
k_{0})^{2}= (k_{0}^{'2} + n_{0}^{'2}) ( k_{0}^{2} + n_{0}^{2})
 . \nonumber
\end{gather*}
Second term is
\begin{gather*}
( 1 - {\boldsymbol W}'{\boldsymbol W}) ^{2} + ({\boldsymbol W} ' +
{\boldsymbol W} -
 {\boldsymbol W}' \times {\boldsymbol W})^{2} = (1 + {\boldsymbol W}'^{2}) ( 1 + {\boldsymbol
W}^{2}) . \nonumber
\end{gather*}
Therefore, (\ref{B12b}) takes the form
\begin{gather*}
(k_{0}'^{\,2} + n_{0}'^{\,2}) ( k_{0}^{2} + n_{0}^{2}) (1 +
{\boldsymbol W}'^{2}) ( 1 + {\boldsymbol W}^{2}) =1 \nonumber
\end{gather*}
which is identity due to equalities
\begin{gather*}
(k_{0}'^{\,2} + n_{0}'^{\,2}) (1 + {\boldsymbol W}'^{2}) =1 ,
\qquad
 ( k_{0}^{2} + n_{0}^{2}) ( 1 + {\boldsymbol W}^{2}) =1 .
\nonumber
\end{gather*}

It is matter of simple calculation to introduce curvilinear
parameters for such an unitary subgroup:
\begin{gather*}
{\boldsymbol e} = ( \sin \theta \cos \phi , \sin \theta \sin \phi,
\cos \theta ) , \nonumber
\\
k_{0} = \cos \alpha \cos \rho , \qquad K = \cos \alpha \sin \rho
 , \qquad
 n_{0} = \sin \alpha \cos \rho , \qquad N = - \sin
\alpha \sin \rho , 
\end{gather*}
and $G$ looks as follows
\begin{gather*}
G = \left | \begin{array}{rr}
\Delta & \Sigma \\
 - \Sigma & \Delta
\end{array} \right | ,\qquad
\Delta =\left | \begin{array}{rr}
\cos \alpha( \cos \rho + i \sin \rho \cos \theta ) & i \cos \alpha \sin \rho \sin \theta e^{-i\phi} \\
i \cos \alpha \sin \rho \sin \theta e^{i\phi} & \cos \alpha( \cos
\rho - i \sin \rho \cos \theta
\end{array} \right | ,
\nonumber
\\
\Sigma =\left | \begin{array}{rr}
 \sin \alpha ( \cos \rho + i \sin \rho \cos \theta ) &
 +i \sin \alpha \sin \rho \sin \theta e^{-i\phi} \\
 i\sin \alpha \sin \rho \sin \theta e^{i\phi} &
 \sin \alpha( \cos \rho - i \sin \rho \cos \theta)
\end{array} \right |.
\end{gather*}

It should be noted that one one may factorize 4-parametric element
into two unitary factors, 1-parametric and 3-parametric. Indeed,
let us consider the product of commuting unitary groups,
isomorphic to Abelian
 group $G_{0}$ and $SU(2)$:
\begin{gather*}
G = G_{0} \otimes SU(2) = SU(2) \otimes G_{0} = \left |
\begin{array}{cc}
 k'_{0} & n_{0}' \\
- n'_{0} & k'_{0}
\end{array} \right |
\left | \begin{array}{cc} a_{0} + i {\boldsymbol a} \vec{\sigma} &
0
\\
0 & a_{0} + i {\boldsymbol a} \vec{\sigma}
\end{array} \right |
\nonumber
\\
\phantom{G=}{}=\left | \begin{array}{cc}
k'_{0} a_{0} + i k'_{0} {\boldsymbol a} \vec{\sigma} & n_{0}' a_{0} + i n_{0}' {\boldsymbol a} \vec{\sigma} \\
-n_{0} 'a_{0} - i n_{0}' {\boldsymbol a} \vec{\sigma} & k'_{0}
a_{0} + i k'_{0} {\boldsymbol a} \vec{\sigma}
\end{array} \right |, \qquad k_{0}'^{\,2} + n_{0}'^{\,2} = 1 , \qquad a_{0}^{2} + {\boldsymbol a}^{2} = 1,
\end{gather*}
with the notation
\begin{gather*}
k'_{0} a_{0} = k_{0} , \qquad k'_{0} {\boldsymbol a} = k_{0}
{\boldsymbol W} , \qquad
 n'_{0} a_{0} = n_{0} , \qquad n'_{0} {\boldsymbol a} = n_{0} {\boldsymbol W} ,
\nonumber
\\
(k_{0}^{2} + n_{0}^{2})(1 +W^{2}) = (k_{0}'^{\,2}  a_{0}^{2} +
n_{0}'^{\,2}  a_{0}^{2}) (1 +W^{2})
 = (k_{0}'^{\,2} + n_{0}'^{\,2})(a_{0}^{2} + {\boldsymbol a}^{2} ) =1 ,
\nonumber
\end{gather*}
takes the form
\begin{gather}
 G_{0} \otimes SU(2) = SU(2) \otimes G_{0}
 = \left | \begin{array}{rr}
k_{0}(1 +i {\boldsymbol W} \vec{\sigma} )& n_{0} (1 + i {\boldsymbol W} \vec{\sigma} ) \\
 - n_{0}(1 + i {\boldsymbol W} \vec{\sigma} )& k_{0} (1 + i {\boldsymbol W} \vec{\sigma})
\end{array} \right | = G .
\label{B.15}
\end{gather}

 Let us summarize the main results of the previous sections:
\begin{quotation}
Parametrization of $ 4 \times 4 $ matrices $G$ of the complex
linear group $GL(4,C)$ in terms of four complex vector-parameters
$G=G(k,m,n,l)$ is developed and the problem of inverting any
$4\times 4$ matrix $G$ is solved. Expression for determinant of
any matrix $G$ is found: $\det G = F(k,m,n,l)$. Unitarity
conditions have been formulated in the form of non-linear cubic
algebraic equations including complex conjugation. Several
simplest solutions of these unitarity equations have been found:
three 2-pa\-rametric subgroups $G_{1}$, $G_{2}$, $G_{3}$ -- each
of subgroups consists of two commuting Abelian unitary groups;
4-parametric unitary subgroup consisting of a product of
a~3-parametric group isomorphic $SU(2)$ and 1-parametric Abelian
group.
\end{quotation}

The task of full solving of the unitarity conditions seems to be
rather complicated and it will be considered elsewhere. In the
remaining part of the present paper we describe some relations of
the above treatment to other considerations of the problem in the
literature. The relations described give grounds to hope that the
full general solution of the unitary equations obtained can be
constructed on the way of combining dif\/ferent techniques used in
the theory of the unitary group $SU(4)$.

\section[On subgroups $GL(3,C)$ and $SU(3)$, expressions for Gell-Mann matrices
 through the Dirac basis]{On subgroups $\boldsymbol{GL(3,C)}$ and $\boldsymbol{SU(3)}$, expressions\\ for Gell-Mann matrices
 through the Dirac basis}\label{sec6}

In this section the main question is
 how in the Dirac parametrization one can
distinguish $GL(3,C)$, subgroup in $GL(4,C)$. To this end, let us
turn to the explicit form of the Dirac basis (the Weyl
representation is used; at some elements the imaginary unit $i$ is
added)
\begin{gather}
 \gamma^{5} = \left | \begin{array}{cccc}
-1 & 0 & 0 & 0 \\
0 & -1 & 0 & 0 \\
0 & 0 & 1 & 0 \\
0 & 0 & 0 & 1
\end{array} \right | , \qquad
\gamma^{0} = \left | \begin{array}{cccc}
0 & 0 & 1 & 0 \\
0 & 0 & 0 & 1 \\
1 & 0 & 0 & 0 \\
0 & 1 & 0 & 0
\end{array} \right | , \qquad
 i\gamma^{5} \gamma^{0} =
\left | \begin{array}{cccc}
0 & 0 & -i & 0 \\
0 & 0 & 0 & -i \\
i & 0 & 0 & 0 \\
0 & i & 0 & 0
\end{array} \right | ,
\nonumber
\\
 i\gamma^{1} =
 \left | \begin{array}{cccc}
0 & 0 & 0 & -i \\
0 & 0 & -i & 0 \\
0 & i & 0 & 0 \\
i & 0 & 0 & 0
\end{array} \right | , \qquad
\gamma^{5} \gamma^{1} = \left | \begin{array}{cccc}
0 & 0 & 0 & 1 \\
0 & 0 & 1 & 0 \\
0 & 1 & 0 & 0 \\
1 & 0 & 0 & 0
\end{array} \right | ,
\qquad i\gamma^{2} = \left | \begin{array}{cccc}
0 & 0 & 0 & -1 \\
0 & 0 & 1 & 0 \\
0 & 1 & 0 & 0 \\
-1 & 0 & 0 & 0
\end{array} \right | ,
\label{A.1a}
\\
\gamma^{5} \gamma^{2} = \left | \begin{array}{cccc}
0 & 0 & 0 & -i \\
0 & 0 & i & 0 \\
0 & -i & 0 & 0 \\
i & 0 & 0 & 0
\end{array} \right | , \qquad
i \gamma^{3} = \left | \begin{array}{cccc}
0 & 0 & -i & 0 \\
0 & 0 & 0 & i \\
i & 0 & 0 & 0 \\
0 & -i & 0 & 0
\end{array} \right | ,\qquad
\gamma^{5} \gamma^{3} = \left | \begin{array}{cccc}
0 & 0 & 1 & 0 \\
0 & 0 & 0 & -1 \\
1 & 0 & 0 & 0 \\
0 & -1 & 0 & 0
\end{array} \right |,
\nonumber
\\
2 \sigma^{01} = \left | \begin{array}{cccc}
0 & 1 & 0 & 0 \\
1 & 0 & 0 & 0 \\
0 & 0 & 0 & -1 \\
0 & 0 & -1 & 0
\end{array} \right | , \qquad
2 \sigma^{02} = \left | \begin{array}{cccc}
0 & -i & 0 & 0 \\
i & 0 & 0 & 0 \\
0 & 0 & 0 & i \\
0 & 0 & -i & 0
\end{array} \right | ,
\qquad 2 \sigma^{03} = \left | \begin{array}{cccc}
1 & 0 & 0 & 0 \\
0 & -1 & 0 & 0 \\
0 & 0 & -1 & 0 \\
0 & 0 & 0 & 1
\end{array} \right | ,
\nonumber
\\
2i\sigma^{12} = \left | \begin{array}{cccc}
1 & 0 & 0 & 0 \\
0 & -1 & 0 & 0 \\
0 & 0 & 1 & 0 \\
0 & 0 & 0 & -1
\end{array} \right | ,\qquad
2i\sigma^{23} = \left | \begin{array}{cccc}
0 & 1 & 0 & 0 \\
1 & 0 & 0 & 0 \\
0 & 0 & 0 & 1 \\
0 & 0 & 1 & 0
\end{array} \right | , \qquad
2i\sigma^{31} = \left | \begin{array}{cccc}
0 & -i & 0 & 0 \\
i & 0 & 0 & 0 \\
0 & 0 & 0 & -i \\
0 & 0 & i & 0
\end{array} \right | .
\nonumber
\end{gather}
All these 15 matrices $\Lambda_{i}$ are of Gell-Mann type: they
have a zero-trace, they are Hermitian, besides, their squares are
unite:
\begin{gather*}
 \mbox{Sp}\,\Lambda =0 , \qquad
 (\Lambda )^{2}= I , \qquad (\Lambda )^{+} = \Lambda , \qquad \Lambda \in \{ \Lambda_{k}:
 k =1,\dots,15 \} .
\end{gather*}
Exponential function of any of them equals to
\begin{gather*}
U = e^{ia\Lambda} = \cos a + i \sin a \Lambda , \qquad \det
e^{ia\Lambda} = +1 , \qquad U^{+} = U^{-1} , \qquad a \in R .
\end{gather*}
Evidently, multiplying of such 15 elementary unitary matrices (at
real parameters $x_{i}$) results in an unitary matrix
\begin{gather*}
U = e^{ia_{1}\Lambda_{1}} e^{ia_{2}\Lambda_{2}} \cdots
e^{ia_{14}\Lambda_{14}} e^{ia_{l5}\Lambda_{l5}} . 
\end{gather*}

At this there arise 15 generalized angle-variables $a_{1} , \dots,
a_{15}$. Evident advantage of this approach is its simplicity, and
evident defect consists in the following: we do not know any
simple group multiplication rule for these angles.

It should be noted that the basis $\lambda_{i}$ used in
\cite{Tilma-1} substantially dif\/fers from the above Dirac basis
$\Lambda_{i}$ -- this peculiarity is closely connected with
distinguishing $ SU(3)$ in $SU(4)$. This problem is evidently
related to the task of distinguishing $GL(3,C)$ in $GL(4,C)$ as
well.

In order to have possibility to compare two approaches we need
exact connection between $\lambda_{i}$ and $\Lambda_{i}$. In
\cite{Tilma-1} the following Gell-Mann basis for $SU(4)$ were
used:
\begin{gather}
\lambda_{1} = \left | \begin{array}{cccc}
0 & 1 & 0 & 0 \\
1 & 0 & 0 & 0 \\
0 & 0 & 0 & 0 \\
0 & 0 & 0 & 0
\end{array} \right |, \qquad
\lambda_{2} = \left | \begin{array}{cccc}
0 &-i & 0 & 0 \\
i & 0 & 0 & 0 \\
0 & 0 & 0 & 0 \\
0 & 0 & 0 & 0
\end{array} \right |, \qquad
\lambda_{3} = \left | \begin{array}{cccc}
1 & 0 & 0 & 0 \\
0 & -1 & 0 & 0 \\
0 & 0 & 0 & 0 \\
0 & 0 & 0 & 0
\end{array} \right |,
\nonumber
\\
\lambda_{4} = \left | \begin{array}{cccc}
0 & 0 & 1 & 0 \\
0 & 0 & 0 & 0 \\
1 & 0 & 0 & 0 \\
0 & 0 & 0 & 0
\end{array} \right |, \qquad
\lambda_{5} = \left | \begin{array}{cccc}
0 & 0 & -i & 0 \\
0 & 0 & 0 & 0 \\
i & 0 & 0 & 0 \\
0 & 0 & 0 & 0
\end{array} \right |, \qquad
\lambda_{6} = \left | \begin{array}{cccc}
0 & 0 & 0 & 0 \\
0 & 0 & 1 & 0 \\
0 & 1 & 0 & 0 \\
0 & 0 & 0 & 0
\end{array} \right |,
\nonumber
\\
\lambda_{7} = \left | \begin{array}{cccc}
0 & 0 & 0 & 0 \\
0 & 0 & -i & 0 \\
0 & i & 0 & 0 \\
0 & 0 & 0 & 0
\end{array} \right |, \qquad
\lambda_{8} = {1 \over \sqrt{3}} \left | \begin{array}{cccc}
1 & 0 & 0 & 0 \\
0 & 1 & 0 & 0 \\
0 & 0 & -2 & 0 \\
0 & 0 & 0 & 0
\end{array} \right |, \qquad
\lambda_{9} = \left | \begin{array}{cccc}
0 & 0 & 0 & 1 \\
0 & 0 & 0 & 0 \\
0 & 0 & 0 & 0 \\
1 & 0 & 0 & 0
\end{array} \right |,
\nonumber
\\
\lambda_{10} = \left | \begin{array}{cccc}
0 & 0 & 0 & -i \\
0 & 0 & 0 & 0 \\
0 & 0 & 0 & 0 \\
i & 0 & 0 & 0
\end{array} \right |, \qquad
\lambda_{11} = \left | \begin{array}{cccc}
0 & 0 & 0 & 0 \\
0 & 0 & 0 & 1 \\
0 & 0 & 0 & 0 \\
0 & 1 & 0 & 0
\end{array} \right |,
\lambda_{12} = \left | \begin{array}{cccc}
0 & 0 & 0 & 0 \\
0 & 0 & 0 & -i \\
0 & 0 & 0 & 0 \\
0 & i & 0 & 0
\end{array} \right |,
\nonumber
\\
\lambda_{13} = \left | \begin{array}{cccc}
0 & 0 & 0 & 0 \\
0 & 0 & 0 & 0 \\
0 & 0 & 0 & 1 \\
0 & 0 & 1 & 0
\end{array} \right |, \qquad
\lambda_{14} = \left | \begin{array}{cccc}
0 & 0 & 0 & 0 \\
0 & 0 & 0 & 0 \\
0 & 0 & 0 & -i \\
0 & 0 & i & 0
\end{array} \right |, \qquad
\lambda_{15} =\tfrac{1}{\sqrt{6}} \left | \begin{array}{cccc}
1 & 0 & 0 & 0 \\
0 & 1 & 0 & 0 \\
0 & 0 & 1 & 0 \\
0 & 0 & 0 & -3
\end{array} \right | .
\label{A.3a}
\end{gather}
All the $\lambda$ excluding $\lambda_{8},
 \lambda_{15}$ possess the same property:
\begin{gather*}
\lambda^{3}_{i} = + \lambda_{i} , \qquad i \neq 8, 15 .
\end{gather*}

The minimal polynomials for $\lambda_{8}$, $\lambda_{15}$ can be
easily found. Indeed,
\begin{gather*}
(\lambda_{8})^{2} = \tfrac{1}{3} \left | \begin{array}{cccc}
1 & 0 & 0 & 0 \\
0 & 1 & 0 & 0 \\
0 & 0 & 4 & 0 \\
0 & 0 & 0 & 0
\end{array} \right | , \qquad
(\lambda_{8})^{3} = \tfrac{1}{3 \sqrt{3}} \left |
\begin{array}{cccc}
1 & 0 & 0 & 0 \\
0 & 1 & 0 & 0 \\
0 & 0 & -8 & 0 \\
0 & 0 & 0 & 0
\end{array} \right |,
\nonumber
\end{gather*}
therefore
\begin{gather*}
(\lambda_{8}) ^{3} = \tfrac{2}{3} \lambda_{8} - \tfrac{1}{\sqrt{3}} ( \lambda_{8})^{2} . 
\end{gather*}

Analogously, for $\lambda_{15}$ we have
\begin{gather*}
(\lambda_{15})^{2} = \tfrac{1}{6} \left | \begin{array}{cccc}
1 & 0 & 0 & 0 \\
0 & 1 & 0 & 0 \\
0 & 0 & 1 & 0 \\
0 & 0 & 0 & 9
\end{array} \right | ,
\qquad (\lambda_{15})^{3} =\tfrac{1}{6\sqrt{6}} \left |
\begin{array}{cccc}
1 & 0 & 0 & 0 \\
0 & 1 & 0 & 0 \\
0 & 0 & 1 & 0 \\
0 & 0 & 0 & -27
\end{array} \right | ,
\nonumber
\end{gather*}
and
\begin{gather*}
(\lambda_{15}) ^{3} = \tfrac{1}{2} \lambda_{15} - \tfrac{2}{\sqrt{6}} ( \lambda_{15})^{2} . 
\end{gather*}

Comparing $\Lambda_{i}$ and $\lambda_{i}$ one can readily derive
the linear combinations:
\begin{gather*}
\gamma^{0} + \gamma^{5}\gamma^{3} = 2 \lambda_{4} , \qquad
\gamma^{0} - \gamma^{5}\gamma^{3} = 2 \lambda_{11} , \qquad i
\gamma^{5} \gamma^{0} + i \gamma^{3} = 2 \lambda_{5} ,
\\ i \gamma^{5} \gamma^{0} - i \gamma^{3} = 2 \lambda_{12}
 , \qquad
 \gamma^{5} \gamma^{1} + i \gamma^{2} = 2 \lambda_{6} , \qquad
 \gamma^{5} \gamma^{1} - i \gamma^{2} = 2 \lambda_{9} ,
\nonumber
\\
i \gamma^{1} + \gamma^{5} \gamma^{2} = 2 \lambda_{10} , \qquad i
\gamma^{1} - \gamma^{5} \gamma^{2} = 2 \lambda_{7}
 , \qquad
2\sigma^{01} + 2i\sigma^{23} = 2 \lambda_{1} , \\
2\sigma^{01} - 2i\sigma^{23} = -2 \lambda_{13} , \qquad
2\sigma^{02} + 2i\sigma^{31} = 2 \lambda_{2} , \qquad
2\sigma^{02} - 2i\sigma^{31} = -2 \lambda_{14} , 
\end{gather*}
and additional six combinations
\begin{gather*}
2\sigma^{03} + 2i\sigma^{12} = \left | \begin{array}{cccc}
2 & 0 & 0 & 0 \\
0 & -2 & 0 & 0 \\
0 & 0 & 0 & 0 \\
0 & 0 & 0 & 0
\end{array} \right |
, \qquad 2\sigma^{03} - 2i\sigma^{12} = \left |
\begin{array}{rrrr}
0 & 0 & 0 & 0 \\
0 & 0 & 0 & 0 \\
0 & 0 & -2 & 0 \\
0 & 0 & 0 & +2 \\
\end{array} \right |
 ,
\nonumber
\\
\gamma^{5} + 2\sigma^{03} = \left | \begin{array}{rrrr}
0 & 0 & 0 & 0 \\
0 & -2 & 0 & 0 \\
0 & 0 & 0 & 0 \\
0 & 0 & 0 & 2 \\
\end{array} \right | , \qquad
\gamma^{5} - 2\sigma^{03} = \left | \begin{array}{rrrr}
-2 & 0 & 0 & 0 \\
0 & 0 & 0 & 0 \\
0 & 0 & 2 & 0 \\
0 & 0 & 0 & 0 \\
\end{array} \right | ,
\nonumber
\\
\gamma^{5} + 2i\sigma^{12} = \left | \begin{array}{rrrr}
0 & 0 & 0 & 0 \\
0 & -2 & 0 & 0 \\
0 & 0 & 2 & 0 \\
0 & 0 & 0 & 0 \\
\end{array} \right | , \qquad
\gamma^{5} - 2i\sigma^{12} = \left | \begin{array}{rrrr}
-2 & 0 & 0 & 0 \\
0 & 0 & 0 & 0 \\
0 & 0 & 0 & 0 \\
0 & 0 & 0 & 2 \\
\end{array} \right | ,
\end{gather*}
they should contain three linearly independent matrices. Those
 three linearly independent matrices might be chosen in dif\/ferent ways.
Let us introduce the notation:
\begin{gather*}
a= \left | \begin{array}{cccc}
1 & 0 & 0 & 0 \\
0 & -1 & 0 & 0 \\
0 & 0 & 0 & 0 \\
0 & 0 & 0 & 0
\end{array} \right |
 , \qquad
 b= \left | \begin{array}{rrrr}
0 & 0 & 0 & 0 \\
0 & -1 & 0 & 0 \\
0 & 0 & 1 & 0 \\
0 & 0 & 0 & 0 \\
\end{array} \right | , \qquad
 c = \left | \begin{array}{rrrr}
-1 & 0 & 0 & 0 \\
0 & 0 & 0 & 0 \\
0 & 0 & 1 & 0 \\
0 & 0 & 0 & 0 \\
\end{array} \right | ,
\nonumber
\\
A = \left | \begin{array}{rrrr}
0 & 0 & 0 & 0 \\
0 & 0 & 0 & 0 \\
0 & 0 & -1 & 0 \\
0 & 0 & 0 & +1 \\
\end{array} \right | ,\qquad\
B = \left | \begin{array}{rrrr}
-1 & 0 & 0 & 0 \\
0 & 0 & 0 & 0 \\
0 & 0 & 0 & 0 \\
0 & 0 & 0 & 1 \\
\end{array} \right | , \qquad
C = \left | \begin{array}{rrrr}
0 & 0 & 0 & 0 \\
0 & -1 & 0 & 0 \\
0 & 0 & 0 & 0 \\
0 & 0 & 0 & 1 \\
\end{array} \right | .
\end{gather*}
The matrices $a$, $b$, $c$ have the $3 \times 3$ blocks
dif\/ferent from zero,
 so they could be generators for~$SU(3)$ transformations; whereas $A$, $B$, $C$ may be generators only
of the group $SU(4)$. All six matrices $a$, $b$, $c$, $A$, $B$,
$C$ have the same minimal polynomial:
\begin{gather*}
\lambda^{3} = \lambda . 
\end{gather*}
Linear space to which these six matrices $a$, $b$, $c$, $A$, $B$,
$C$ belong is 3-dimensional. Indeed, one easily obtains
\begin{gather*}
c = b - a , \qquad C-A = b , \qquad C-B = a , \qquad B-A = c ,
\nonumber
\\
\mbox{basis} \ \{ a,b,C \}
 \qquad \Longrightarrow \qquad c = b - a
, \qquad A = C -b , \qquad B = C -a .
\end{gather*}

These relations can be rewritten dif\/ferently
\begin{gather*}
a = b - c , \qquad C-A = b , \qquad C-B = b-c , \qquad B-A = c ,
\nonumber
\\
 \mbox{basis} \ \{ b,c,A \}  \qquad \Longrightarrow
\qquad a = b - c , \qquad C = A +b , \qquad B = A +c
\end{gather*}
 or
\begin{gather*}
b = a + c , \qquad C-A = b , \qquad C-B = b-c , \qquad B-A = c ,
\nonumber
\\
 \mbox{basis} \ \{ a,c,B \}  \qquad \Longrightarrow
\qquad b = a+c , \qquad C = B +a , \qquad A = B -c .
\end{gather*}

One should note that in the basis $\lambda_{i}$ (\ref{A.3a}) the
corresponding three linearly independent
 elements $\lambda_{3}$, $\lambda_{8}$, $\lambda_{15}$ are taken as follows:
\begin{gather*}
 \lambda_{3} = \tfrac{1}{2} 2\sigma^{03} + \tfrac{1}{2} 2i\sigma^{12} ,
\qquad \lambda_{8} =-\tfrac{1}{\sqrt{3}} \gamma^{5} +
\tfrac{1}{2\sqrt{3}} 2\sigma^{03} - \tfrac{1}{2 \sqrt{3}}
2i\sigma^{12}
 , \nonumber
\\
\lambda_{15} =-\tfrac{1}{\sqrt{6}} \gamma^{5} -
\tfrac{1}{\sqrt{6}} 2\sigma^{03} +
\tfrac{1}{\sqrt{6}} 2i\sigma^{12}, 
\end{gather*}
their minimal polynomials look
\begin{gather*}
(\lambda_{3})^{3} = \lambda_{3} , \qquad (\lambda_{8}) ^{3} =
\tfrac{2}{3} \lambda_{8} - \tfrac{1}{\sqrt{3}} ( \lambda_{8})^{2}
, \qquad
(\lambda_{15}) ^{3} = \tfrac{1}{2} \lambda_{15} - \tfrac{2}{\sqrt{6}} ( \lambda_{15})^{2} .
\end{gather*}
It is the matter of simple calculation to f\/ind relationships
between $\lambda_{3}$, $\lambda_{8}$, $\lambda_{15}$ and the basis
$ \{ a,b,C \}:$
\begin{gather}
\lambda_{3} = a , \qquad \lambda_{8} = \tfrac{1}{\sqrt{3}} a -
\tfrac{2}{\sqrt{3}} b , \qquad \lambda_{15} = \tfrac{1}{\sqrt{6}}
a + \tfrac{1}{\sqrt{6}} b - \tfrac{3}{\sqrt{6}} C . \label{A.8a}
\end{gather}

In the following we will use the notation
\begin{gather*}
a = \lambda_{3} , \qquad b = \lambda_{8}' , \qquad C =
\lambda_{15}' , 
\end{gather*}
so the previous formulas (\ref{A.8a}) will read ($\{ \lambda_{3},
\lambda_{8}, \lambda_{15} \} \Longleftrightarrow \{ \lambda_{3},
\lambda_{8}', \lambda_{15}' \}$)
\begin{gather*}
\lambda_{3} = \lambda_{3} , \qquad \lambda_{8} =
\tfrac{1}{\sqrt{3}} \lambda_{3} - \tfrac{2}{\sqrt{3}} \lambda'_{8}
, \qquad \lambda_{15} = \tfrac{1}{\sqrt{6}} \lambda_{3} +
\tfrac{1}{\sqrt{6}} \lambda'_{8} - \tfrac{3}{\sqrt{6}}
\lambda'_{15}
. 
\end{gather*}
The inverse relations are
\begin{gather}
\lambda_{3} = \lambda_{3} , \qquad \lambda_{8} ' = \tfrac{1}{2 }
 \lambda_{3} - \tfrac{\sqrt{3}}{2} \lambda_{8} , \qquad
\lambda_{15}' = - \tfrac{\sqrt{6}}{3 } \lambda_{15} + \tfrac{1}{2
} \lambda_{3} - \tfrac{1}{2\sqrt{3}} \lambda_{8} . \label{A.8d}
\end{gather}

Now, starting with the linear decomposition of $G \in GL(4,C)$ the
in Dirac basis (\ref{1}):
\begin{gather*}
G= a_{0} I + ib_{0} \gamma^{5} + i A_{0} \gamma^{0} + i A_{k}
\gamma^{k} + B_{0} \gamma^{0} \gamma^{5} + B_{k} \gamma^{k}
\gamma^{5} \nonumber
\\
\phantom{G=}{}  + a_{k} 2\sigma_{0k} + b_{1} 2\sigma_{23} + b_{2}
2\sigma_{31}+ b_{3} 2\sigma_{12} \nonumber
\\
\phantom{G}{}= a_{0} I + ib_{0} \gamma^{5} + i A_{0} \gamma^{0} +
A_{k} (i \gamma^{k}) + iB_{0} (i \gamma^{5} \gamma^{0}) - B_{k}
(\gamma^{5}\gamma^{k} ) \nonumber
\\
\phantom{G=}{} + a_{k} (2\sigma_{0k}) -i b_{1} ( 2i\sigma_{23}) -i
b_{2} (2i\sigma_{31})
 - i b_{3} (2i\sigma_{12} ) ,
\end{gather*}
 with the help of the formulas
\begin{gather*}
\gamma^{0} + \gamma^{5}\gamma^{3} = 2 \lambda_{4} , \qquad
\gamma^{0} - \gamma^{5}\gamma^{3} = 2 \lambda_{11} , \qquad
i \gamma^{5} \gamma^{0} + i \gamma^{3} = 2 \lambda_{5} ,\\
 i \gamma^{5} \gamma^{0} - i \gamma^{3} = 2 \lambda_{12}
 ,\qquad
 \gamma^{5} \gamma^{1} + i \gamma^{2} = 2 \lambda_{6} ,
 \qquad
 \gamma^{5} \gamma^{1} - i \gamma^{2} = 2 \lambda_{9} ,
\nonumber
\\
i \gamma^{1} + \gamma^{5} \gamma^{2} = 2 \lambda_{10} , \qquad i
\gamma^{1} - \gamma^{5} \gamma^{2} = 2 \lambda_{7}
 , \qquad
2\sigma^{01} + 2i\sigma^{23} = 2 \lambda_{1} , \\
2\sigma^{01} - 2i\sigma^{23} = -2 \lambda_{13} , \qquad
2\sigma^{02} + 2i\sigma^{31} = 2 \lambda_{2} , \qquad
2\sigma^{02} - 2i\sigma^{31} = -2 \lambda_{14} , \\
2\sigma^{03} + 2i\sigma^{12} = 2 \lambda_{3} , \qquad 2\sigma^{03}
- 2i\sigma^{12} = 2(\lambda_{15}' -\lambda_{8}')
 ,
\qquad \gamma^{5} + 2\sigma^{03} = 2\lambda_{15}'
 , \\
\gamma^{5} - 2\sigma^{03} = 2(\lambda_{8}'- \lambda_{3}) , \qquad
\gamma^{5} + 2i\sigma^{12} = 2\lambda_{8}' , \qquad \gamma^{5} -
2i\sigma^{12} = 2(\lambda_{15}' -\lambda_{3}) 
\end{gather*}
and inverse ones
\begin{gather*}
\gamma^{0} = \lambda_{4} + \lambda_{11} , \qquad \gamma^{5}
\gamma^{3} = \lambda_{4} - \lambda_{11} , \qquad i\gamma^{5}
\gamma^{0} = \lambda_{5} + \lambda_{12} , \qquad i \gamma^{3} =
\lambda_{5} - \lambda_{12} , \nonumber
\\
\gamma^{5} \gamma^{1} = \lambda_{6} + \lambda_{9} , \qquad i
\gamma^{2} = \lambda_{6} - \lambda_{9} , \qquad
 i \gamma^{1} = \lambda_{10} + \lambda_{7}
, \qquad \gamma^{5} \gamma^{2} = \lambda_{10} - \lambda_{7} ,
\nonumber
\\
2i\sigma^{23} = \lambda_{1} + \lambda_{13} , \qquad 2\sigma^{01} =
\lambda_{1} - \lambda_{13} , \qquad 2i\sigma^{31} = \lambda_{2} +
\lambda_{14} , \qquad 2\sigma^{02} = \lambda_{2} - \lambda_{14} ,
\nonumber
\\
2\sigma^{03} = \lambda_{3} - (\lambda_{8}' - \lambda_{15}') ,
\qquad 2i\sigma^{12} = \lambda_{3} + (\lambda_{8}' -
\lambda_{15}') , \qquad \gamma^{5} = -\lambda_{3} + (\lambda_{8}'
+ \lambda_{15}'),
\end{gather*}
we will arrive at
\begin{gather*}
G= a_{0} I +
 ( a_{1} -i b_{1}) \lambda_{1} + (a_{2} -i b_{2}) \lambda_{2}
 +( i A_{0}-B_{3}) \lambda_{4} + (A_{3} + iB_{0} ) \lambda_{5} +( A_{2} - B_{1} ) \lambda_{6}\\
\phantom{G=}{}+ ( A_{1} + B_{2} ) \lambda_{7}  +
 (a_{3}- i b_{3}- ib_{0} ) \lambda_{3} +( -i b_{3} + ib_{0} - a_{3} ) \lambda_{8}'
 + (ib_{0} +a_{3} +i b_{3} ) \lambda_{15}'
\nonumber
\\
\phantom{G=}{} + (-B_{1} - A_{2} ) \lambda_{9} + (A_{1} - B_{2} )
\lambda_{10} + (i A_{0} + B_{3}) \lambda_{11} + (- A_{3} + iB_{0}
) \lambda_{12} \nonumber
\\
\phantom{G=}{}+ (- a_{1} - i b_{1} ) \lambda_{13} + ( - a_{2} - i
b_{2}
) \lambda_{14} . 
\end{gather*}
In variables $(k,m,l,n)$ (see (\ref{5}), (\ref{6}))
\begin{gather*}
B_{0} -iA_{0} = l_{0} , \qquad B_{j} -iA_{j} = l_{j} , \qquad
B_{0} +iA_{0} = n_{0} , \qquad B_{j} +iA_{j} = n_{j} , \nonumber
\\
a_{0} -ib_{0} = k_{0} , \qquad a_{j} -ib_{j} = k_{j} , \qquad
a_{0} +ib_{0} = m_{0} , \qquad a_{j} +ib_{j} = m_{j}
\end{gather*}
the previous expansion looks
\begin{gather*}
G= \tfrac{1}{2} (k_{0} + m_{0}) I +
 k_{1} \lambda_{1} + k_{2} \lambda_{2}\!
 + \tfrac{1}{2} [ ( n_{0}-n_{3}) - (l_{0} +l_{3}) ] \lambda_{4}\! +
 \tfrac{1}{2 i} [ -(n_{0}-n_{3}) -(l_{0}+l_{3} ) ] \lambda_{5}\!
\nonumber
\\
\phantom{G=}{} + \tfrac{1}{2} [ -(n_{1}+in_{2} ) - (l_{1}-il_{2})
] \lambda_{6} + \tfrac{1}{2i} [ (n_{1} +in_{2}) - (l_{1}-il_{2}) ]
\lambda_{7}
 +
 \left[ k_{3} + \tfrac{1}{2} (k_{0}-m_{0} ) \right] \lambda_{3}\nonumber\\
\phantom{G=}{}  +
 \left[ - m_{3} + \tfrac{1}{2} ( m_{0} - k_{0} ) \right] \lambda_{8}'
 +
 \left[ m_{3} + \tfrac{1}{2 } (m_{0} - k_{0} ) \right] \lambda_{15}'
+\tfrac{1}{2} [ -(n_{1}-in_{2} ) - (l_{1}+il_{2}) ]
\lambda_{9}\nonumber
\\ \phantom{G=}{}+
 \tfrac{1}{2i} [ (n_{1} -in_{2}) - (l_{1}+il_{2}) ] \lambda_{10}
+
 \tfrac{1}{2} [ ( n_{0}+n_{3}) - (l_{0} -l_{3}) ] \lambda_{11}
 \nonumber
\\
\phantom{G=}{}  +
 \tfrac{1}{2 i} [ -(n_{0}+n_{3}) -(l_{0}-l_{3} ) ] \lambda_{12}
-m_{1} \lambda_{13} - m_{2} \lambda_{14} .
\end{gather*}

Let coef\/f\/icients at $ \lambda_{9}$, $\lambda_{10}$,
$\lambda_{11}$, $\lambda_{12}$, $\lambda_{13}$, $\lambda_{14} $ be
equal to zero:
\begin{gather}
(n_{1}-in_{2} ) + (l_{1}+il_{2}) = 0 , \qquad (n_{1} -in_{2}) -
(l_{1}+il_{2}) =0 ,\qquad ( n_{0}+n_{3}) - (l_{0} -l_{3}) =0 ,
\nonumber
\\ (n_{0}+n_{3})
+(l_{0}-l_{3} ) =0 , \qquad m_{1} =0 , \qquad m_{2} =0 .
\label{A.13a}
\end{gather}
Note that we do not require vanishing of the coef\/f\/icient at
 $ \lambda_{15}'$:
\begin{gather*}
 \lambda_{15}' \left( m_{3} + {m_{0} - k_{0} \over 2} \right) \neq 0 .
\end{gather*}
As a result we have a subgroup of $4\times 4$ matrices def\/ined
by 10 complex parameters. At this four elements are diagonal
matrices:
\begin{gather*}
I , \qquad \lambda_{3} , \qquad \lambda_{8}' , \qquad
\lambda_{15}' , 
\end{gather*}
all other matrices have on the diagonal only zeros. Equations
(\ref{A.13a}) give
\begin{gather}
i n_{2} = n_{1} , \qquad n_{3} =-n_{0} , \qquad
 il_{2} = - l_{1} ,
 \qquad
 l_{3} =l_{0} , \qquad
 m_{1} =0 , \qquad m_{2} =0 ,
\label{A.14b}
\end{gather}
so that any matrix $
 G(k_{a}, n_{0},n_{1}, l_{0},l_{1},m_{0},m_{3})
$ is decomposed according to
\begin{gather}
G =
 k_{1} \lambda_{1}\! + k_{2} \lambda_{2}\!
 +
 ( n_{0}\! - l_{0} ) \lambda_{4}\!
+ i ( n_{0}\! + l_{0} ) \lambda_{5}\! +
 ( -n_{1}\! - l_{1}) \lambda_{6}\!
 + i ( - n_{1}\! + l_{1}) \lambda_{7}\!+ \tfrac 12 (k_{0} + m_{0}) I \nonumber
\\
\phantom{G =}{}   +
 \left[ k_{3} + \tfrac{1}{2}(k_{0}-m_{0} ) \right] \lambda_{3}
 + \left[ - m_{3} + \tfrac{1}{2}( m_{0} - k_{0} ) \right] \lambda_{8}' +
 \left[ m_{3} + \tfrac{1}{2 } (m_{0} - k_{0} )\right ] \lambda_{15}' .
 \label{A.15}
 \end{gather}
 Explicit form of the matrices parameterized by
(\ref{A.15}) can be obtained from representation for arbitrary
element of $GL(4,C)$ (\ref{9a})
\begin{gather*}
G(k,m,n,l ) = \left | \begin{array}{cccc}
+(k_{0} + k_{3}) & +(k_{1} - ik_{2}) &   +(n_{0} - n_{3}) & -(n_{1} - in_{2}) \\
+(k_{1} + ik_{2} ) & + (k_{0} - k_{3} ) &  -(n_{1} + in_{2}) & + (n_{0} + n_{3} ) \\
-(l_{0} + l_{3}) & -(l_{1} - il_{2}) &   +(m_{0} - m_{3}) & -(m_{1} - im_{2}) \\
-(l_{1} + il_{2}) & -(l_{0} - l_{3}) &   -(m_{1} + im_{2}) &
+(m_{0} + m_{3})
\end{array} \right |
\nonumber
\end{gather*}
with additional restrictions (\ref{A.14b}):
\begin{gather}
G
 = \left | \begin{array}{cccc}
k_{0} + k_{3} & k_{1} - ik_{2} & +2n_{0} & 0 \\
k_{1} + ik_{2} & k_{0} - k_{3} & -2n_{1} & 0 \\
-2l_{0} & -2l_{1} & m_{0} - m_{3} & 0 \\
0 & 0 & 0 & m_{0} + m_{3}
\end{array} \right | .
\label{A.16b}
\end{gather}
If additionally one requires $
 m_{0} + m_{3} =1$,
then
\begin{gather*}
G
 = \left | \begin{array}{cccc}
k_{0} + k_{3} & k_{1} - ik_{2} & +2n_{0} & 0 \\
k_{1} + ik_{2} & k_{0} - k_{3} & -2n_{1} & 0 \\
-2l_{0} & -2l_{1} & 1 - 2m_{3} & 0 \\
0 & 0 & 0 & 1
\end{array} \right |
\end{gather*}
with decomposition rule
\begin{gather}
G= k_{1} \lambda_{1} + k_{2} \lambda_{2}
 +
 ( n_{0} - l_{0} ) \lambda_{4}
+ i ( n_{0} + l_{0} ) \lambda_{5} +
 ( -n_{1} - l_{1}) \lambda_{6}
 + i ( - n_{1} + l_{1}) \lambda_{7} \nonumber
\\
\phantom{G=}{} + \tfrac 12 (1 + k_{0} -m_{3}) I +
 \left[ k_{3} +\tfrac 12( k_{0} +m_{3} -1 ) \right] \lambda_{3}
\nonumber
\\
\phantom{G=}{}  + \left[ - m_{3} + \tfrac 12( 1 -m_{3} - k_{0} )
\right] \lambda_{8}' + \tfrac 12 (1 -k_{0} + m_{3}) \lambda_{15}'
.
 \label{A.17c}
 \end{gather}

 In the diagonal part of (\ref{A.17c}), there are four independent matrices because equation
(\ref{A.17c}) represents $4 \times 4$ matrix with the structure
\begin{gather*}
G \sim \left | \begin{array}{cc}
GL(3,C) & 0\\
0 & 1
\end{array} \right | .
 \end{gather*}

To deal with the matrices from $GL(3,C)$, in the diagonal part of
 (\ref{A.17c}) one should separate only a $3 \times 3$ block:
\begin{gather*}
\mbox{Diag} = \tfrac 12 (1 + k_{0} -m_{3}) I^{(3)} + \left[ k_{3}
+ \tfrac 12 ( k_{0} +m_{3} -1 ) \right] \lambda_{3}^{(3)}
\nonumber
\\
\phantom{\mbox{Diag} =}{}  + \left[ - m_{3} + \tfrac{1}{2}( 1
-m_{3} - k_{0} ) \right] \lambda_{8}'^{(3)} +\tfrac 12 (1 -k_{0} +
m_{3}) \lambda_{15}'^{(3)} = \nonumber
\\
\phantom{\mbox{Diag} =}{}= \tfrac 12 (1 + k_{0} -m_{3}) \left |
\begin{array}{ccc}
1 & 0 & 0 \\
 0 & 1 & 0 \\
 0 & 0 & 1
 \end{array} \right | +
 \left[ k_{3} + \tfrac 12 ( k_{0} +m_{3} -1 ) \right]
 \left | \begin{array}{ccc}
1 & 0 & 0 \\
 0 & -1 & 0 \\
 0 & 0 & 0
 \end{array} \right |
\nonumber
\\
\phantom{\mbox{Diag} =}{} + \left[ - m_{3} + \tfrac 12( 1 -m_{3} -
k_{0} )\right ]
 \left | \begin{array}{ccc}
0 & 0 & 0 \\
 0 & -1 & 0 \\
 0 & 0 & 1
 \end{array} \right | +
 \tfrac 12 (1 -k_{0} + m_{3})
 \left | \begin{array}{ccc}
0 & 0 & 0 \\
 0 & -1 & 0 \\
 0 & 0 & 0
 \end{array} \right | .
\end{gather*}
Resolving $\lambda_{15}'^{(3)}$ in terms of $ I^{(3)}$,
 $\lambda_{3}^{(3)}$, $\lambda_{8}'^{(3)}$:
\begin{gather*}
\lambda_{15}'^{(3)} =
 - \tfrac 13 I^{(3)} + \tfrac 13 \lambda_{3}^{(3)} + \tfrac 13 \lambda_{8}^{'(3)} ,
 \end{gather*}
we arrive at a 3-term relation:
\begin{gather*}
 \mbox{Diag} =\tfrac 13 (1 + 2k_{0} -2 m_{3}) I^{(3)} +
 \left[ k_{3} + \tfrac 13 ( k_{0} +2m_{3} -1)\right] \lambda_{3}^{(3)} +
\tfrac 13 (-4 m_{3} - 2 k_{0} +2) \lambda_{8}'^{(3)} .
\end{gather*}

The group law for parameters of $SL(3,C)$ has the form (the
notation $M = 1 -2 m_{3}$ is used)
\begin{gather*}
k_{0}'' = k_{0}' k_{0} + {\boldsymbol k}' {\boldsymbol k}
 + 2 ( -n'_{0} l_{0} + n'_{1} l_{1}) ,
\nonumber
\\
({\boldsymbol k}'' ) _{1} = (k'_{0} {\boldsymbol k} + {\boldsymbol
k}' k_{0} + i
 {\boldsymbol k}' \times {\boldsymbol k})_{1} + 2 (-n_{0}' l_{1} + n_{1}'
 l_{0}) , \nonumber
\\
({\boldsymbol k}'' )_{2} = (k'_{0} {\boldsymbol k} + {\boldsymbol
k}' k_{0} + i {\boldsymbol k}' \times {\boldsymbol k})_{2} + 2 (-i
n_{0}' l_{1} - i n_{1}'l_{0}) , \nonumber
\\
({\boldsymbol k}'' )_{3} = (k'_{0} {\boldsymbol k} + {\boldsymbol
k}' k_{0} + i {\boldsymbol k}' \times {\boldsymbol k})_{3} + 2 (-
n_{0}' l_{0} - n_{1}' l_{1}) , \nonumber
\\
n_{0}'' = (k'_{0} +k'_{3}) n_{0} - (k'_{1} -ik'_{2}) n_{1} +
n'_{0} M , \nonumber
\\
n''_{1} = (k'_{0} - k'_{3}) n_{1} - (k'_{1} + i k'_{2}) n_{0} +
n'_{1} M , \nonumber
\\
l_{0}'' = l_{0}' ( k_{0} + k_{3}) + l'_{1} ( k_{1} + i k_{2}) + M'
l_{0} , \nonumber
\\
l''_{1} = l'_{0} (k_{1} -ik_{2}) + l'_{1} (k_{0}-k_{3}) + M'
 l_{1} , \nonumber
\\
M'' = M' M - 4 (l'_{0} n_{0} - l'_{1} n_{1}) .
\end{gather*}
These rules determine multiplication of the matrices
\begin{gather*}
G = \left | \begin{array}{llll}
k_{0} + k_{3} & k_{1} - ik_{2} & +2n_{0} & 0 \\
k_{1} + ik_{2} & k_{0} - k_{3} & -2n_{1} & 0 \\
-2l_{0} & -2l_{1} & M & 0 \\
0 & 0 & 0 & 1
\end{array} \right | .
\nonumber
\end{gather*}

If additionally, in equation (\ref{A.16b}) one requires
\begin{gather*}
n_{0} = 0 , \qquad n_{1}= 0 , \qquad l_{0} = 0 , \qquad l_{1}= 0 ,
\qquad m_{3} =0 , \qquad m_{0} =1 ,
\end{gather*}
then
\begin{gather*}
G = \left | \begin{array}{cccc}
k_{0} + k_{3} & k_{1} - ik_{2} & 0 & 0 \\
k_{1} + ik_{2} & + k_{0} - k_{3} & 0 & 0 \\
0 & 0 & 1 & 0 \\
0 & 0 & 0 & 1
\end{array} \right | ,
\end{gather*}
with the decomposition rule
\begin{gather*}
G =
 k_{1} \lambda_{1} + k_{2} \lambda_{2}
 + \tfrac 12 (1 + k_{0}) I +
 \left[ k_{3} - \tfrac 12( 1 -k_{0} ) \right] \lambda_{3} +
 \tfrac 12 ( 1 - k_{0} ) \lambda_{8}' +
\tfrac 12 (1 -k_{0} ) \lambda_{15}' .
 \end{gather*}
One can readily verify that the $2 \times 2$ block is given by
\begin{gather*}
G^{(2)}(k_{a}) =
 k_{0} I^{(2)} +
 k_{1} \lambda_{1}^{(2)} + k_{2} \lambda_{2}^{(2)}
 + k_{3} \lambda_{3}^{(2)} .
\end{gather*}

\section[On the multiplication law for $GL(4,C)$ in Dirac basis]{On the multiplication law for $\boldsymbol{GL(4,C)}$ in Dirac basis}\label{sec7}

In the Gell-Mann basis $\lambda_{i}$, an element of $GL(4,C)$ is
\begin{gather*}
G= a_{0} I +
 ( a_{1} -i b_{1}) \lambda_{1} + (a_{2} -i b_{2}) \lambda_{2}
 +( i A_{0}-B_{3}) \lambda_{4} + (A_{3} + iB_{0} ) \lambda_{5}+ ( A_{2} - B_{1} ) \lambda_{6}
\nonumber
\\
\phantom{G=}{}+ ( A_{1} + B_{2} ) \lambda_{7}
 +
 (a_{3}- i b_{3}- ib_{0} ) \lambda_{3} +( -i b_{3} + ib_{0} - a_{3} ) \lambda_{8}'
+ (-B_{1} - A_{2} ) \lambda_{9}\\
\phantom{G=}{} + (A_{1} - B_{2} ) \lambda_{10} + (i A_{0} + B_{3})
\lambda_{11} + (- A_{3} + iB_{0} ) \lambda_{12}
+ (- a_{1} - i b_{1} ) \lambda_{13} \\
\phantom{G=}{}+ ( - a_{2} - i b_{2} ) \lambda_{14}
 + (ib_{0} +a_{3} +i b_{3} ) \lambda_{15}' ,
\end{gather*}
or in variables $(k,m,l,n)$:
\begin{gather*}
G= \tfrac 12 (k_{0} + m_{0}) I\! +
 k_{1} \lambda_{1} + k_{2} \lambda_{2}\!
 + \tfrac 12 [ ( n_{0}-n_{3}) - (l_{0} +l_{3}) ] \lambda_{4}\! +
 \tfrac{1}{2 i} [ -(n_{0}-n_{3}) -(l_{0}+l_{3} ) ] \lambda_{5}\!
\nonumber
\\
 \phantom{G=}{}+ \tfrac 12 [ -(n_{1}+in_{2} ) - (l_{1}-il_{2}) ] \lambda_{6}
+ \tfrac{1}{2i} [ (n_{1} +in_{2}) - (l_{1}-il_{2}) ] \lambda_{7}
 +
 \left[ k_{3} + \tfrac 12 (k_{0}-m_{0} ) \right] \lambda_{3}\\
\phantom{G=}{}  +
 \left[ - m_{3} + \tfrac 12 ( m_{0} - k_{0} ) \right] \lambda_{8}'
+\tfrac 12  [ -(n_{1}-in_{2} ) - (l_{1}+il_{2}) ]
\lambda_{9} \\
\phantom{G=}{}+ \tfrac{1}{2i} [ (n_{1} -in_{2}) - (l_{1}+il_{2}) ]
\lambda_{10}
 +
\tfrac{1}{2} [ ( n_{0}+n_{3}) - (l_{0} -l_{3}) ] \lambda_{11}\\
\phantom{G=}{} +
 \tfrac{1}{2 i} [ -(n_{0}+n_{3}) -(l_{0}-l_{3} ) ] \lambda_{12}
-m_{1} \lambda_{13} - m_{2} \lambda_{14} + \left[ m_{3} +
\tfrac{1}{2 } (m_{0} - k_{0} )\right ] \lambda_{15}' .
\end{gather*}

The problem is to establish the multiplication rule $G'' = G' G $
in $\lambda$-basis:
\begin{gather*}
 x_{k}'' \lambda_{k} = x'_{m} \lambda_{m} x_{n} \lambda_{n}=
x'_{m} x_{n} \lambda_{m} \lambda_{n} . \nonumber
\end{gather*}
As by def\/inition the relationships $\lambda_{m} \lambda_{n} =
e_{mnk} \lambda_{k} $
 must hold, the multiplication rule is
\begin{gather}
x_{k}'' = e_{mnk} x'_{m} x_{n} . \label{B.2b}
\end{gather}

\noindent The main claim is that the all properties of the
$GL(4,C)$ with all its subgroups are determined by the bilinear
function (\ref{B.2b}), the latter is described by structure
constants $e_{mnk}$. It is evident that these group constants
should be simpler in the Dirac basis $\Lambda_{i}$ than in the
basis $\lambda_{i}$. Our next task is to establish the
multiplication law
 $G'' = G' G $ in $\Lambda$-basis:
 \begin{gather*}
\Lambda_{m} \Lambda_{n} = E_{mnk} \Lambda_{k} , \qquad
X_{k}'' = E_{mnk} X'_{m} X_{n} . 
\end{gather*}

Before searching for structural constants $E_{mnk} $, let us
introduce a special way to list the Dirac basis $\Lambda_{i}$:
\begin{gather*}
 \alpha_{1} = \gamma^{0}\gamma^{2} , \qquad
 \alpha_{2} = i \gamma^{0}\gamma^{5} , \qquad \alpha_{3} = \gamma^{5}\gamma^{2}
\qquad
 \alpha_{i}^{2} = I, \qquad  \alpha_{1} \alpha_{2} =i\alpha_{3} ,
 \alpha_{2} \alpha_{1} =-i\alpha_{1} ,
\\
 \beta_{1} = i \gamma^{3}\gamma^{1} , \qquad
 \beta_{2} = i \gamma^{3} , \qquad
 \beta_{3} = i\gamma^{1} ,
\qquad
 \beta_{i}^{2} = I, \qquad  \beta_{1} \beta_ {2} =i\beta_{3} ,
 \beta_{2} \beta_ {1} = - i\beta_{3} ,
\end{gather*}
these two set commute with each others $\alpha_{j} \beta_{k} =
\beta_{k} \alpha_{j} $, and their multiplications provides us with
9 remaining basis elements of $\{ \Lambda_{k}\}$:
\begin{alignat}{4}
& A_{1}= \alpha_{1} \beta_{1}= - \gamma^{5} , \qquad &&
 B_{1}= \alpha_{1} \beta_{2}= \gamma^{5}\gamma^{1} ,\qquad &&
 C_{1}=\alpha_{1} \beta_{3} = \gamma^{3} \gamma^{5} , & \nonumber\\
& A_{2}= \alpha_{2} \beta_{1}= -i \gamma^{2} , \qquad &&
 B_{2}= \alpha_{2} \beta_{2}= - i \gamma^{1}\gamma^{2} , \qquad &&
 C_{2}= \alpha_{2} \beta_{3}= - i \gamma^{2}\gamma^{3} ,&\nonumber\\
& A_{3}= \alpha_{3} \beta_{1}= \gamma^{0} ,\qquad &&
 B_{3}= \alpha_{3} \beta_{2}= \gamma^{0}\gamma^{1} , \qquad&&
C_{3}= \alpha_{3} \beta_{3}= \gamma^{0}\gamma^{3} .& \label{B.6}
\end{alignat}
The multiplication rules for basic elements
\begin{gather*}
\alpha_{1} , \ \alpha_{2}, \ \alpha_{3}, \qquad \beta_{1} ,\
\beta_{2}, \ \beta_{3}, \qquad A_{1} , \ A_{2}, \ A_{3}, \qquad
B_{1} , \ B_{2}, \ B_{3}, \qquad C_{1} ,\  C_{2}, \ C_{3},
\nonumber
\end{gather*}
 are
 \begin{gather}
 \begin{array}{c|ccc}
 & \alpha_{1} & \alpha_{2} & \alpha_{3} \\
 \hline
\alpha_{1} & I & i\alpha_{3} & -i\alpha_{2} \\
\alpha_{2} & -i\alpha_{3} & I & i\alpha_{1} \\
\alpha_{3} & i\alpha_{2} & -i\alpha_{1} & I
\end{array}  \qquad
 \begin{array}{c|ccc}
 & \beta_{1} & \beta_{2} & \beta_{3} \\
 \hline
\alpha_{1} & A_{1} & B_{1} & C_{1} \\
\alpha_{2} & A_{2} & B_{2} & C_{2} \\
\alpha_{3} & A_{3} & B_{3} & C_{3}
\end{array}
\qquad \begin{array}{c|ccc}
 & A_{1} & A_{2} & A_{3} \\
 \hline
\alpha_{1} & \beta_{1} & iA_{3} & -i A_{2} \\
\alpha_{2} & -iA_{3} & \beta_{1} & iA_{1} \\
\alpha_{3} & iA_{2} & -iA_{1} & \beta_{1}
\end{array} \nonumber\\[2mm]
 \begin{array}{c|ccc}
 & B_{1} & B_{2} & B_{3} \\
 \hline
\alpha_{1} & \beta_{2} & iB_{3} & -i B_{2} \\
\alpha_{2} & -iB_{3} & \beta_{2} & iB_{1} \\
\alpha_{3} & iB_{2} & -iB_{1} & \beta_{2}
\end{array} \qquad
 \begin{array}{c|ccc}
 & C_{1} & C_{2} & C_{3} \\
 \hline
\alpha_{1} & \beta_{3} & iC_{3} & -i C_{2} \\
\alpha_{2} & -iC_{3} & \beta_{3} & iC_{1} \\
\alpha_{3} & iC_{2} & -iC_{1} & \beta_{3}
\end{array} \qquad \begin{array}{c|ccc}
 & \alpha_{1} & \alpha_{2} & \alpha_{3} \\
 \hline
\beta_{1} & A_{1} & A_{2} & A_{3} \\
\beta_{2} & B_{1} & B_{2} & B_{3} \\
\beta_{3} & C_{1} & C_{2} & C_{3}
\end{array}
\nonumber\\[2mm]
\begin{array}{c|ccc}
 & \beta_{1} & \beta_{2} & \beta_{3} \\
 \hline
\beta_{1} & I & i\beta_{3} & -i\beta_{2} \\
\beta_{2} & -i\beta_{3} & I & i\beta_{1} \\
\beta_{3} & i\beta_{2} & -i\beta_{1} & I
\end{array} \qquad
\begin{array}{c|ccc}
& A_{1} & A_{2} & A_{3} \\
\hline
\beta_{1} & \alpha_{1} & \alpha_{2} & \alpha_{3} \\
\beta_{2} & -iC_{1} & -iC_{2} & -iC_{3} \\
\beta_{3} & iB_{1} & iB_{2} & iB_{3}
\end{array}\qquad
 \begin{array}{c|ccc}
& B_{1} & B_{2} & B_{3} \\
\hline
\beta_{1} & iC_{1} & iC_{2} & iC_{3} \\
\beta_{2} & \alpha_{1} & \alpha_{2} & \alpha_{3} \\
\beta_{3} & -iA_{1} & -iA_{2} & -iA_{3}
\end{array} \!\!\!\!\!\nonumber\\[2mm]
 \begin{array}{c|ccc}
& C_{1} & C_{2}& C_{3}\\
\hline
\beta_{1} & -iB_{1} & -iB_{2} & -iB_{3} \\
\beta_{2} & iA_{1} & iA_{2} & iA_{3} \\
\beta_{3} & \alpha_{1} & \alpha_{2} & \alpha_{3}
\end{array}
\qquad
\begin{array}{c|ccc}
 & \alpha_{1} & \alpha_{2} & \alpha_{3} \\
 \hline
A_{1} & \beta_{1} & i A_{3} & -iA_{2} \\
A_{2} & -iA_{3} & \beta_{1} & iA_{1} \\
A_{3} & iA_{2} & -iA_{1} & \beta_{1}
\end{array} \qquad
 \begin{array}{c|ccc}
 & \beta_{1} & \beta_{2} & \beta_{3} \\
 \hline
 A_{1} & \alpha_{1}& i C_{1} & -iB_{1} \\
A_{2} & \alpha_{2} & iC_{2} & -iB_{2} \\
A_{3} & \alpha_{3} & iC_{3} & -iB_{3}
\end{array}
\nonumber\\[2mm]
 \begin{array}{c|ccc}
 & A_{1} & A_{2} & A_{3} \\
 \hline
A_{1} & I & i\alpha_{3} & -i\alpha_{2} \\
A_{2} & -i\alpha_{3} & I & i\alpha_{1} \\
A_{3} & i\alpha_{2} & -i\alpha_{1} & I
\end{array} \qquad
\begin{array}{c|ccc}
 & B_{1} & B_{2} & B_{3} \\
 \hline
A_{1} & i\beta_{3} & -C_{3} & C_{2} \\
A_{2} & C_{3} & i\beta_{3} & -C_{1} \\
A_{3} & -C_{2} & C_{1} & i\beta_{3}
\end{array} \qquad
\begin{array}{c|ccc}
 & C_{1} & C_{2} & C_{3}\\
 \hline
A_{1} & -i\beta_{2} & B_{3} & -B_{2} \\
A_{2} & -B_{3} & -i\beta_{2} & B_{1} \\
A_{3} & B_{2} & -B_{1} & -i\beta_{2}
\end{array}
\nonumber\\[2mm]
 \begin{array}{c|ccc}
 & \alpha_{1} & \alpha_{2} & \alpha_{3} \\
 \hline
B_{1} & \beta_{2} & i B_{3} & -iB_{2} \\
B_{2} & -iB_{3} & \beta_{2} & iB_{1} \\
B_{3} & iB_{2} & -iB_{1} & \beta_{2}
\end{array} \qquad
\begin{array}{c|ccc}
 & \beta_{1} & \beta_{2} & \beta_{3} \\
 \hline
B_{1} & -i C_{1} & \alpha_{1} & iA_{1} \\
B_{2} & -i C_{2} & \alpha_{2} & iA_{2} \\
B_{3} & -i C_{3} & \alpha_{3} & iA_{3}
\end{array}
\qquad \begin{array}{c|ccc}
 & A_{1} & A_{2} & A_{3}\\
 \hline
B_{1}& -i\beta_{3} & C_{3} & -C_{2} \\
B_{2} & -C_{3} & -i\beta_{3} & C_{1} \\
B_{3} & C_{2} & -C_{1} & -i\beta_{3}
\end{array} \nonumber\\[2mm]
 \begin{array}{c|ccc}
 & B_{1} & B_{2} & B_{3} \\
 \hline
B_{1} & I & i\alpha_{3} & -i\alpha_{2} \\
B_{2} & -i\alpha_{3} & I & i\alpha_{1} \\
B_{3} & i\alpha_{2} & -i\alpha_{1} & I
\end{array} \qquad \begin{array}{c|ccc}
 & C_{1} & C_{2} & C_{3} \\
 \hline
B_{1} & i\beta_{1} & -A_{3} & A_{2} \\
B_{2} & A_{3} & i\beta_{1} & - A_{1} \\
B_{3} & -A_{2} & A_{1} & i\beta_{1}
\end{array} \qquad
 \begin{array}{c|ccc}
 & \alpha_{1} & \alpha_{2} & \alpha_{3} \\
 \hline
C_{1} & \beta_{3} & i C_{3} & -iC_{2} \\
C_{2} & -iC_{3} & \beta_{3} & iC_{1} \\
C_{3} & iC_{2} & -iC_{1} & \beta_{3}
\end{array}\!\!\!\!\! \nonumber\\[2mm]
 \begin{array}{c|ccc}
 & \beta_{1} & \beta_{2} & \beta_{3} \\
 \hline
C_{1} & iB_{1} & -iA_{1} & \alpha_{1} \\
C_{2} & iB_{2} & -iA_{2} & \alpha_{2} \\
C_{3} & iB_{3} & -iA_{3} & \alpha_{3}
\end{array} \qquad \begin{array}{c|ccc}
 & A_{1} & A_{2} & A_{3} \\
 \hline
C_{1} & i\beta_{2} & -B_{3} & B_{2} \\
C_{2} & B_{3} & i\beta_{2} & -B_{1} \\
C_{3} & -B_{2} & B_{1} & i\beta_{2}
\end{array} \qquad \begin{array}{c|ccc}
 & B_{1} & B_{2}& B_{3} \\
 \hline
C_{1} & -i\beta_{1} & A_{3} & -A_{2} \beta_{1} \\
C_{2} & -A_{3} & -i\beta_{1} & A_{1} \\
C_{3} & A_{2} & -A_{1} & -i\beta_{1}
\end{array} \nonumber\\[2mm]
 \begin{array}{c|ccc}
 & C_{1} & C_{2} & C_{3} \\
 \hline
C_{1} & I & i\alpha_{3} & -i\alpha_{2} \\
C_{2} & -i\alpha_{3} & I & i\alpha_{1} \\
C_{3} & i\alpha_{2} & -i\alpha_{1} & I
\end{array}
\label{B.7}
\end{gather}
These relations provide us with simple formulas for f\/ifteen
coordinates of the element of $GL(4,C)$
\begin{gather*}
G = \gamma I + a_{j} \alpha_{j} + b_{j}\beta_{j} + X_{j} A_{j} +
Y_{j}B_{j} + Z_{j} C_{j} , \nonumber
\\
\gamma = \tfrac 14 \mbox{Sp}\, G , \qquad a_{j} = \tfrac 14
\mbox{Sp} \,\alpha_{j} G , \qquad b_{j} =
\tfrac 14 \mbox{Sp}\, \beta_{j} G , \nonumber\\
X_{j} = \tfrac 14 \mbox{Sp}\, A_{j} G , \qquad Y_{j} =
\tfrac 14 \mbox{Sp} B_{j} G , \qquad Z_{j} = \tfrac 14 \mbox{Sp}\, C_{j} G . 
\end{gather*}
With the use of relations (\ref{B.7}) an explicit form of the
group law for $(15+1)$ parameters can be found:
\begin{gather*}
(\gamma' I + a_{i}' \alpha_{i} + b_{i}'\beta_{i} + X_{i}' A_{i} +
Y_{i}'B_{i} + Z_{i}' C_{i})
 (\gamma I + a_{j} \alpha_{j} + b_{j}\beta_{j} + X_{j} A_{j} + Y_{j}B_{j} + Z_{j} C_{j})
\\
\qquad {}= \gamma' \gamma I + ( \gamma' a_{j} + a_{j}'\gamma)
\alpha_{j} + ( \gamma' b_{j} + b_{j}' \gamma ) \beta_{j} +
(\gamma' X_{j} + X_{j}'\gamma ) A_{j} + (\gamma'Y_{j} +
Y_{i}' \gamma) B_{j} \\
\qquad{}+ (\gamma'Z_{j} + Z_{j}'\gamma ) C_{j} + a_{1}' ( a_{1} +
a_{2} i\alpha_{3} -a_{3} i \alpha_{2}) + a_{2}' ( - a_{1} i
\alpha_{3} +a_{2} +a_{3} i
\alpha_{1}) \\
\qquad{}+ a_{3}' ( a_{1} i\alpha_{2} -a_{2} i \alpha_{1} + a_{3} )
+a_{1}' ( b_{1} A_{1} + b_{2} B_{1} + b_{3} C_{1}) + a_{2}'
( b_{1} A_{2} + b_{2} B_{2} + b_{3} C_{2}) \\
\qquad{}+ a_{3}' ( b_{1} A_{3} + b_{2} B_{3} + b_{3} C_{3} )
+a_{1}' ( X_{1} \beta_{1} + X_{2} i A_{3} - X_{3} i A_{2})\\
\qquad{} + a_{2}' ( - X_{1} i A_{3} +X_{2} \beta_{1} + X_{3} i
A_{1}) + a_{3}' ( X_{1} iA_{2} - X_{2} i A_{1} + X_{3} \beta_{1} )
\\
\qquad{}+a_{1}' ( Y_{1} \beta_{2} + Y_{2} i B_{3} - Y_{3} i B_{2})
+
a_{2}' ( - Y_{1} i B_{3} +Y_{2} \beta_{2} + Y_{3} i B_{1})\\
\qquad{} + a_{3}' ( Y_{1} iB_{2} - Y_{2} i B_{1} + Y_{3} \beta_{2}
)
+a_{1}' ( Z_{1} \beta_{3} + Z_{2} i C_{3} - Z_{3} i C_{2})\\
\qquad{} + a_{2}' ( - Z_{1} i C_{3} +Z_{2} \beta_{3} +Z_{3} i
C_{1}) + a_{3}' ( Z_{1} iC_{2} - Z_{2} i C_{1} + Z_{3} \beta_{3} )
\\
\qquad {}+b_{1}' ( a_{1} A_{1} + a_{2} A_{2} + a_{3} A_{3}) +
b_{2}' ( a_{1} B_{1} + a_{2} B_{2} + a_{3} B_{3}) + b_{3}' ( a_{1}
C_{1} + a_{2} C_{2} + a_{3} C_{3} )
\\
\qquad{}+ b_{1}' ( b_{1} + b_{2} i\beta_{3} -b_{3} i \beta_{2}) +
b_{2}' ( - b_{1} i \beta_{3} +b_{2} +b_{3} i \beta_{1}) + b_{3}' (
b_{1} i\beta_{2} -b_{2} i \beta_{1} + b_{3} ) \nonumber
\\
\qquad{}+b_{1}' ( X_{1} \alpha_{1} + X_{2} \alpha_{2} + X_{3}
\alpha_{3}) + b_{2}' ( - X_{1} iC_{1} - X_{2} iC_{2} -
X_{3} iC_{3}) \\
\qquad{}+ b_{3}' ( X_{1} i B_{1} + X_{2} iB_{2} + X_{3} iB_{3} )
+b_{1}' ( Y_{1} iC_{1} + Y_{2} iC_{2} + Y_{3} iC_{3})\\
\qquad{} + b_{2}' ( Y_{1} \alpha_{1} + Y_{2} \alpha_{2} + Y_{3}
\alpha_{3}) + b_{3}' ( -Y_{1} i A_{1} - Y_{2} iA_{2} - Y_{3}
iA_{3} )\\
\qquad{}+b_{1}' ( - Z_{1} iB_{1} - Z_{2} iB_{2} - Z_{3} iB_{3}) +
b_{2}' ( Z_{1} iA_{1} + Z_{2} iA_{2} + Z_{3} iA_{3}) \\
\qquad{}+ b_{3}' ( Z_{1} \alpha_{1} + Z_{2} \alpha_{2} + Z_{3}
\alpha_{3} )
+X_{1}' ( a_{1} \beta_{1} + a_{2} iA_{3} - a_{3} iA_{2})\\
\qquad{} + X_{2}' ( -a_{1} iA_{3} + a_{2} \beta_{1} + a_{3}
iA_{1}) + X_{3}' ( a_{1} iA_{2} - a_{2} iA_{1} + a_{3} \beta_{1} )
\\
\qquad{}+X_{1}' ( b_{1} \alpha_{1} + b_{2} iC_{1} - b_{3}
iB_{1})\! + X_{2}' ( b_{1} \alpha_{2} + b_{2} iC_{2} - b_{3}
iB_{2})\! + X_{3}' ( b_{1} \alpha_{3} + b_{2} iC_{3} - b_{3} i
B_{3} )
\\
\qquad{}+X_{1}' ( X_{1}\! + X_{2} i\alpha_{3}\! - X_{3}
i\alpha_{2})\! + X_{2}' ( -X_{1} i\alpha _{3}\! + X_{2} + X_{3}
i\alpha_{1})\! + X_{3}' ( X_{1} i\alpha_{2}\! - X_{2} i\alpha _{1}
+ X_{3} )\! \nonumber
\\
\qquad{}+X_{1}' ( Y_{1} i\beta_{3}\! - Y_{2} C_{3}\! + Y_{3}
C_{2})\! + X_{2}' ( Y_{1} C _{3}\! + Y_{2} i \beta_{3}\! - Y_{3}
C_{1})\! + X_{3}' ( -Y_{1} C_{2}\! + Y_{2} C_{1}\! + Y_{3} i
\beta_{3} ) \nonumber
\\
\qquad{}+X_{1}' ( - Z_{1} i\beta_{2} + Z_{2} B_{3} - Z_{3} B_{2})
+ X_{2}' ( -Z_{1} B _{3} - Z_{2} i \beta_{2} + Z_{3}
B_{1})\\
\qquad{} + X_{3}' ( Z_{1} B_{2} - Z_{2} B_{1} - Z_{3} i \beta_{2}
)
+Y_{1}' ( a_{1} \beta_{2} + a_{2} iB_{3} - a_{3} iB_{2}) \\
\qquad{}+ Y_{2}' ( -a_{1} iB_{3} + a_{2} \beta_{2} + a_{3} iB_{1})
+ Y_{3}' ( a_{1} iB_{2} - a_{2} iB_{1} + a_{3} \beta_{2} )
\\
\qquad{}+Y_{1}' ( -b_{1} iC_{1} + b_{2} \alpha_{1} + b_{3} iA_{1})
+ Y_{2}' ( -b_{1} iC_{2} + b_{2} \alpha_{2} + b_{3}
iA_{2}) \\
\qquad{}+ Y_{3}' ( -b_{1} iC_{3} + b_{2} \alpha_{3} + b_{3} i
A_{3} )
+Y_{1}' (- X_{1} i\beta_{3} + X_{2} C_{3} - X_{3} C_{2}) \\
\qquad{}+ Y_{2}' ( -X_{1} C_{3} - X_{2} i\beta_{3} + X_{3} C_{1})
+ Y_{3}' ( X_{1} C_{2} - X_{2} C_{1} - X_{3} i \beta_{3} )
\\
\qquad{}+Y_{1}' ( Y_{1} + Y_{2} i\alpha_{3} - Y_{3} i\alpha_{2}) +
Y_{2}' ( - Y_{1} i\alpha_{3} + Y_{2} + Y_{3} i\alpha_{1}) + Y_{3}'
( Y_{1} i\alpha_{2} - Y_{2} i\alpha_{1} + Y_{3} )
\\
\qquad{}+Y_{1}' ( Z_{1}i\beta_{1} - Z_{2} A_{3} + Z_{3} A_{2}) +
Y_{2}' ( Z_{1} A_{3} + Z_{2} i\beta_{1} - Z_{3} A_{1}) \\
\qquad{}+ Y_{3}' ( -Z_{1} A_{2} + Z_{2} A_{1} + Z_{3}i\beta_{1} )
+Z_{1}' ( a_{1} \beta_{3} + a_{2} iC_{3} - a_{3} iC_{2})\\
\qquad{} + Z_{2}' ( -a_{1} iC_{3} + a_{2} \beta_{3} + a_{3}
iC_{1}) + Z_{3}' ( a_{1} iC_{2} - a_{2} iC_{1} + a_{3} \beta_{3} )
\\
\qquad{}+Z_{1}' ( b_{1} iB_{1} - b_{2} iA_{1} + b_{3} \alpha_{1})
+ Z_{2}' ( b_{1} iB_{2} - b_{2} iA_{2} + b_{3} \alpha_{2}) +
Z_{3}' ( b_{1} iB_{3} - b_{2} i A_{3} + b_{3} \alpha_{3} )
\\
\qquad{}+Z_{1}' ( X_{1} i\beta_{2} - X_{2} B_{3} + X_{3} B_{2}) +
 Z_{2}' ( X_{1} B_{3} + X_{2} i\beta_{2} - X_{3} B_{1})\\
 \qquad{} +
 Z_{3}' ( -X_{1} B_{2} + X_{2} B_{1} + X_{3} i \beta_{2} )
+Z_{1}' ( -Y_{1} i\beta_{1} + Y_{2} A_{3} - Y_{3} A_{2})\\
\qquad{} +
 Z_{2}' ( -Y_{1} A_{3} - Y_{2} i\beta_{1} + Y_{3} A_{1}) +
 Z_{3}' ( Y_{1} A_{2} - Y_{2} A_{1} - Y_{3} i \beta_{1} )
\\
\qquad{}+Z_{1}' ( Z_{1} + Z_{2} i\alpha_{3} - Z_{3} i\alpha_{2}) +
Z_{2}' ( -Z_{1} i\alpha _{3} + Z_{2} + Z_{3} i\alpha_{1}) + Z_{3}'
( Z_{1} i\alpha_{2} - Z_{2} i\alpha _{1} + Z_{3} ) .
\end{gather*}

From these relations we arrive at the following composition rules:
\begin{gather}
\gamma'' = \gamma ' \gamma + (a'_{1} a_{1} + a'_{2} a_{2} + a'_{3}
a_{3} )+ (b'_{1} b_{1} + b'_{2} b_{2} + b'_{3} b_{3} ) \nonumber
\\
\phantom{\gamma'' =}{}+ (X'_{1} X_{1} + X'_{2} X_{2} + X'_{3}
X_{3} ) + (Y'_{1} Y_{1} + Y'_{2} Y_{2} + Y'_{3} Y_{3} ) + (Z'_{1}
Z_{1} + Z'_{2} Z_{2} + Z'_{3} Z_{3} ) , \nonumber
\\
a_{1}'' = (\gamma' a_{1} + a_{1}' \gamma) +
 (b'_{1} X_{1} + b'_{2} Y_{1} + b'_{3} Z_{1}) +
 ( X_{1}' b_{1} + Y'_{1} b_{2} + Z_{1}' b_{3})
\nonumber
\\
 \phantom{a_{1}'' =}{} + i (a_{2}' a_{3} - a_{3}' a_{2}) +
i ( X_{2} ' X_{3} - X_{3} ' X_{2}) +i ( Y_{2} ' Y_{3} - Y_{3} '
Y_{2}) +
 i ( Z_{2} ' Z_{3} - Z_{3} ' Z_{2}) ,
\nonumber
\\
a_{2}'' = (\gamma' a_{2} + a_{2}' \gamma) + (b'_{1} X_{2} + b'_{2}
Y_{2}+ b'_{3} Z_{2} ) +
 ( X_{2}' b_{1} + Y'_{2} b_{2} + Z_{2}' b_{3})
\nonumber
\\
\phantom{a_{2}'' =}{} + i (a_{3}' a_{1} - a_{1}' a_{3}) + i (
X_{3} ' X_{1} - X_{1} ' X_{3}) +
 i ( Y_{3} ' Y_{1} - Y_{1} ' Y_{3}) + i ( Z_{3} ' Z_{1} - Z_{1} ' Z_{3}) ,
\nonumber
\\
a_{3}'' = (\gamma' a_{3} + a_{3}' \gamma ) +
 (b'_{1} X_{3} + b'_{2} Y_{3} + b'_{3} Z_{3} ) +
 ( X_{3}' b_{1} + Y'_{3} b_{2} + Z_{3}' b_{3})
\nonumber
\\
\phantom{a_{3}'' =}{} + i (a_{1}' a_{2} - a_{2}' a_{1}) + i (
X_{1} ' X_{2} - X_{2} ' X_{1}) + i ( Y_{1} ' Y_{2} - Y_{2} '
Y_{1}) + i ( Z_{1} ' Z_{2} - Z_{2} ' Z_{1}) , \nonumber
\\
b_{1}'' = \gamma' b_{1} + b_{1}' \gamma + i (b_{2}' b_{3} - b_{3}'
b_{2}) + (a'_{1} X_{1} + a'_{2} X_{2} +a'_{3} X_{3} ) +
 (X'_{1} a_{1} + X'_{2} a_{2} +X'_{3} a_{3} )
\nonumber
\\
\phantom{b_{1}'' =}{}+ i ( Y_{1}' Z_{1} + Y_{2}' Z_{2}+ Y_{3}'
Z_{3}) - i
 ( Z_{1}' Y_{1} + Z_{2}' Y_{2}+ Z_{3}' Y_{3}) ,
\nonumber
\\
b_{2}'' = \gamma' b_{2} + b_{2}' \gamma + i (b_{3}' b_{1} - b_{1}'
b_{3}) + (a'_{1} Y_{1} + a'_{2} Y_{2} +a'_{3} Y_{3} ) + (Y'_{1}
 a_{1} + Y'_{2} a_{2} +Y'_{3} a_{3} ) \nonumber
\\
\phantom{b_{2}'' =}{} + i ( Z_{1}' X_{1} + Z_{2}' X_{2}+ Z_{3}'
X_{3}) - i
 ( X_{1}' Z_{1} + X_{2}' Z_{2}+ X_{3}' Z_{3}) ,
\nonumber
\\
b_{3}'' = \gamma' b_{3} + \gamma b_{3}' + i (b_{1}'
 b_{2} - b_{2}' b_{1})
+ (a'_{1} Z_{1} + a'_{2} Z_{2} +a'_{3} Z_{3} ) + (Z'_{1}
 a_{1} + Z'_{2} a_{2} +Z'_{3} a_{3} ) \nonumber
\\
\phantom{b_{3}'' =}{} + i ( X_{1}' Y_{1} + X_{2}' Y_{2}+ X_{3}'
Y_{3}) - i
 ( Y_{1}' X_{1} + Y_{2}' X_{2}+ Y_{3}' X_{3}) ,
\nonumber
\\
X_{1} '' = (\gamma' X_{1} + \gamma X'_{1} ) + (a_{1}' b_{1} +
a_{1} b'_{1} ) + i (Y'_{1} b_{3} - Y_{1} b'_{3}) + i (b_{2}' Z_{1}
- b_{2} Z'_{1}) \nonumber
\\
\phantom{X_{1} '' =}{} + i (a'_{2} X_{3} - a'_{3} X_{2}) -i (a_{2}
X'_{3} - a_{3} X'_{2})+ (Z_{2} Y'_{3} - Z_{3} Y'_{2} ) + (Z_{2}'
Y_{3} - Z_{3}' Y_{2} ) , \nonumber
\\
X_{2} '' = (\gamma' X_{2} + \gamma X'_{2} ) + (a_{2}' b_{1} +
a_{2} b'_{1} ) +i (Y'_{2} b_{3} - Y_{2} b'_{3}) + i (b_{2}' Z_{2}
- b_{2} Z'_{2}) \nonumber
\\
\phantom{X_{2} '' =}{} + i (a'_{3} X_{1} - a'_{1} X_{2}) -i (
a_{3} X'_{1} - a_{1} X'_{3})+ (Z_{3} Y'_{1} - Z_{1} Y'_{3} ) +
(Z_{3}' Y_{1} - Z_{1}' Y_{3} ) , \nonumber
\\
X_{3} '' = (\gamma' X_{3} + \gamma X'_{3} ) + (a_{3}' b_{1} +
a_{3} b'_{1} ) +i (Y'_{3} b_{3} - Y_{3} b'_{3}) + i (b_{2}' Z_{3}
- b_{2} Z'_{3}) \nonumber
\\
\phantom{X_{3} '' =}{} + i (a'_{1} X_{2} - a'_{2} X_{1}) - i (
a_{1} X'_{2} - a_{2} X'_{1} )+ (Z_{1} Y'_{2} - Z_{2} Y'_{1} ) +
(Z_{1}' Y_{2} - Z_{2}' Y_{1} ) , \nonumber
\\
Y_{1} '' = (\gamma' Y_{1} + \gamma Y'_{1} ) + (a_{1}' b_{2} +
a_{1} b'_{2} ) +i (Z'_{1} b_{1} - Z_{1} b'_{1}) + i (b_{3}' X_{1}
- b_{3} X'_{1}) \nonumber
\\
\phantom{Y_{1} '' =}{} + i (a'_{2} Y_{3} - a'_{3} Y_{2}) -i (a_{2}
Y'_{3} - a_{3} Y'_{2})+ (X_{2} Z'_{3} - X_{3} Z'_{2} ) + (X_{2}'
 Z_{3} - X_{3}' Z_{2} ) , \nonumber
\\
Y''_{2} = (\gamma' Y_{2} + \gamma Y'_{2} ) + (a_{2}' b_{2} + a_{2}
b'_{2} ) +i (Z'_{2} b_{1} - Z_{2} b'_{1}) + i (b_{3}' X_{2} -
b_{3} X'_{2}) \nonumber
\\
\phantom{Y''_{2} =}{} + i (a'_{3} Y_{1} - a'_{1} Y_{3}) -i (a_{3}
Y'_{1} - a_{1} Y'_{3})+ (X_{3} Z'_{1} - X_{1} Z'_{3} ) + (X_{3}'
 Z_{1} - X_{1}' Z_{3} ) , \nonumber
\\
Y''_{3} = (\gamma' Y_{3} + \gamma Y'_{3} ) + (a_{3}' b_{2} + a_{3}
b'_{2} ) +i (Z'_{3} b_{1} - Z_{3} b'_{1}) + i (b_{3}' X_{3} -
b_{3} X'_{3}) \nonumber
\\
\phantom{Y''_{3} = }{} + i (a'_{1} Y_{2} - a'_{2} Y_{1}) - i
(a_{1} Y'_{2} - a_{2} Y'_{1})+ (X_{1} Z'_{2} - X_{2} Z'_{1} ) +
(X_{1}' Z_{2} - X_{2}' Z_{1} ) , \nonumber
\\
Z_{1} '' = (\gamma' Z_{1} + \gamma Z'_{1} ) + (a_{1}' b_{3} +
a_{1} b'_{3} ) +i (Y_{1} b_{1}' - Y'_{1} b_{1} ) + i (X_{1}'
b_{2}- X_{1} b_{2}' ) \nonumber
\\
\phantom{Z_{1} '' =}{} + i (a'_{2} Z_{3} - a'_{3} Z_{2}) -i (a_{2}
Z'_{3} - a_{3} Z'_{2})+ (Y_{2} X'_{3} - Y_{3} X'_{2} ) + (Y_{2}'
 X_{3} - Y_{3}' X_{2} ) , \nonumber
\\
Z_{2} '' = (\gamma' Z_{2} + \gamma Z'_{2} ) + (a_{2}' b_{3} +
a_{2} b'_{3} ) +i (Y_{2} b_{1}' - Y'_{2} b_{1} ) + i (X_{2}'
b_{2}- X_{2} b_{2}' ) \nonumber
\\
\phantom{Z_{2} '' =}{} + i (a'_{3} Z_{1} - a'_{1} Z_{3}) -i (a_{3}
Z'_{1} - a_{1} Z'_{3})+ (Y_{3} X'_{1} - Y_{1} X'_{3} ) + (Y_{3}'
 X_{1} - Y_{1}' X_{3} ) , \nonumber
\\
Z_{3} '' = (\gamma' Z_{3} + \gamma Z'_{3} ) + (a_{3}' b_{3} +
a_{3} b'_{3} ) +i (Y_{3} b_{1}' - Y'_{3} b_{1} ) + i (X_{3}' b_{2}
- X_{3} b_{2}' ) \nonumber
\\
\phantom{Z_{3} '' =}{} + i (a'_{1} Z_{2} - a'_{2} Z_{1}) - i
(a_{1} Z'_{2} - a_{2} Z'_{1})+ (Y_{1} X'_{2} - Y_{2} X'_{1} ) +
(Y_{1}' X_{2} - Y_{2}' X_{1} ) . \label{B.7'}
\end{gather}
 With the help of the index notation
\begin{gather*}
{\boldsymbol X} = {\boldsymbol C} ^{(1)} , \qquad {\boldsymbol Y}
= {\boldsymbol C}^{(2)} , \qquad {\boldsymbol Z} = {\boldsymbol
C}^{(3)} , \nonumber
\end{gather*}
it is easy to see a cyclic symmetry in the above relationships:
\begin{gather}
\gamma'' = \gamma ' \gamma + a'_{k} a_{k} + b_{k}' b_{k} +
 C^{(1)'}_{k} C^{(1)}_{k} + C^{(2)'}_{k} C^{(2)}_{k}
+ C^{(3)'}_{k} C^{(3)}_{k} , \nonumber
\\
 a''_{k} = \gamma' a_{k} + \gamma a'_{k}
 + (b_{1}' C^{(1)}_{k} + b_{1} C^{(1)'}_{k} ) + (b_{2}' C^{(2)}_{k} +b_{2} C^{(2)'}_{k})+
(b_{3}' C^{(3)}_{k} ) + b_{3} C^{(3)'} _{k} ) \nonumber
\\
\phantom{a''_{k} =}{} + i \epsilon_{kln} a_{l} ' a_{n} + i
\epsilon_{kln} C^{(1)'}_{l} C^{(1)}_{n} + i \epsilon_{kln}
C^{(2)'}_{l} C^{(2)}_{n} + i \epsilon_{kln} C^{(3)'}_{l}
C^{(3)}_{n} , \nonumber
\\
b_{k}'' = \gamma' b_{k} + \gamma b'_{k} + i \epsilon_{kln} b' _{l}
b_{n} + (a_{1}' C^{(k)}_{1} +a_{1} C^{(k)'}_{1} ) + ( a_{2}'
C^{(k)}_{2} + a_{2} C^{(k)'}_{2} ) \nonumber
\\
\phantom{b_{k}'' =}{} +( a_{3}' C^{(k)}_{3}
 + a_{3} C^{(k)'}_{3})
 + i \epsilon_{kln} C^{(l)'}_{m} C^{(n)}_{m} ,
\nonumber
\\
C^{(1)}_{k} = \gamma' C^{(1)}_{k} + \gamma C^{(1)'}_{k} + (a'_{k}
b_{1} + a_{k} b'_{1}) + i \epsilon_{(1)ln} ( C^{(l)'}_{k} b_{n} -
C^{(l)}_{k} b'_{n}) \nonumber
\\
\phantom{C^{(1)}_{k} =}{}+ i \epsilon_{kln} (a_{l} 'C_{n}^{(1)} -
a_{l} C_{n}^{(1)'}) + \epsilon_{kln} (C^{(2)'}_{l} C^{(3)}_{n} +
C^{(2)}_{l} C^{(3)'}_{n} ) , \nonumber
\\
C^{(2)}_{k} = \gamma' C^{(2)}_{k} + \gamma C^{(2)'}_{k} + (a'_{k}
b_{2} + a_{k} b'_{2}) + i \epsilon_{(2)ln} ( C^{(l)'}_{k} b_{n} -
C^{(l)}_{k} b'_{n}) \nonumber
\\
\phantom{C^{(2)}_{k} =}{} + i \epsilon_{kln} (a_{l}' C_{n}^{(2)} -
a_{l} C_{n}^{(2)'} ) + \epsilon_{kln} ( C^{(3)'}_{l} C^{(1)}_{n} +
C^{(3)}_{l} C^{(1)'}_{n}) , \nonumber
\\
C^{(3)}_{k} = \gamma' C^{(3)}_{k} + \gamma C^{(3)'}_{k} + (a'_{k}
b_{3} + a_{k} b'_{3}) + i \epsilon_{(3)ln} ( C^{(l)'}_{k} b_{n} -
C^{(l)}_{k} b'_{n}) \nonumber
\\
\phantom{C^{(3)}_{k} =}{} + i \epsilon_{kln} ( a_{l}' C_{n}^{(3)}
- a_{l} C_{n}^{(3)'} ) + \epsilon_{kln} (C^{(1)'}_{l} C^{(2)}_{n}
+ C^{(1)}_{l} C^{(2)'}_{n}) . \label{B.7''}
\end{gather}

It is readily seen that these group multiplication laws
(\ref{B.7'}), (\ref{B.7''})
 permit 15 two-parametric subgroups:
\begin{gather*}
(\gamma, a ) \in
 \{ (\gamma, a_{1}) , (\gamma, a_{2}) ,
(\gamma, a_{3}) , \ (\gamma, b_{1}) , (\gamma, b_{2}) , \gamma,
b_{3}) , \ (\gamma, X_{1}) , (\gamma, X_{2}) , (\gamma, X_{3})
 ,\\
 \phantom{(\gamma, a ) \in{}}{}   (\gamma, Y_{1}) , (\gamma, Y_{2}) , (\gamma,
Y_{3}) , \ (\gamma, Z_{1}) , (\gamma, Z_{2}) , (\gamma, Z_{3}) \}
\nonumber
\end{gather*}
with the same composition law:
\begin{gather*}
\gamma'' = \gamma ' \gamma + a' a , \qquad a'' = \gamma
' a + \gamma a' , 
\end{gather*}
which in variables $ \gamma = W \cos \phi$, $a = i W \sin \phi $
takes the form
\begin{gather*}
W'' = W'W , \qquad \alpha'' = \alpha ' + \alpha .
\end{gather*}
The variable $W$ is determined by $\det G(W, \alpha) = W^{4}  $,
the choice $W=1$ guarantees $\det G = +1 $.

All 15 basis elements $\Lambda_{(\rho)} \in \{ \alpha_{k} ,
\beta_{k} , A_{k} , B_{k} , C_{k} \} $ possess the same
properties:
\begin{gather*}
\Lambda^{+}_{(\rho)}= \Lambda_{(\rho)} , \qquad
\Lambda^{2}_{(\rho)} = I. 
\end{gather*}
Therefore, one can construct 15 dif\/ferent elementary unitary (at
real valued parameters) matrices by one the same recipe:
\begin{gather*}
U_{(\rho)} = e^{i \phi_{(\rho)} \Lambda_{(\rho)}} = \cos
\phi_{(\rho)} + i \sin \phi_{(\rho)} \Lambda_{(\rho)} , \nonumber
\\
 U^{+}_{(\rho)} = U^{-1}_{(\rho)} = e^{-i \phi_{(\rho)} \Lambda_{(\rho)}}
 =\cos \phi_{(\rho)} - i \sin \phi_{(\rho)} \Lambda_{(\rho)} .
\end{gather*}
The whole set of unitary matrices $SU(4)$ may be constructed on
the basis of a simple factorized formula:
\begin{gather*}
U = e^{i \phi_{(1)} \Lambda_{(1)}} \cdots e^{i
\phi_{(15)} \Lambda_{(15)}} . 
\end{gather*}
The order of the factors is important. Every such order leads us
to a def\/inite parametrization for the group $SU(4)$ -- all them
seem to be equivalent.

In the end of the section let us write down the explicit form of
these 15 elementary unitary transformations:
\begin{gather}
\alpha_{1} = \left | \begin{array}{cc}
\sigma_{2} & 0 \\
0 & -\sigma_{2}
\end{array} \right | ,\qquad
\alpha_{2} = \left | \begin{array}{cc}
0 & i \\
-i & 0
\end{array} \right | ,\qquad
\alpha_{3} = \left | \begin{array}{cc}
0 & \sigma_{2} \\
\sigma_{2} & 0
\end{array} \right | ,
\nonumber
\\
U_{1}^{\alpha} =
 \left | \begin{array}{cc}
\cos \phi +i \sin \phi \sigma_{2} & 0 \\
0 & \cos \phi -i \sin \phi \sigma_{2}
\end{array} \right | ,
\qquad U_{2}^{\alpha} =
 \left | \begin{array}{cc}
\cos \phi & -\sin \phi \\
\sin \phi & \cos \phi
\end{array} \right |, \nonumber\\
U_{3}^{\alpha} =
 \left | \begin{array}{cc}
\cos \phi & i\sin \phi \sigma_{2}\\
i \sin \phi \sigma_{2} & \cos \phi
\end{array} \right | ,
\qquad \beta_{1} = \left | \begin{array}{cc}
\sigma_{2} & 0 \\
0 & \sigma_{2}
\end{array} \right | ,\qquad
\beta_{2} = \left | \begin{array}{cc}
0 & -i\sigma_{3} \\
i \sigma_{3} & 0
\end{array} \right | ,\nonumber\\
\beta_{3} = \left | \begin{array}{cc}
0 & -i\sigma_{1} \\
i \sigma_{1} & 0
\end{array} \right | ,
\qquad U_{1} ^{\beta} =
 \left | \begin{array}{cc}
\cos \phi +i \sin \phi \sigma_{2} & 0 \\
0 & \cos \phi +i \sin \phi \sigma_{2}
\end{array} \right |,
\nonumber
\\
U_{2}^{\beta} =
 \left | \begin{array}{cc}
\cos \phi & \sin \phi \sigma_{3}\\
- \sin \phi \sigma_{3} & \cos \phi
\end{array} \right |, \qquad
U_{3}^{\beta} =
 \left | \begin{array}{cc}
\cos \phi & \sin \phi \sigma_{1}\\
- \sin \phi \sigma_{1} & \cos \phi
\end{array} \right |,
\qquad A_{1} = \left | \begin{array}{cc}
I & 0 \\
0 & -I
\end{array} \right | ,\nonumber\\
A_{2} =
 \left | \begin{array}{cc}
0 & i\sigma_{2} \\
-i \sigma_{2} & 0
\end{array} \right | ,\qquad
A_{3} = \left | \begin{array}{cc}
0 & I \\
I & 0
\end{array} \right | ,
\qquad U_{1}^{A} =
 \left | \begin{array}{cc}
\cos \phi +i \sin \phi & 0\\
0& \cos \phi -i \sin \phi
\end{array} \right |,\nonumber\\
U_{2}^{A} =
 \left | \begin{array}{cc}
\cos \phi & - \sin \phi \sigma_{2}\\
 \sin \phi \sigma_{2} & \cos \phi
\end{array} \right |,
\qquad U_{3}^{A} =
 \left | \begin{array}{cc}
\cos \phi & i\sin \phi \\
i\sin \phi & \cos \phi
\end{array} \right |,
\qquad B_{1} =
 \left | \begin{array}{cc}
0 & \sigma_{1} \\
 \sigma_{1} & 0
\end{array} \right | ,\nonumber\\
B_{2} =
 \left | \begin{array}{cc}
- \sigma_{3} & 0 \\
 0& - \sigma_{3}
\end{array} \right | ,\qquad
B_{3} =
 \left | \begin{array}{cc}
- \sigma_{1} & 0 \\
 0& \sigma_{1}
\end{array} \right | ,
\qquad U_{1}^{B} =
 \left | \begin{array}{cc}
\cos \phi & i \sin \phi \sigma_{1}\\
i \sin \phi \sigma_{1} & \cos \phi
\end{array} \right |,
\nonumber
\\
U_{2} ^{B} =
 \left | \begin{array}{cc}
\cos \phi -i \sin \phi \sigma_{3} & 0 \\
0 & \cos \phi -i \sin \phi \sigma_{3}
\end{array} \right |,
\nonumber\\
U_{3}^{B} =
 \left | \begin{array}{cc}
\cos \phi -i \sin \phi \sigma_{1} & 0 \\
0 & \cos \phi +i \sin \phi \sigma_{1}
\end{array} \right |,
\qquad C_{1} =
 \left | \begin{array}{cc}
0 & -\sigma_{3} \\
 -\sigma_{3} & 0
\end{array} \right | ,\nonumber\\
C_{2} =
 \left | \begin{array}{cc}
- \sigma_{1} & 0 \\
 0& - \sigma_{1}
\end{array} \right | ,\qquad
C_{3} =
 \left | \begin{array}{cc}
 \sigma_{3} & 0 \\
 0& -\sigma_{3}
\end{array} \right | ,
\qquad U_{1}^{C} =
 \left | \begin{array}{cc}
\cos \phi & -i \sin \phi \sigma_{3}\\
-i \sin \phi \sigma_{3} & \cos \phi
\end{array} \right |,
\nonumber
\\
U_{2}^{C} = \left | \begin{array}{cc}
\cos \phi -i \sin \phi \sigma_{1} & 0 \\
0 & \cos \phi -i \sin \phi \sigma_{1}
\end{array} \right |,
\nonumber
\\
U_{3}^{C} =
 \left | \begin{array}{cc}
\cos \phi +i \sin \phi \sigma_{3} & 0 \\
0 & \cos \phi -i \sin \phi \sigma_{3}
\end{array} \right |.
\label{B.11}
\end{gather}

\noindent Certainly, these relations provide us with 15 elementary
solutions of the unitarity equations (\ref{1.3}). For instance,
the generator $\alpha_{2}$ gives rise to the above 1-parametric
Abelian
 subgroup $G^{0}(\alpha )$; whereas the above 4-parametric subgroup $G_{0} \times SU(2)$ (\ref{B.15})
 is generated by $(\alpha_{2}; \beta_{1}, B_{2}, C_{2})$.

The question is how one could describe all combinations of the
above 15 simple sub-solutions by a single unifying formula~-- the
latter should evidently exist.

\section[On factorization $SU(4)$ and the group fine-structure]{On factorization $\boldsymbol{SU(4)}$ and the group f\/ine-structure}\label{sec8}

On the basis of 9 matrices (\ref{B.6}) one can construct six
3-dimensional sub-sets:
\begin{gather*}
{\boldsymbol K} = \{ A_{1}= \alpha_{1} \beta_{1} , \ B_{2}=
\alpha_{2} \beta_{2} , \  C_{3}= \alpha_{3} \beta_{3}
 \} , \nonumber
\\
{\boldsymbol L} = \{ C_{1}= \alpha_{1} \beta_{3} , \ A_{2}=
\alpha_{2} \beta_{1} , \
 B_{3}= \alpha_{3} \beta_{2} \} ,
\nonumber
\\
{\boldsymbol M} = \{ B_{1} = \alpha_{1} \beta_{2} , \  C_{2} =
\alpha_{2} \beta_{3} , \  A_{3} = \alpha_{3} \beta_{1} \} ,
\nonumber
\\
{\boldsymbol K} ' = \{ -C'_{1} = -\alpha_{1} \beta_{3} , \ -B'_{2}
= -\alpha_{2} \beta_{2} , \  -C'_{3} = -\alpha_{3} \beta_{3} \} ,
\nonumber
\\
{\boldsymbol L}' = \{ -B'_{1} = -\alpha_{1} \beta_{2} , \ -A'_{2}
= -\alpha_{2} \beta_{1} , \  -B'_{3} = -\alpha_{3} \beta_{2} \} ,
\nonumber
\\
{\boldsymbol M}' = \{ -A'_{1} = -\alpha_{1} \beta_{1} , \ -C'_{2}
= -\alpha_{2} \beta_{3} , \  -B'_{3} =
-\alpha_{3} \beta_{2} \} , 
\end{gather*}
(one may recall the rule to calculate the determinant of a
$3\times 3$ matrix) with the same commutation relations:
\begin{gather}
\Gamma_{1} \Gamma_{2} = - \Gamma_{3} , \qquad \Gamma_{2}
\Gamma_{1} = - \Gamma_{3} , \qquad \Gamma_{1} \Gamma_{2} -
\Gamma_{2} \Gamma_{1} =0 , \qquad \Gamma_{1} \Gamma_{2} +
\Gamma_{2} \Gamma_{1} = -2 \Gamma_{3}, \label{C.3}
\end{gather}
and analogous by cyclic symmetry. The whole set of the above 9
matrices coincides with
\begin{gather}
\vec{\alpha} , \quad \vec{\beta} , \quad {\boldsymbol K} , \quad
{\boldsymbol L} , \quad {\boldsymbol M} , \label{C.4a}
\end{gather}
or
\begin{gather*}
\vec{\alpha} , \quad \vec{\beta} , \quad {\boldsymbol K} ', \quad
{\boldsymbol L}' , \quad {\boldsymbol M}' . \nonumber
\end{gather*}
It suf\/f\/ices to consider one variant, let it be (\ref{C.4a}).
It seem reasonable to suppose that arbitrary element from
$GL(4,C)$ can be factorized as follows
\begin{gather}
S = e^ {i \vec{a} \vec{\alpha} } e^ {i \vec{b} \vec{\beta} } e^{i
{\boldsymbol k} {\boldsymbol K} } e^{i {\boldsymbol l}
 {\boldsymbol L} } e^{i {\boldsymbol m} {\boldsymbol M} } . \label{C.5}
\end{gather}

\noindent When all parameters are real-valued, the formula
provides us with the rule to construct elements from $SU(4)$
group\footnote{Just such a structure was described in
\cite{Kihlberg}.}. The order of factors might be dif\/ferent. Let
us specify the group law for these 5 subsets. First are the two
groups:
\begin{gather*}
e^ {i \vec{a} \vec{\alpha} } = \cos a + i \sin a ( n_{1}
\alpha_{1} + n_{2} \alpha_{2} + n_{3} \alpha_{3} ) , \qquad e^ {i
\vec{b} \vec{\beta} } = \cos b + i \sin b ( n_{1} \beta_{1} +
n_{2} \beta_{2} + n_{3} \beta _{3} ) . \nonumber
\end{gather*}
They are isomorphic, so one can consider only the f\/irst one:
\begin{gather*}
e^ {i \vec{a} \vec{\alpha} } = \cos a + i \sin a ( n_{1}
\alpha_{1} + n_{2} \alpha_{2} + n_{3} \alpha_{3} ) = x_{0} - i
x_{1} \alpha_{1} -i x_{2} \alpha_{2} - i
x_{3} \alpha_{3} . 
\end{gather*}
 Multiplying two matrices we arrive at
\begin{gather}
x''_{0} = x'_{0}x_{0} - x'_{1}x_{1}- x'_{2}x_{2}- x'_{3}x_{3} ,
\qquad x''_{1} = x'_{0}x_{1} + x'_{1} x_{0} + ( x'_{2} x_{3} -
x'_{2} x_{3}) , \nonumber
\\
x''_{2} = x'_{0} x_{2} + x'_{2} x_{0} + ( x'_{3} x_{1} - x'_{1}
x_{3}) , \qquad x''_{3} = x'_{0}x_{3} + x'_{3} x_{0} + ( x'_{1}
x_{2} - x'_{2} x_{1}) . \label{C.6b}
\end{gather}
Parameters $(x_{0}, x_{i})$ should obey
\begin{gather*}
x_{0}^{2} + x_{1}^{2} + x_{2}^{2} + x_{3}^{2} = 1 \qquad
\Longleftrightarrow \qquad
 \det e^ {i \vec{a} \vec{\alpha} } = + 1 .
\end{gather*}
The inverse matrix looks
\begin{gather*}
(x_{0}, {\boldsymbol x})^{-1} = (x_{0}, -{\boldsymbol x}) . 
\end{gather*}
With real $(x_{0}, x_{i})$ we have a group isomorphic to $SU(2)$,
spinor covering for $SO(3,R)$:
\begin{gather*}
{\boldsymbol c} ={ {\boldsymbol x} \over x_{0}} , \qquad
{\boldsymbol c}'' = { {\boldsymbol c}' + {\boldsymbol c} +
{\boldsymbol c}' \times {\boldsymbol c} \over 1 - {\boldsymbol c}'
{\boldsymbol
c}} . 
\end{gather*}
At complex $(x_{0}, x_{i})$ we have a group isomorphic to
$GL(2,C)$, spinor covering
 for $SO(3,C)$ or Lorentz group.

Now let us turn to f\/inite transformations from remaining
subsets. It is readily verif\/ied that these 1-parametric f\/inite
elements
\begin{gather*}
e^ { i y_{1} \Gamma _{1} } = \cos y_{1} + i\sin y_{1} \Gamma_{1} ,
\qquad e^ { i y_{2} \Gamma _{1} } = \cos y_{2} + i\sin y_{2}
\Gamma_{2} , \qquad e^ { i y_{3} \Gamma _{3} } = \cos y_{3} +
i\sin y_{3}
\Gamma_{3} , 
\end{gather*}
commute with each other:
\begin{gather*}
e^ { i y_{1} \Gamma _{1} } e^ { i y_{2} \Gamma _{2} } = (\cos
y_{1} + i\sin y_{1} \Gamma_{1})(\cos y_{2} + i\sin y_{2}
\Gamma_{2})= \nonumber
\\
\phantom{e^ { i y_{1} \Gamma _{1} } e^ { i y_{2} \Gamma _{2} }}{}
= \cos y_{1} \cos y_{2} +i \cos y_{1} \sin y_{2} \Gamma _{2} + i
\cos y_{2} \sin y_{1} \Gamma _{1}+ \sin y_{1} \sin y_{2}
\Gamma_{3} , \nonumber
\\
e^ { i y_{2} \Gamma _{2} } e^ { i y_{1} \Gamma _{1} } = (\cos
y_{2} + i\sin y_{2} \Gamma_{2})(\cos y_{1} + i\sin y_{1}
\Gamma_{1})= \nonumber
\\
\phantom{e^ { i y_{2} \Gamma _{2} } e^ { i y_{1} \Gamma _{1} }}{}
= \cos y_{2} \cos y_{1} +i \cos y_{2} \sin y_{1} \Gamma _{1} + i
\cos y_{1} \sin y_{2} \Gamma _{2}+ \sin y_{2} \sin y_{1}
\Gamma_{3} , \nonumber
\end{gather*}
that is $ e^ { i y_{1} \Gamma _{1} } e^ { i y_{2} \Gamma _{2} } =
e^ { i y_{2} \Gamma _{2} } e^ { i y_{1} \Gamma _{1}
}$, 
and so on. Evidently, this property correlates with the
commutative relations (\ref{C.3}). Thus, each of tree subgroups
can be constructed as multiplying of elementary 1-parametric
commuting transformations. Their explicit forms are:

subgroup $K$
\begin{gather*}
{\boldsymbol K} = \{ A_{1}= \alpha_{1} \beta_{1} , \  B_{2}=
\alpha_{2} \beta_{2} , \  C_{3}= \alpha_{3} \beta_{3}
 \} , \nonumber
\\
A_{1} = \left | \begin{array}{cc} I & 0 \\0 & -I \end{array}
\right |,
 \qquad
 B_{2} = \left | \begin{array}{cc}
-\sigma_{3} & 0 \\0 & -\sigma_{3} \end{array} \right |,
 \qquad
 C_{3} = \left | \begin{array}{cc}
\sigma_{3} & 0 \\0 & - \sigma_{3} \end{array} \right |, \nonumber
\\
e^ { i k_{1} K_{1} } = \cos k_{1} + i\sin k_{1} A_{1}
 , \qquad
e^ { i k_{2} K _{1} } = \cos k_{2} + i\sin k_{2} B_{2}
 , \\
 e^ { i k_{3} K _{3} } = \cos k_{3} + i\sin k_{3} C_{3} ; 
\end{gather*}

subgroup $L$
\begin{gather*}
{\boldsymbol L} = \{ C_{1}= \alpha_{1} \beta_{3} , \ A_{2}=
\alpha_{2} \beta_{1} , \
 B_{3}= \alpha_{3} \beta_{2} \} ,
\nonumber
\\
C_{1} = \left | \begin{array}{cc}
0 & - \sigma_{3} \\
-\sigma_{3} & 0
\end{array} \right |, \qquad
A_{2} = \left | \begin{array}{cc}
0 & i \sigma_{2} \\
-i\sigma_{2} & 0
\end{array} \right |, \qquad
 B_{3} = \left | \begin{array}{cc}
-\sigma_{1} & 0 \\0 & \sigma_{1} \end{array} \right |, \nonumber
\\
e^ { i l_{1} L_{1} } = \cos l_{1} + i\sin l_{1} C_{1}
 , \qquad
e^ { i l_{2} L _{1} } = \cos l_{2} + i\sin l_{2} A_{2}
 , \qquad
e^ { i l_{3} L _{3} } = \cos l_{3} + i\sin l_{3} B_{3}
 ; 
\end{gather*}

subgroup $M$
\begin{gather*}
{\boldsymbol M} = \{ B_{1} = \alpha_{1} \beta_{2} , \  C_{2} =
\alpha_{2} \beta_{3} , \  A_{3} = \alpha_{3} \beta_{1} \} ,
\nonumber
\\
B_{1} = \left | \begin{array}{cc}
0 & \sigma_{1} \\
\sigma_{1} & 0
\end{array} \right |, \qquad
 C_{2} = \left | \begin{array}{cc}
-\sigma_{1} & 0 \\0 & -\sigma_{1} \end{array} \right |, \qquad
A_{3} = \left | \begin{array}{cc}
0 & I \\
I & 0
\end{array} \right |,
\nonumber
\\
e^ { i m_{1} M_{1} } = \cos m_{1} + i\sin m_{1} B_{1}
 , \qquad
e^ { i m_{2} M _{1} } = \cos m_{2} + i\sin m_{2} C_{2}
 , \nonumber
\\
e^ { i m_{3} M _{3} } = \cos m_{3} + i\sin m_{3} A_{3}
 . 
\end{gather*}

One additional note should be made. In the recent paper by
A.~Gsponer \cite{Gsponer-1} on the quaternion approach to the
problem of building the f\/inite transformations from $SU(3)$ and
$SU(4)$ an important point was to divide 15 basis $4\times4$
matrices into three sets:
\begin{enumerate}
\itemsep=0pt \item[]set $A$ of antisymmetrical matrices,
\item[]set $S$ of symmetrical matrices, \item[]set $D$ of diagonal
traceless ones.
\end{enumerate}

 It is easily seen that
\begin{enumerate}
\itemsep=0pt \item[]set $A = \{ \alpha_{i} \oplus \beta_{i} \} $ ;
\item[]set $S = \{ A_{2}, A_{3}, B_{1} , B_{3}, C_{1} , C_{2} \} =
\{ {\boldsymbol L} \oplus {\boldsymbol M} \} $; \item[] set $D =
\{ A_{1}, B_{2} , C_{3} = {\boldsymbol K} \} $.
\end{enumerate}

Turning again to relationship (\ref{C.5}), let us rewrite it as
follows
\begin{gather*}
S = e^ {i \vec{a} \vec{\alpha} } [ e^{i {\boldsymbol k}
{\boldsymbol K} } e^{i {\boldsymbol l} {\boldsymbol L} } e^{i
{\boldsymbol m} {\boldsymbol M} }
 ] e^ {i \vec{b} \vec{\beta} }
 .
\end{gather*}
which exactly corresponds to the structure used in
\cite{Gsponer-1} in connection with the
 Lanczos decomposition theorem \cite{Lanczos}.

Several last comments should be made. On the basis of 15 matrices
\begin{gather*}
 \alpha_{1} , \qquad \alpha_{2}, \qquad \alpha_{3} , \qquad
\beta_{1} , \qquad \beta_{2}, \qquad \beta_{3} , \\
 A_{1}= \alpha_{1} \beta_{1} , \qquad
 B_{1}= \alpha_{1} \beta_{2} ,\qquad
 C_{1}=\alpha_{1} \beta_{3} , \\
 A_{2}= \alpha_{2} \beta_{1} , \qquad
 B_{2}= \alpha_{2} \beta_{2} , \qquad
 C_{2}= \alpha_{2} \beta_{3} ,\\
A_{3}= \alpha_{3} \beta_{1} ,\qquad
 B_{3}= \alpha_{3} \beta_{2} , \qquad
C_{3}= \alpha_{3} \beta_{3}
\end{gather*}
one can easily see the following 20 ways to separate $SU(2)$
subgroups (certainly, those arise at real-valued parameters;
complex-valued parameters give rise to linear subgroups
$GL(2,C)$):
\begin{gather}
(\alpha_{1} , \alpha_{2}, \alpha_{3} ) , \qquad (\beta_{1} ,
\beta_{2}, \beta_{3} ) , \nonumber
\\
(\alpha_{1}, A_{2}, A_{3}) , \qquad (A_{1}, \alpha_{2}, A_{3})
 , \qquad (A_{1}, A_{2} , \alpha_{3}) , \nonumber
\\
(\alpha_{1}, B_{2}, B_{3}) , \qquad (B_{1}, \alpha_{2}, B_{3})
 , \qquad (B_{1},B_{2}, \alpha_{3}) , \nonumber
\\
(\alpha_{1}, C_{2}, C_{3}) , \qquad (C_{1}, \alpha_{2}, C_{3})
 , \qquad (C_{1},C_{2}, \alpha_{3}) , \nonumber
\\
(\beta_{1}, B_{1}, C_{1}) , \qquad (\beta_{1}, B_{2}, C_{2}) ,
\qquad (\beta_{1}, B_{3}, C_{3}) , \nonumber
\\
(A_{1}, \beta_{2}, C_{1}) , \qquad (A_{2}, \beta_{2}, C_{2}) ,
\qquad (A_{3}, \beta_{2}, C_{3}) , \nonumber
\\
(A_{1}, B_{1},\beta_{3} ) , \qquad (A_{2}, B_{2}, \beta_{3}) ,
\qquad (A_{3}, B_{3}, \beta_{3}) . \label{su-2}
\end{gather}
Certainly, they provide us with twenty dif\/ferent
 3-parametric solutions of the unitarity equations~(\ref{1.3}).
 Such 3-subgroups might be used as bigger elementary blocks in constructing of a general transformation [25, 28].

For instance, for the variant from (\ref{su-2}): $ (\alpha_{1},
A_{2}, A_{3}) \Longrightarrow (a_{1}, X_{2},X_{3}) $
 the general multiplication law (\ref{B.7'}) gives
\begin{gather*}
\gamma'' = \gamma ' \gamma + a'_{1} a_{1} +
 X'_{2} X_{2} + X'_{3} X_{3}
 , \qquad
a_{1}'' = \gamma' a_{1} + a_{1}' \gamma + i ( X_{2} '
 X_{3} - X_{3} ' X_{2}) , \nonumber
\\
a_{2}'' = 0 , \qquad
 a_{3}'' = 0 , \qquad b_{1}'' = 0 , \qquad b_{2}'' = 0 , \qquad b_{3}'' = 0 , \qquad
X_{1} '' = 0 ,
\\
X_{2} '' = \gamma' X_{2} + \gamma X'_{2} + + i ( - a'_{1} X_{2}) +
a_{1} X'_{3}) , \qquad X_{3} '' = \gamma' X_{3} + \gamma X'_{3} +
i (a'_{1} X_{2} - a_{1} X'_{2} ) , \nonumber
\\
Y_{1} '' = 0 , \qquad Y_{2} '' = 0 , \qquad Y_{3} '' = 0 , \qquad
Z_{1} '' = 0 , \qquad Z_{2} '' = 0 , \qquad Z_{3} '' = 0 ,
\nonumber
\end{gather*}
that is
\begin{gather*}
\gamma'' = \gamma ' \gamma + a'_{1} a_{1} +
 X'_{2} X_{2} + X'_{3} X_{3}
 , \qquad
a_{1}'' = \gamma' a_{1} + a_{1}' \gamma + i ( X_{2} '
 X_{3} - X_{3} ' X_{2}) ,   \nonumber
\\
X_{2} '' = \gamma' X_{2} + \gamma X'_{2} + i ( X'_{3} a_{1} -
a'_{1} X_{2}) , \qquad X_{3} '' = \gamma' X_{3} + \gamma X'_{3} +
i (a'_{1} X_{2} - a_{1} X'_{2} ) , \nonumber
\end{gather*}
which coincides with equation (\ref{C.6b}). The same can be done
for any other representative from~(\ref{su-2}).

\section[On pseudo-unitary group $SU(2,2)$]{On pseudo-unitary group $\boldsymbol{SU(2,2)}$}\label{sec9}

As said, the Dirac basis was used previously \cite{Barut-2,Kihlberg,Mack',Mack} 
in studying the exponentials for $SU(2,2)$ matrices. Let us show
how the above formalism can apply to this pseudo-unitary group
$SU(2,2)$. Transformations from $SU(2,2)$ should leave invariant
the following form
\begin{gather*}
(z_{1}^{*} , z_{2}^{*}, z_{3}^{*}, z_{4}^{*} ) \left |
\begin{array}{cccc}
+1 & 0 & 0 & 0 \\
0 &+1 & 0 & 0 \\
0 & 0 & -1 & 0 \\
0 & 0 & 0 & -1
\end{array} \right |
\left | \begin{array}{c}
z_{1} \\
z_{2} \\
z_{3} \\
z_{4}
\end{array} \right | , \qquad z^{+} \eta z = z^{'+} \eta z' ,
\end{gather*}
which leads to
\begin{gather*}
z' = U z , \qquad \mbox{where} \qquad U^{+} \eta = \eta U^{-1} ,
\qquad \eta = \left | \begin{array}{cccc}
+1 & 0 & 0 & 0 \\
0 &+1 & 0 & 0 \\
0 & 0 & -1 & 0 \\
0 & 0 & 0 & -1
\end{array} \right | .
\end{gather*}
Any generator $\Lambda'$ of those transformations must obey
relation
\begin{gather*}
U_{k} = e^{ia \Lambda'_{k} } , \qquad (\Lambda' _{k})^{+} \eta =
\eta \Lambda' _{k} , \qquad k \in \{ 1, \dots, 15 \}
\end{gather*}
or allowing for identity $\eta = - \gamma^{5}$
\begin{gather*}
(\Lambda' _{k})^{+} \gamma^{5} = \gamma^{5}
\Lambda'_{k}, \qquad k \in \{ 1, \dots, 15 \} . 
\end{gather*}

All generators $\Lambda'_{k} $ of the group $SU(2,2)$ can be
readily constructed on the basis of the known generators
$\Lambda_{k}$ of $SU(4)$ (see (\ref{A.1a}))
\begin{gather}
\Lambda'_{1} = \Lambda_{1} = \gamma^{5} , \qquad \Lambda'_{1} = i
\Lambda_{1} = i \gamma^{0} , \qquad \Lambda'_{3} = i \Lambda_{3} =
-\gamma^{5} \gamma^{0} , \nonumber
\\
\Lambda'_{4} = i \Lambda_{4} = -\gamma^{1} , \qquad \Lambda'_{5} =
i \Lambda_{5} = i \gamma^{5} \gamma^{1} , \qquad \Lambda'_{6} = i
\Lambda_{6} = -\gamma^{2} , \nonumber
\\
\Lambda'_{7} = i \Lambda_{7} = i\gamma^{5} \gamma^{2} , \qquad
\Lambda'_{8} = i \Lambda_{8} = - \gamma^{3} , \qquad \Lambda'_{9}
= i \Lambda_{9} = i \gamma^{5} \gamma^{3} , \nonumber
\\
\Lambda'_{10} = \Lambda_{10} =2 \sigma^{01} , \qquad \Lambda'_{11}
= \Lambda_{11} = 2 \sigma^{02} , \qquad \Lambda'_{12} =
\Lambda_{12} = 2 \sigma^{03} , \nonumber
\\
\Lambda'_{13} = \Lambda_{13} = 2i\sigma^{12} , \qquad
\Lambda'_{14} = \Lambda_{14} =2i\sigma^{23} , \qquad \Lambda'_{15}
= \Lambda_{15} = 2i\sigma^{31} . \label{D.5}
\end{gather}

Basis elements may be listed as follows:
\begin{gather*}
 \alpha' _{1} = \alpha_{1} = \gamma^{0}\gamma^{2} , \qquad
 \alpha'_{2} = i \alpha_{2} = - \gamma^{0}\gamma^{5} , \qquad \alpha'_{3} = i \alpha_{3} = i \gamma^{5}\gamma^{2}
\nonumber
\\
 (\alpha'_{1})^{2} = I, \qquad (\alpha'_{2})^{2} = -I, \qquad(\alpha'_{3})^{2} = -I,
\nonumber
\\
 \alpha_{2}' \alpha_{3} ' = - i\alpha_{1} , \qquad
 \alpha_{3}' \alpha_{1}' =i\alpha_{2}' , \qquad \alpha_{1} ' \alpha_{2}' = i\alpha_{3} ' ;
\\
\beta'_{1} = \beta_{1} = i \gamma^{3}\gamma^{1} , \qquad
\beta'_{2} = i \beta_{2} = - \gamma^{3} , \qquad \beta'_{3} = i
\beta_{3} = -\gamma^{1} , \nonumber
\\
 (\beta'_{1})^{2} = I, \qquad (\beta'_{2})^{2} = -I, \qquad(\beta'_{3})^{2} = -I,
 \nonumber
 \\
 \beta_{2}' \beta_{3} ' = - i\beta_{1} , \qquad
 \beta_{3}' \beta_{1}' =i\beta_{2}' , \qquad \beta_{1} ' \beta_{2}' = i\beta_{3} ' .
\end{gather*}
These two sets commute with each other: $\alpha_{j} \beta_{k} =
\beta_{k} \alpha_{j} ; $ and their multiplications give remaining
9 elements:
\begin{gather*}
 A_{1}'= A_{1} = \alpha_{1}' \beta_{1}'= - \gamma^{5} , \!\!\qquad
B_{1}' = i B_{1}= \alpha_{1}' \beta_{2}'= i \gamma^{5}\gamma^{1} ,
\!\!\qquad
C_{1}' = i C_{1}=\alpha_{1}' \beta_{3}' = i \gamma^{3} \gamma^{5} , \\
 A_{2}' = i A_{2}= \alpha_{2}' \beta_{1}'= \gamma^{2} , \!\!\qquad
 B_{2}' = - B_{2}= \alpha_{2}' \beta_{2}'= + i \gamma^{1}\gamma^{2} , \!\!\qquad
C_{2} '= - C_{2}= \alpha_{2}' \beta_{3}'= + i \gamma^{2}\gamma^{3} ,\\
A_{3}' = i A_{3}= \alpha_{3}' \beta_{1}'= i \gamma^{0} ,
\!\!\qquad
 B_{3}' = - B_{3}= \alpha_{3}' \beta_{2}'= - \gamma^{0}\gamma^{1} , \!\!\qquad
C_{3}' = -C_{3}= \alpha_{3}' \beta_{3}'= - \gamma^{0}\gamma^{3} .
\end{gather*}

Making in relations (\ref{B.11}) a formal change in accordance
with
\begin{gather*}
e^{ia\Lambda} = \cos a + i \sin a \Lambda , \qquad \Lambda' = i
\Lambda , \nonumber
\\
\cos a + i \sin a \Lambda = e^{ia\Lambda}\quad
\Longrightarrow\quad
 e^{ia\Lambda'}
 = \cos ia + i \sin ia \Lambda = \cosh a - \sinh a \Lambda ,
\nonumber
\end{gather*}

\noindent we arrive at explicit form of elementary pseudo-unitary
$SU(2,2)$-transformations:
\begin{gather}
 \alpha_{1}' = \alpha_{1} = \left | \begin{array}{cc}
\sigma_{2} & 0 \\
0 & -\sigma_{2}
\end{array} \right | , \qquad
U_{1}^{\alpha} =
 \left | \begin{array}{cc}
\cos \phi +i \sin \phi \sigma_{2} & 0 \\
0 & \cos \phi -i \sin \phi \sigma_{2}
\end{array} \right |,
\nonumber
\\
\alpha_{2}' = i \alpha_{2} = \left | \begin{array}{cc}
0 & -1 \\
1 & 0
\end{array} \right | , \qquad
U_{2}^{\alpha} =
 \left | \begin{array}{cc}
\cosh  \chi & - i \sinh  \chi \\
i \sinh  \chi & \cosh  \chi
\end{array} \right |,
\nonumber
\\
\alpha_{3} ' = i \alpha_{3} = \left | \begin{array}{cc}
0 & i \sigma_{2} \\
i \sigma_{2} & 0
\end{array} \right | ,
 \qquad
U_{3}^{\alpha} =
 \left | \begin{array}{cc}
\cosh  \chi & - \sinh  \chi \sigma_{2}\\
\sinh  \chi \sigma_{2} & \cosh  \chi
\end{array} \right | ,
\nonumber
\\
\beta_{1}' = \beta_{1} = \left | \begin{array}{cc}
\sigma_{2} & 0 \\
0 & \sigma_{2}
\end{array} \right | , \qquad
U_{1}^{\beta} =
 \left | \begin{array}{cc}
\cos \phi +i \sin \phi \sigma_{2} & 0 \\
0 & \cos \phi +i \sin \phi \sigma_{2}
\end{array} \right |,
\nonumber
\\
\beta'_{2} = i \beta_{2} = \left | \begin{array}{cc}
0 & \sigma_{3} \\
- \sigma_{3} & 0
\end{array} \right | , \qquad
U_{2}^{\beta} =
 \left | \begin{array}{cc}
\cosh  \chi & i \sinh  \chi \sigma_{3}\\
- i \sinh  \chi \sigma_{3} & \cosh  \chi
\end{array} \right |, \qquad
\nonumber
\\
\beta'_{3} = i\beta_{3} = \left | \begin{array}{cc}
0 & \sigma_{1} \\
- \sigma_{1} & 0
\end{array} \right | ,\qquad
U_{3}^{\beta} =
 \left | \begin{array}{cc}
\cosh  \chi & i \sinh  \chi \sigma_{1}\\
- i \sinh  \chi \sigma_{1} & \cosh  \chi
\end{array} \right | ,
\nonumber
\\
A_{1}' = A_{1} = \left | \begin{array}{cc}
I & 0 \\
0 & -I
\end{array} \right | , \qquad
U_{1}^{A} =
 \left | \begin{array}{cc}
\cos \phi +i \sin \phi & 0\\
0& \cos \phi -i \sin \phi
\end{array} \right |,\qquad
\nonumber
\\
A_{2}' = i A_{2} =
 \left | \begin{array}{cc}
0 & -\sigma_{2} \\
 \sigma_{2} & 0
\end{array} \right | , \qquad
U_{2}^{A} =
 \left | \begin{array}{cc}
\cosh  \chi & - i \sinh  \chi \sigma_{2}\\
i \sinh  \chi \sigma_{2} & \cosh  \chi
\end{array} \right |,
\nonumber
\\
A_{3}' = iA_{3} = \left | \begin{array}{cc}
0 & i \\
i & 0
\end{array} \right | ,\qquad
U_{3}^{A} =
 \left | \begin{array}{cc}
\cosh  \chi & -\sinh  \chi\\
- \sinh  \chi & \cosh  \chi
\end{array} \right | ,
\nonumber
\\
B'_{1} = i B_{1} =
 \left | \begin{array}{cc}
0 & i \sigma_{1} \\
 i \sigma_{1} & 0
\end{array} \right | , \qquad
U_{1}^{B} =
 \left | \begin{array}{cc}
\cosh  \chi & - \sinh  \chi \sigma_{1}\\
- \sinh  \chi \sigma_{1} & \cosh  \chi
\end{array} \right |,
\nonumber
\\
B'_{2} = - B_{2} =
 \left | \begin{array}{cc}
 \sigma_{3} & 0 \\
 0& \sigma_{3}
\end{array} \right | , \qquad
U_{2}^{B} =
 \left | \begin{array}{cc}
\cos \phi +i \sin \phi \sigma_{3} & 0 \\
0 & \cos \phi +i \sin \phi \sigma_{3}
\end{array} \right |,
\nonumber
\\
B'_{3} = -B_{3} =
 \left | \begin{array}{cc}
 \sigma_{1} & 0 \\
 0& -\sigma_{1}
\end{array} \right | ,\qquad
U_{3}^{B} =
 \left | \begin{array}{cc}
\cos \phi +i \sin \phi \sigma_{1} & 0 \\
0 & \cos \phi -i \sin \phi \sigma_{1}
\end{array} \right | ,
\nonumber
\\
C'_{1} = i C_{1} =
 \left | \begin{array}{cc}
0 & -i\sigma_{3} \\
 -i\sigma_{3} & 0
\end{array} \right | , \qquad
U_{1}^{C} =
 \left | \begin{array}{cc}
\cosh  \chi & \sinh  \chi \sigma_{3}\\
\sinh  \chi \sigma_{3} & \cosh  \chi
\end{array} \right |,
\nonumber
\\
C'_{2} = - C_{2} =
 \left | \begin{array}{cc}
 \sigma_{1} & 0 \\
 0& \sigma_{1}
\end{array} \right | , \qquad
U_{2}^{C} =
 \left | \begin{array}{cc}
\cos \phi i \sin \phi \sigma_{1} & 0 \\
0 & \cos \phi +i \sin \phi \sigma_{1}
\end{array} \right |,
\nonumber
\\
C'_{3} = - C_{3} =
 \left | \begin{array}{cc}
 -\sigma_{3} & 0 \\
 0& +\sigma_{3}
\end{array} \right | ,\qquad
U_{3}^{C} =
 \left | \begin{array}{cc}
\cos \phi -i \sin \phi \sigma_{3} & 0 \\
0 & \cos \phi +i \sin \phi \sigma_{3}
\end{array} \right |.
\label{D.9}
\end{gather}

We can easily obtain unitarity equations for $SU(2,2)$ group,
simple solutions to which are given by~(\ref{D.9}). Indeed, taking
into account the formulas (see~(\ref{3a})
 \begin{gather*}
 G^{+} \eta =
 \left | \begin{array}{rr}
k_{0}^{*} + {\boldsymbol k}^{*} \vec{\sigma} & - l_{0}^{*} - {\boldsymbol l}^{*} \vec{\sigma} \\[1mm]
n_{0}^{*} - {\boldsymbol n} ^{*} \vec{\sigma} & m_{0}^{*} -
 {\boldsymbol m}^{*} \vec{\sigma}
\end{array} \right |
\left | \begin{array}{cc} -I & 0 \\0 & +I
\end{array} \right | =
\left | \begin{array}{rr}
-k_{0}^{*} - {\boldsymbol k}^{*} \vec{\sigma} & - l_{0}^{*} - {\boldsymbol l}^{*} \vec{\sigma} \\[1mm]
-n_{0}^{*} + {\boldsymbol n} ^{*} \vec{\sigma} & m_{0}^{*} -
 {\boldsymbol m}^{*} \vec{\sigma}
\end{array} \right |
 ,
\nonumber
\\
\eta G ^{-1} = \left| \begin{array}{cc} -I & 0 \\0 & I
\end{array} \right |
 \left | \begin{array}{rr}
k'_{0} + {\boldsymbol k}' \vec{\sigma} & n'_{0} - {\boldsymbol n}' \vec{\sigma} \\[1mm]
- l'_{0} - {\boldsymbol l}' \vec{\sigma} & m'_{0} - {\boldsymbol
m} ' \vec{\sigma}
\end{array} \right |=
 \left | \begin{array}{rr}
-k'_{0} - {\boldsymbol k}' \vec{\sigma} & -n'_{0} + {\boldsymbol n}' \vec{\sigma} \\[1mm]
- l'_{0} - {\boldsymbol l}' \vec{\sigma} & m'_{0} - {\boldsymbol
m} ' \vec{\sigma}
\end{array} \right |
 ,
\nonumber
\end{gather*}
 from $G^{+} \eta = \eta G^{-1}$ we produce
\begin{gather*}
k_{0}^{*} = k'_{0} , \qquad {\boldsymbol k}^{*} = {\boldsymbol k}'
, \qquad m_{0}^{*} = m'_{0} , \qquad {\boldsymbol m}^{*} =
{\boldsymbol m}' , \nonumber
\\
l_{0}^{*} = n'_{0} , \qquad {\boldsymbol l}^{*} = -{\boldsymbol
n}' , \qquad n_{0}^{*}= l'_{0} , \qquad {\boldsymbol n}^{*} =
-{\boldsymbol
l}' . 
\end{gather*}
These relations dif\/fer from analogous ones (\ref{1.2}) for
$SU(4)$ group only in all signs of the second line. Therefore, the
unitarity conditions for the group $SU(2,2)$ are (compare with
(\ref{1.3}))
\begin{gather*}
k_{0}^{*} = + k_{0} (mm) + m_{0} (ln) + l_{0} (nm)
 - n_{0} (lm) + i {\boldsymbol l} ({\boldsymbol m} \times {\boldsymbol n} ) ,
\nonumber
\\
m_{0}^{*} = + m_{0} (kk) + k_{0} (nl) + n_{0} (lk)
 - l_{0} (nk) - i {\boldsymbol n} ( {\boldsymbol k} \times {\boldsymbol l} ) ,
\nonumber
\\
{\boldsymbol k}^{*} =
 - {\boldsymbol k} (mm) - {\boldsymbol m} (ln) - {\boldsymbol l} (nm) + {\boldsymbol n } (lm) +
 2 {\boldsymbol l} \times ({\boldsymbol n} \times {\boldsymbol m})
 \nonumber
 \\
\phantom{{\boldsymbol k}^{*} =}{} +
 i m_{0} ( {\boldsymbol n} \times {\boldsymbol l} ) + i l_{0} ( {\boldsymbol n} \times {\boldsymbol m} ) + i
n_{0} ( {\boldsymbol l} \times {\boldsymbol m} ) , \nonumber
\\
{\boldsymbol m}^{*} = - {\boldsymbol m} (kk) - {\boldsymbol k}
(nl) - {\boldsymbol n} (lk) + {\boldsymbol l } (nk) +
 2 {\boldsymbol n} \times ( {\boldsymbol l} \times {\boldsymbol k})
 \nonumber
 \\
\phantom{{\boldsymbol m}^{*} =}{} - i k_{0} ({\boldsymbol l}
\times {\boldsymbol n}) - in_{0} ({\boldsymbol l} \times
{\boldsymbol k}) - i l_{0} ( {\boldsymbol n} \times {\boldsymbol
k}) , \nonumber
\\
 -l_{0}^{*} = + k_{0} (nm) - m_{0}
(kn) + l_{0} (nn) + n_{0} (km) +
 i {\boldsymbol k} ({\boldsymbol n} \times {\boldsymbol m} ) ,
 \nonumber
 \\
-n_{0}^{*} = + m_{0} (lk) - k_{0} (ml) + n_{0} (ll) + l_{0} (mk) -
i {\boldsymbol m} ( {\boldsymbol l} \times {\boldsymbol k} ) ,
\nonumber
\\
-{\boldsymbol l}^{*} = - {\boldsymbol k} (nm) + {\boldsymbol m}
(kn) - {\boldsymbol l} (nn) - {\boldsymbol n} (km) + 2
{\boldsymbol k} \times ( {\boldsymbol m} \times {\boldsymbol n} )
\nonumber
\\
\phantom{-{\boldsymbol l}^{*} =}{} + i k_{0} ({\boldsymbol m}
\times {\boldsymbol n}) + i m_{0} ({\boldsymbol k} \times
{\boldsymbol n}) + i n_{0} ({\boldsymbol m} \times {\boldsymbol
k}) , \nonumber
\\
- {\boldsymbol n}^{*} = - {\boldsymbol m} (kl) + {\boldsymbol k}
(ml) - {\boldsymbol n} (ll) - {\boldsymbol l} (mk) + 2{\boldsymbol
m} \times ( {\boldsymbol k} \times {\boldsymbol l} ) - i m_{0}
({\boldsymbol k} \times {\boldsymbol l} ) \nonumber
\\
\phantom{- {\boldsymbol n}^{*} =}{} - i k_{0} ({\boldsymbol m}
\times {\boldsymbol l} ) - i l_{0} ({\boldsymbol k} \times
{\boldsymbol
m}) . 
\end{gather*}

Several words on factorization $SU(2,2)= SU(1,1) \times [ K \times
 L \times M ]\times SU(1,1)$ and a further group f\/ine-structure for $SU(2,2)$.
On the basis of 9 matrices
\begin{gather*}
 A_{1}' = A_{1}= \alpha_{1}'
\beta_{1}'= - \gamma^{5} , \qquad B_{1}' = i B_{1}= \alpha_{1}'
 \beta_{2}' = i \gamma^{5}\gamma^{1} ,\qquad
C_{1}' = i C_{1}=\alpha_{1}' \beta_{3}' =i \gamma^{3} \gamma^{5} , \\
A_{2}' = i A_{2}= \alpha_{2}' \beta_{1}'= \gamma^{2} , \qquad
B_{2}' = - B_{2}= \alpha_{2}' \beta_{2}'= + i \gamma^{1}\gamma^{2}
, \qquad
C_{2}' = - C_{2}= \alpha_{2}' \beta_{3}'= i \gamma^{2}\gamma^{3} ,\\
A_{3}' = i A_{3}= \alpha_{3}' \beta_{1}' = i\gamma^{0} , \!\qquad
B_{3}' = - B_{3}= \alpha_{3}' \beta_{2}'= - \gamma^{0}\gamma^{1} ,
\!\qquad C_{3}'= - C_{3}= \alpha_{3}' \beta_{3}'= -
\gamma^{0}\gamma^{3}
\end{gather*}
one can construct three 3-dimensional subsets (omitting three
others):
\begin{gather*}
{\boldsymbol K}' = \{ A_{1}' = A_{1}= \alpha_{1}' \beta_{1}' , \
B_{2}' = -B_{2}= \alpha_{2}' \beta_{2}' , \ C_{3}' = - C_{3}=
\alpha_{3}' \beta_{3} ' \} , \nonumber
\\
{\boldsymbol L}' = \{ C_{1}' = i C_{1} = \alpha_{1}' \beta_{3}' ,
\ A_{2}' = i A_{2}= \alpha_{2}' \beta_{1}' , \ B_{3} = - B_{3}=
\alpha_{3}' \beta_{2}' \} , \nonumber
\\
{\boldsymbol M}' = \{ B_{1}' = i B_{1} = \alpha_{1}' \beta_{2}' ,
\ C_{2}'= - C_{2} = \alpha_{2}' \beta_{3}' , \ A_{3}' = i A_{3} =
\alpha_{3}' \beta_{1}' \} ,
\end{gather*}
with the same commutation relations:
\begin{gather*}
\Gamma_{1} \Gamma_{2} = + \Gamma_{3} , \qquad \Gamma_{2}
\Gamma_{1} = + \Gamma_{3} , \qquad \Gamma_{1} \Gamma_{2} -
\Gamma_{2} \Gamma_{1} =0 , \qquad \Gamma_{1} \Gamma_{2} +
\Gamma_{2} \Gamma_{1} = +2 \Gamma_{3} ,
\end{gather*}
and analogous ones by cyclic symmetry. Arbitrary element from
$SU(2,2)$ can be factorized as follows
\begin{gather*}
S = e^ {i \vec{a} \vec{\alpha}' } e^ {i \vec{b} \vec{\beta}' }
e^{i {\boldsymbol k} {\boldsymbol K}' } e^{i {\boldsymbol l}
 {\boldsymbol L}' } e^{i {\boldsymbol m} {\boldsymbol M} '} , 
\end{gather*}
 all parameters are real-valued.
Let us specify the group law for these 5 sub-sets. Two groups $e^
{i \vec{a} \vec{\alpha}' }$, $e^ {i \vec{b}
 \vec{\beta} '} $
 are isomorphic so one can consider only the f\/irst one:
\begin{gather*}
e^ {i \vec{a} \vec{\alpha}' } = I + i (a_{1} \alpha'_{1} + a_{2}
\alpha'_{2} + a_{3} \alpha'_{3}) - \tfrac{1}{2!} (a_{1}^{2} -
a_{2} ^{2} - a_{3} ^{2} ) \nonumber
\\
\phantom{e^ {i \vec{a} \vec{\alpha}' } =}{} -i \tfrac{1}{3!}
(a_{1}^{2} - a_{2} ^{2} - a_{3} ^{2} ) (a_{1} \alpha'_{1} + a_{2}
\alpha'_{2} + a_{3} \alpha'_{3}) + \tfrac{1}{4 !} (a_{1}^{2} -
a_{2} ^{2} - a_{3} ^{2} )^{2} +\cdots
 . \nonumber
\end{gather*}
In the variables
\begin{gather*}a_{i} = a n_{i} , \qquad
n_{1}^{2} - n_{2} ^{2} - n_{3} ^{2} = 1 \nonumber
\end{gather*}
we have
\begin{gather*}
e^ {i \vec{a} \vec{\alpha}' } =
 I + i (n_{1} \alpha'_{1} + n_{2} \alpha'_{2} + n_{3} \alpha'_{3}) a -
\tfrac{1}{2!} a^{2} -i \tfrac{1}{3!} (n_{1} \alpha'_{1} + n_{2}
\alpha'_{2} +
n_{3} \alpha'_{3}) a^{3}\\
\phantom{e^ {i \vec{a} \vec{\alpha}' } =}{} + \tfrac{1}{4 !} a^{4}
+ \tfrac{1}{5!} (n_{1} \alpha'_{1} + n_{2} \alpha'_{2} + i n_{3}
\alpha'_{3}) a^{5} - \tfrac{1}{6!} a^{6} + \cdots, \nonumber
\end{gather*}
that is
\begin{gather*}
e^ {i \vec{a} \vec{\alpha}' }
 = \cos a + i \sin a (n_{1} \alpha'_{1} + n_{2} \alpha'_{2} +
n_{3} \alpha'_{3})  = x_{0} - i x_{1} \alpha_{1}' -i x_{2}
\alpha_{2}' - i x_{3} \alpha_{3}' . \nonumber
\end{gather*}
Multiplying two matrices we arrive at
\begin{gather*}
x''_{0} = x'_{0}x_{0} - x'_{1}x_{1} + x'_{2}x_{2} + x'_{3}x_{3} ,
\qquad x''_{1} = x'_{0}x_{1} + x'_{1} x_{0} - ( x'_{2} x_{3} -
x'_{2} x_{3}) , \nonumber
\\
x''_{2} = x'_{0} x_{2} + x'_{2} x_{0} + ( x'_{3} x_{1} - x'_{1}
x_{3}) , \qquad x''_{3} = x'_{0}x_{3} + x'_{3} x_{0} + ( x'_{1}
x_{2} - x'_{2}
x_{1}) . 
\end{gather*}
The inverse matrix looks
\begin{gather*}
(x_{0}, {\boldsymbol x})^{-1} = (x_{0}, -{\boldsymbol x}) .
\nonumber
\end{gather*}
Parameters $(x_{0}, x_{i})$ should obey the following condition
\begin{gather*}
x_{0}^{2} + x_{1}^{2} - x_{2}^{2} - x_{3}^{2} = 1. \nonumber
\end{gather*}
In the variables $a$, $n_{i}$ it will look
\begin{gather*}
\cos^{2} a + \sin^{2} a ( n_{1}^{2} - n_{2}^{2} - n_{3}^{2}) = 1 ,
\qquad n_{1}^{2} - n_{2}^{2} - n_{3}^{2}=1 . \nonumber
\end{gather*}

For three particular cases we will have:
\begin{gather*}
{\boldsymbol n} = (1, 0 , 0 ), \qquad \cos^{2} a+ \sin^{2} a = 1 ,
\qquad e^{ia\alpha'_{1}} = \cos a + i \sin a \alpha'_{1} , \qquad
a \in R ; \nonumber
\\
{\boldsymbol n} = (0, i , 0 ), \qquad \cos^{2}a - \sin^{2} a = 1 ,
\nonumber
\\
 a = ib   ,\qquad \cos ib = \cosh  b , \qquad
\sin ib = i \sinh  b , \qquad b \in R , \nonumber
\\
e^{ia\alpha'_{2}} = \cos a + i \sin a n_{2} \alpha_{2} ' = \cosh
b - i \sinh  b \alpha_{2} ' ;
\\
{\boldsymbol n} = (0, 0 , i ), \qquad \cos^{2}a - \sin^{2} a = 1 ,
\nonumber
\\
a = ib ,\qquad \cos ib = \cosh  b , \qquad \sin ib = i \sinh  b ,
\qquad b \in R , \nonumber
\\
e^{ia\alpha'_{3}} = \cos a + i \sin a n_{3} \alpha_{3} ' = \cosh
b - i \sinh  b \alpha_{3} ' .
\end{gather*}

Now let us turn to f\/inite transformations from remaining
sub-sets ${\boldsymbol K}'$, ${\boldsymbol L}'$, ${\boldsymbol M}'
$. Each of tree subgroups can be constructed as multiplying of
elementary 1-parametric commuting transformations. Their explicit
forms are:

subgroup $K'$
\begin{gather*}
{\boldsymbol K}' = \{ A_{1}' = A_{1}= \alpha_{1}' \beta_{1}' , \
B_{2}' = -B_{2}= \alpha_{2}' \beta_{2}' , \ C_{3}' = - C_{3}=
\alpha_{3}' \beta_{3} ' \} , \nonumber
\\
A_{1}' = \left | \begin{array}{cc} I & 0 \\0 & -I \end{array}
\right |,
 \qquad
 B_{2}' = \left | \begin{array}{cc}
\sigma_{3} & 0 \\0 & \sigma_{3} \end{array} \right |,
 \qquad
 C_{3} ' = \left | \begin{array}{cc}
-\sigma_{3} & 0 \\0 & \sigma_{3} \end{array} \right | ;
\end{gather*}

subgroup $L'$
\begin{gather*}
{\boldsymbol L}' = \{ C_{1}' = i C_{1}= \alpha_{1}' \beta_{3}' , \
A_{2}' = i A_{2}= \alpha_{2}' \beta_{1}' , \ B_{3}' = -B_{3}=
\alpha_{3}' \beta_{2}' \} , \nonumber
\\
C_{1} ' = \left | \begin{array}{cc}
0 & - i\sigma_{3} \\
-i\sigma_{3} & 0
\end{array} \right |, \qquad
A_{2} '= \left | \begin{array}{cc}
0 & - \sigma_{2} \\
\sigma_{2} & 0
\end{array} \right |, \qquad
 B_{3}' = \left | \begin{array}{cc}
\sigma_{1} & 0 \\0 & -\sigma_{1} \end{array} \right | ;
\end{gather*}

subgroup $M'$
\begin{gather*}
{\boldsymbol M}' = \{ B_{1} '= i B_{1} = \alpha_{1} \beta_{2} , \
C_{2}' = - C_{2} = \alpha_{2} \beta_{3} , \ A_{3}' = i A_{3} =
\alpha_{3} \beta_{1} \} , \nonumber
\\
B_{1}' = \left | \begin{array}{cc}
0 & i\sigma_{1} \\
i\sigma_{1} & 0
\end{array} \right |, \qquad
 C_{2}' = \left | \begin{array}{cc}
\sigma_{1} & 0 \\0 & \sigma_{1} \end{array} \right |, \qquad
A_{3}' = \left | \begin{array}{cc}
0 & i \\
i & 0
\end{array} \right |.
\end{gather*}

On the basis of 15 matrices
\begin{gather*}
 \alpha_{1}' , \qquad \alpha_{2}', \qquad \alpha_{3} ' , \qquad
\beta_{1}' , \qquad \beta_{2}', \qquad \beta_{3} ' , \nonumber
\\
 A_{1}'= \alpha_{1}' \beta_{1}' , \qquad
 B_{1}'= \alpha_{1}' \beta_{2}' ,\qquad
 C_{1}'=\alpha_{1}' \beta_{3}' , \\
 A_{2}'= \alpha_{2}' \beta_{1}' , \qquad
 B_{2}'= \alpha_{2}' \beta_{2}' , \qquad
 C_{2}'= \alpha_{2}' \beta_{3}' ,\\
A_{3}'= \alpha_{3}' \beta_{1}' ,\qquad
 B_{3}'= \alpha_{3}' \beta_{2}' , \qquad
C_{3}'= \alpha_{3}' \beta_{3}'
\end{gather*}
one can easily see the following 20 ways to separate $SU(1,1)$
subgroups:
\begin{gather*}
(\alpha_{1}' , \alpha_{2}', \alpha_{3}' ) , \qquad (\beta_{1}' ,
\beta_{2}', \beta_{3}' ) , \nonumber
\\
(\alpha_{1}', A_{2}', A_{3}') , \qquad (A_{1}', \alpha_{2}',
A_{3}') , \qquad (A_{1}', A_{2}' , \alpha_{3}') , \nonumber
\\
(\alpha_{1}', B_{2}', B_{3}') , \qquad (B_{1}', \alpha_{2}',
B_{3}') , \qquad (B_{1}',B_{2}', \alpha_{3}') , \nonumber
\\
(\alpha_{1}', C_{2}', C_{3}') , \qquad (C_{1}', \alpha_{2}',
C_{3}') , \qquad (C_{1}',C_{2}', \alpha_{3}') , \nonumber
\\
(\beta_{1}', B_{1}', C_{1}') , \qquad (\beta_{1}', B_{2}', C_{2}')
, \qquad (\beta_{1}', B_{3}', C_{3}') , \nonumber
\\
(A_{1}', \beta_{2}', C_{1}') , \qquad (A_{2}', \beta_{2}', C_{2}')
, \qquad (A_{3}', \beta_{2}', C_{3}') , \nonumber
\\
(A_{1}, B_{1},\beta_{3} ) , \qquad (A_{2}, B_{2}, \beta_{3}) ,
\qquad (A_{3}, B_{3}, \beta_{3}) . \label{su-2'}
\end{gather*}
such 3-subgroups might be used as bigger elementary blocks in
constructing a general transformation.

\section[On pseudo-unitary group $SU(3,1)$]{On pseudo-unitary group $\boldsymbol{SU(3,1)}$}\label{sec10}

Let us show how the above formalism can apply to the  pseudo-unitary group $SU(3,1)$.
Transformations from $SU(3,1)$ should leave invariant the following form
\begin{gather*}
(z_{1}^{*} , z_{2}^{*}, z_{3}^{*}, z_{4}^{*} ) \left |
\begin{array}{cccc}
+1 & 0 & 0 & 0 \\
0 &+1 & 0 & 0 \\
0 & 0 & +1 & 0 \\
0 & 0 & 0 & -1
\end{array} \right |
\left | \begin{array}{c}
z_{1} \\
z_{2} \\
z_{3} \\
z_{4}
\end{array} \right | , \qquad z^{+} \eta z = z^{'+} \eta z' ,
\end{gather*}
which leads to
\begin{gather*}
z' = U z , \qquad \mbox{where} \qquad U^{+} \eta = \eta U^{-1} ,
\qquad \eta = \left | \begin{array}{cccc}
+1 & 0 & 0 & 0 \\
0 &+1 & 0 & 0 \\
0 & 0 & +1 & 0 \\
0 & 0 & 0 & -1
\end{array} \right | .
\end{gather*}

\noindent Any generator $\Lambda'$ of those transformations must
obey the relation
\begin{gather*}
U_{k} = e^{ia \Lambda'_{k} } , \qquad (\Lambda' _{k})^{+} \eta =
\eta \Lambda' _{k} . \qquad k \in \{ 1, \dots, 15 \}. \label{D.3}
\end{gather*}
The matrix $\eta$ is a linear combination
\begin{gather*}
\eta = \tfrac 12 ( 2i\sigma^{12} - 2\sigma^{03} - \gamma^{5} + I)
= \tfrac 12 ( i\gamma^{1} \gamma^{2} - \gamma^{0} \gamma^{3}
 + i \gamma^{0} \gamma^{1}\gamma^{2}\gamma^{3}+ I) .
\nonumber
\end{gather*}
All generators of the group $SU(3,1)$ can readily be constructed
on the basis of the known generators $\lambda_{k}$ of $SU(4)$ (see
(\ref{A.3a}))
\begin{gather}
\lambda_{1} , \qquad \lambda_{2} , \qquad \lambda_{3} , \qquad
\lambda_{4} , \qquad \lambda_{5} , \qquad \lambda_{6} , \qquad
\lambda_{7} , \qquad \lambda_{8} , \qquad i \lambda_{9} , \qquad i
\lambda_{10} , \nonumber
\\
i \lambda_{11} , \qquad i \lambda_{12} , \qquad i \lambda_{13} ,
\qquad i \lambda_{14} , \qquad \lambda_{15} ;
\end{gather}
generator $\lambda_{9} , \dots, \lambda_{14}$ are multiplied by
imaginary unit $i$. Instead of $\lambda_{8}$, $\lambda_{15}$ one
can introduce other generator $\lambda_{8}'$, $\lambda_{15}'$
 see (\ref{A.8d}) (diagonal generators are the same for group $SU(4)$, $SU(2,2)$, and $SU(3,1)$.

\section{Discussion}\label{sec11}

Let us summarize the main point of the present treatment.

Parametrization of $4 \times 4$ matrices $G$ of the complex linear
group $GL(4,C)$ in terms of four complex 4-vector parameters
$(k,m,n,l)$ is investigated. Additional restrictions separating
some subgroups of $GL(4,C)$ are given explicitly. In the given
parametrization, the problem of inverting any $4\times 4$ matrix
$G$ is solved. Expression for determinant of any matrix $G$ is
found: $\det G = F(k,m,n,l)$.
 Unitarity conditions $G^{+} = G^{-1}$ have been formulated in the form of
non-linear cubic algebraic equations including complex
conjugation. Several simplest solutions of these unitarity
equations have been found:
 three 2-parametric subgroups $G_{1}$, $G_{2}$, $G_{3}$ --
each of subgroups consists of two commuting Abelian unitary
groups; 4-parametric unitary subgroup consisting of a product of a
3-parametric group isomorphic $SU(2)$ and 1-parametric Abelian
group.

The Dirac basis of generators $\Lambda_{k}$, being of Gell-Mann
type, substantially dif\/fers from the basis
 $\lambda_{i}$ used in the literature on $SU(4)$ group, formulas relating them are found --
 they permit to separate $SU(3)$ subgroup in $SU(4)$.
Special way to list 15 Dirac generators of $GL(4,C)$ can be used
$\{ \Lambda_{k} \} = \{ \alpha_{i} \oplus \beta_{j} \oplus (
\alpha_{i} \beta_{j} =
 {\boldsymbol K} \oplus {\boldsymbol L} \oplus {\boldsymbol M} ) \}$, which permit to factorize $SU(4)$ transformations
 according
 to
 $S =
e^ {i \vec{a} \vec{\alpha} } e^ {i \vec{b} \vec{\beta} } e^{i
{\boldsymbol k} {\boldsymbol K} } e^{i {\boldsymbol l}
{\boldsymbol L} } e^{i {\boldsymbol m}
 {\boldsymbol M} } , $ where two f\/irst factors commute with each
other and are isomorphic to $SU(2)$ group, the three last are
3-parametric groups, each of them consists of three Abelian
commuting unitary subgroups. Besides, the structure of f\/ifteen
Dirac matrices~$\Lambda_{k}$ permits to separate twenty
3-parametric subgroups in $SU(4)$ isomorphic to~$SU(2)$; those
subgroups might be used as bigger elementary blocks in
constructing a general
 transformation~$SU(4)$.
It is shown how one can specify the present approach for the
unitary group~$SU(2,2)$ and~$SU(3,1)$.

In principle, all dif\/ferent approaches used in the literature
are closely related so that any result obtained within one
technique may be easily translated to any other. There is no sense
to persist in exploiting only one representation, thinking that it
is much better than all others. Success should lie in combining
dif\/ferent techniques. For instance, Euler angles-based approach
provides us with the group elements in the separated variables
form, which may be of a supreme importance at calculating matrix
elements of the group. In turn, a factorized subgroup- based
structure is of special interest in the particle physics and gauge
theory of fundamental interaction. Geometrical properties of the
groups, their global structure, dif\/ferences between orthogonal
groups and their double covering, and so on,
 seem to be most easily understood
in terms of bilinear functions in space of linear parameters: $G =
x_{j} \Lambda_{j}$, $x''{j} = e_{jkl} x'_{k}
 x_{l} $.

We have no ground to think that only exponential functions
$e^{i\Lambda}$ are suitable for exploration into group structures.
We may expect that in addition to Euler angles many other
curvilinear coordinates might be of value for studying of the
group structure. For instance, in the case of the group $SO(4,C)$
we have known~34 such coordinate systems owing to Olevskiy
investigation \cite{3} on 3-orthogonal coordinates in real
Lobachevski space.

In conclusion, several words about possible application areas of
the obtained results. The main argument in favor of constructing
the theory of unitary groups $SU(4)$ (and related to it) in terms
of Dirac matrices is the role of spinor methods being widely
adopted in physics. Let us mention several problems most
attractive for authors:

\begin{enumerate}\itemsep=0pt

\item[] $SU(2,2)$ and conformal symmetry, massless particles;

\item[] classical Yang-Mills equations and gauge f\/ields;

\item[] geometric phases for multi-level quantum systems;

\item[] composite structure of quarks and leptons;

\item[] $SU(4)$ gauge models.
\end{enumerate}

In particular, description of the group $SU(2,2)$ in terms of
matrices $\alpha^{j}$, $\beta^{j}$ should be of great benef\/it in
investigation of {\it conformal symmetry in massless particles
theory}. For instance, classical Maxwell equations in a medium can
be presented in 4-dimensional complex matrix form with the use of
two sets of matrices, exploited above:
\[
\left(-i {\partial \over \partial x^{0}} + \alpha^{j} {\partial
\over \partial x^{j}}\right) M +\left(-i {\partial \over \partial
x^{0}} + \beta^{j} {\partial \over \partial x^{j}}\right) N = {1
\over \epsilon_{0}} \left | \begin{array}{c} \rho \\ {\boldsymbol
j}/c
\end{array} \right | ,
\]
where
\begin{gather*}
M = \left | \begin{array}{c} 0 \\ {\boldsymbol M}
\end{array} \right | , \qquad
{\boldsymbol M} = {1 \over \epsilon_{0}} ({\boldsymbol D} +
i{\boldsymbol H}/c ) + ( {\boldsymbol E} + ic{\boldsymbol B}) ,
\\
N = \left | \begin{array}{c} 0 \\ {\boldsymbol N} \end{array}
\right | , \qquad {\boldsymbol N} = {1 \over \epsilon_{0}}
({\boldsymbol D} - i{\boldsymbol H}/c ) - ( {\boldsymbol E} -
ic{\boldsymbol B}) .
\end{gather*}

\subsection*{Acknowledgements}

Authors are grateful to participants of the seminar of Laboratory
of Physics of Fundamental Interaction,
 National Academy of Sciences of Belarus for discussion.
 Authors are grateful to the anonymous reviewer for many comments and advice improving
 the paper.

This work was supported by Fund for Basic Research of Belarus
 F07-314.
 We wish to thank the Organizers of the Seventh International Conference ``Symmetry
in Nonlinear Mathematical Physics'' (June 24--30, 2007, Kyiv) and
ICTP Of\/f\/ice of External Activities for having given us the
opportunity to talk on this subject as well as for local and
travel support.

\pdfbookmark[1]{References}{ref}
\LastPageEnding

\end{document}